\documentclass[article, twocolumn, superscriptaddress,nofootinbib]{revtex4-2}
\usepackage{graphicx}
\graphicspath{ {figures/} }

\usepackage{amsmath}
\usepackage{amssymb}
\usepackage{amsthm}
\usepackage{mathtools}
\usepackage{etoolbox}
\usepackage{dcolumn}  
\usepackage{multirow}
\usepackage{mathrsfs}
\usepackage{braket}
\usepackage{soul} %

\usepackage[version=4]{mhchem}

\usepackage[noend]{algpseudocode}
\usepackage{algorithm}

\usepackage{tabularx}

\newcounter{algosuffix}
\newcounter{algomain}
\newcommand{\thealgowithsuffix}{\arabic{algomain}\ifnum\value{algosuffix}>0\Alph{algosuffix}\fi}

\makeatletter
\newcommand{\algosuffix}[1]{
  \ifnum\value{algosuffix}=0
    \refstepcounter{algomain} %
  \fi
  \setcounter{algosuffix}{#1}
  \edef\@currentlabel{\thealgowithsuffix}
}
\newcommand{\nextalgo}{
  \setcounter{algosuffix}{0}
  \refstepcounter{algomain}
}
\makeatother

\usepackage[svgnames, usenames, dvipsnames]{xcolor}
\usepackage{listings}

\usepackage{hyperref}
\hypersetup{
    hidelinks,
    colorlinks=true,
    allcolors=DarkBlue
}

\AfterEndEnvironment{proposition}{\noindent\ignorespaces}

\newcommand{\pointer}[2]{\left({#1}\rightarrow{#2}\right)}
\newcommand{\AInB}[2]{{#1}\in{#2}}

\newcommand{\Eloc}{\ensuremath{E_{\rm loc}}}
\newcommand{\Elocvar}{\ensuremath{E_{\rm loc}^{\rm var}}}
\newcommand{\Evar}{\ensuremath{E^{\rm var}}}

\newcommand{\Estate}{\ensuremath{\langle E \rangle}}

\newcommand{\xvec}{\ensuremath{\mathbf{x}}}
\newcommand{\xvecp}{\ensuremath{\mathbf{x}^\prime}}

\newcommand{\Hhat}{\ensuremath{\hat{H}}}
\newcommand{\Helem}{\ensuremath{H_{\xvec \xvecp}}}
\newcommand{\Phat}{\ensuremath{\hat{P}}}

\newcommand{\PauliI}{\ensuremath{\hat{I}}}
\newcommand{\PauliX}{\ensuremath{\hat{X}}}
\newcommand{\PauliY}{\ensuremath{\hat{Y}}}
\newcommand{\PauliZ}{\ensuremath{\hat{Z}}}

\newcommand{\Batch}{\ensuremath{\mathcal{B}}}
\newcommand{\Unique}{\ensuremath{\mathcal{U}}}
\newcommand{\UniqueGumbel}{\ensuremath{\mathcal{UG}}}
\newcommand{\UniqueGumbelp}{\ensuremath{\mathcal{UG}^\prime}}

\newcommand{\SampledAndCoupled}{\ensuremath{\mathcal{SC}}}
\newcommand{\SampledAndCoupledx}{\ensuremath{\mathcal{SC}_{\xvec}}}

\newcommand{\PauliSet}{\ensuremath{\mathcal{P}}}
\newcommand{\XYPauliSet}{\ensuremath{\mathcal{P}_{\XYvec}}}
\newcommand{\XYPauliSetSize}{\ensuremath{\left|\mathcal{P}_{\XYvec}\right|}}

\newcommand{\UniqueXYSet}{\ensuremath{\mathcal{XY}}}

\newcommand{\HamYZStarts}{\ensuremath{\pointer{\UniqueXYSet}{\textsc{Index}(\mathcal{P}_{\UniqueXYSet}^{(0)})}}}
\newcommand{\HamYZNums}{\ensuremath{\left|\mathcal{P}_{\UniqueXYSet}\right|}}

\newcommand{\YZStarts}{\ensuremath{\pointer{\XYvec}{\textsc{Index}(\mathcal{P}_{\XYvec}^{(0)})}}}
\newcommand{\YZNums}{\ensuremath{\left|\mathcal{P}_{\XYvec}\right|}}

\newcommand{\SCXYx}{\ensuremath{\mathcal{SCXY}_{\xvec}}}

\newcommand{\Tree}{\ensuremath{\mathcal{T}}}

\newcommand{\UniqueTree}{\ensuremath{\Tree^\Unique}}
\newcommand{\UniqueXYTree}{\ensuremath{\Tree^\UniqueXYSet}}

\newcommand{\UniqueTreeLevel}{\ensuremath{\mathcal{T}^{\Unique}}}    
\newcommand{\UniqueXYTreeLevel}{\ensuremath{\mathcal{T}^{\UniqueXYSet}}}

\newcommand{\FindSampledAndCoupled}{\textsc{FindCoupledPairs}}
\newcommand{\MatrixElement}{\textsc{MatrixElement}}

\newcommand{\CoupleViaUniqueXY}{\textsc{LoopOverTerms}}
\newcommand{\CoupleAllToAll}{\textsc{LoopOverBatch}}
\newcommand{\CoupleViaPrefixTree}{\textsc{LoopOverTrie}}

\newcommand{\ConstructPrefixTree}{\textsc{ConstructPrefixTree}}

\newcommand{\Gather}{\textsc{Gather}}
\newcommand{\ScatterAdd}{\textsc{ScatterAdd}}
\newcommand{\FindAInB}{\textsc{FindAInB}}
\newcommand{\ExpandPointers}{\textsc{ExpandPointers}}

\newcommand{\UniqueFunc}{\textsc{UniqueColumns}}
\newcommand{\src}{\texttt{src}}
\newcommand{\indexttt}{\texttt{index}}
\newcommand{\result}{\texttt{result}}

\newcommand{\Attt}{\texttt{A}}
\newcommand{\Bttt}{\texttt{B}}

\newcommand{\arrttt}{\texttt{array}}

\newcommand{\labelsttt}{\texttt{labels}}
\newcommand{\unqlabelsttt}{\texttt{unq\_labels}}
\newcommand{\unqlabelsinvttt}{\texttt{unq\_labels\_inv}}

\newcommand{\rowonettt}{\texttt{row1}}      
\newcommand{\unqrowonettt}{\texttt{unq\_row1}}      
\newcommand{\unqrowoneinvttt}{\texttt{unq\_row1\_inv}}      

\newcommand{\rowtwottt}{\texttt{row2}}      
\newcommand{\unqrowtwottt}{\texttt{unq\_row2}}      
\newcommand{\unqrowtwoinvttt}{\texttt{unq\_row2\_inv}}  
  	
\newcommand{\candxvecptr}{\ensuremath{\pointer{\xvec?}{\Unique}}}
\newcommand{\xvecptr}{\ensuremath{\pointer{\xvec}{\Unique}}}

\newcommand{\candxvecpptr}{\ensuremath{\pointer{\xvecp?}{\Unique}}}
\newcommand{\xvecpptr}{\ensuremath{\pointer{\xvecp}{\Unique}}}   	

\newcommand{\candxyvecinxy}{\ensuremath{\XYvec?\in\UniqueXYSet}}

\newcommand{\xyvecptr}{\ensuremath{\pointer{\XYvec}{\UniqueXYSet}}}   
\newcommand{\xypaulisettermptr}{\ensuremath{\pointer{\XYPauliSet}{\Phatl}}}  
\newcommand{\xypaulisetxyptr}{\ensuremath{\pointer{\XYPauliSet}{\XYvec}}}    
  	
\newcommand{\UniqueTreeArrOne}[1]{\ensuremath{\pointer{\Unique}{\UniqueTreeLevel_{#1}}}}   
\newcommand{\UniqueTreeArrUp}[1]{\ensuremath{\UniqueTreeLevel_{#1, \uparrow}}}      
\newcommand{\UniqueTreeArrDown}[1]{\ensuremath{\UniqueTreeLevel_{#1, \downarrow}}}     
    
\newcommand{\UniqueXYTreeArrOne}[1]{\ensuremath{\pointer{\UniqueXYSet}{\UniqueXYTreeLevel_{#1}}}}   
      
\newcommand{\UniqueXYTreeArrDown}[1]{\ensuremath{\UniqueXYTreeLevel_{#1, \downarrow}}}    

\newcommand{\LoopTrieArrX}[1]{\ensuremath{\pointer{\xvec_{<#1}}{\Unique}}} 
\newcommand{\LoopTrieArrXp}[1]{\ensuremath{\pointer{\xvecp_{<#1}}{\UniqueTreeLevel_{#1}}}}      \newcommand{\LoopTrieArrXY}[1]{\ensuremath{\pointer{\XYvec_{<#1}}{\UniqueXYTreeLevel_{#1}}}}

\newcommand{\LoopTrieArrXpCand}[1]{\ensuremath{\pointer{\xvecp_{<#1}?}{\UniqueTreeLevel_{#1}}}}      \newcommand{\LoopTrieArrXYCand}[1]{\ensuremath{\pointer{\XYvec_{<#1}?}{\UniqueXYTreeLevel_{#1}}}}  

\newcommand{\memoryttt}{\texttt{memory}}
\newcommand{\startsttt}{\texttt{starts}}
\newcommand{\numsttt}{\texttt{nums}}
\newcommand{\pointersttt}{\texttt{pointers}}
\newcommand{\offsetsttt}{\texttt{offsets}}    
\newcommand{\sawttt}{\texttt{relative}}    
\newcommand{\inclinettt}{\texttt{incline}}    
\newcommand{\platoesttt}{\texttt{platoes}}
 
\newcommand{\PEmptyList}{\ensuremath{\left[\ \right]}}

\newcommand{\PIn}{\ensuremath{\textbf{in}}}

\newcommand{\PFrom}{\ensuremath{\textbf{from}}}
\newcommand{\PTo}{\ensuremath{\textbf{to}}}

\newcommand{\PAppend}[1]{\ensuremath{\texttt{.append}\left(#1\right)}}

\newcommand{\Node}{\textsc{Node}}
\newcommand{\CurNode}{\texttt{current\_node}}
\newcommand{\NewNode}{\texttt{new\_node}}

\newcommand{\Parent}{\texttt{parent}}
\newcommand{\Children}{\texttt{children}}
\newcommand{\Value}{\texttt{value}}
\newcommand{\None}{\texttt{None}}

\newcommand{\UnqNode}{\texttt{u\_node}}
\newcommand{\UnqXYNode}{\texttt{xy\_node}}

\newcommand{\NewUnqNode}{\texttt{new\_u\_node}}
\newcommand{\NewUnqXYNode}{\texttt{new\_xy\_node}}

\newcommand{\ORop}{\oplus}
\newcommand{\ANDop}{\odot}
\newcommand{\XORop}{\otimes}

\newcommand{\PhatN}[1]{\ensuremath{\hat{P}^{(#1)}}}
\newcommand{\Phatl}{\ensuremath{\hat{P}^{(l)}}}
\newcommand{\Phatli}{\ensuremath{\hat{P}_i^{(l)}}}
\newcommand{\Phatlelem}{\ensuremath{\braket{\xvec|\Phatl|\xvecp}}}

\newcommand{\Yvec}{\ensuremath{\mathbf{Y}}}

\newcommand{\XYvec}{\ensuremath{\mathbf{\underline{XY}}}}
\newcommand{\XYvecN}[1]{\ensuremath{\XYvec_{#1}}}
\newcommand{\YZvec}{\ensuremath{\mathbf{\underline{YZ}}}}

\newcommand{\Xvecl}{\ensuremath{\mathbf{X}^{(l)}}}
\newcommand{\Yvecl}{\ensuremath{\mathbf{Y}^{(l)}}}
\newcommand{\Zvecl}{\ensuremath{\mathbf{Z}^{(l)}}}
\newcommand{\XYvecl}{\ensuremath{\mathbf{\underline{XY}}^{(l)}}}
\newcommand{\YZvecl}{\ensuremath{\mathbf{\underline{YZ}}^{(l)}}}

\newcommand{\Xvecli}{\ensuremath{X_i^{(l)}}}

\newcommand{\AnsatzComplexity}{\ensuremath{\mathcal{C}_{\rm NQS}}}
\newcommand{\BitwiseComplexity}{\ensuremath{\mathcal{C}_{\rm bitop}}}

\newcommand{\FindInUComplexity}{\ensuremath{\mathcal{C}_{\in \Unique}}}
\newcommand{\FindInXYComplexity}{\ensuremath{\mathcal{C}_{\in \UniqueXYSet}}}

\newcommand{\psit}{\ensuremath{\psi_{\theta}}}

\newcommand{\Ns}{\ensuremath{N_{\rm s}}}
\newcommand{\Nunq}{\ensuremath{N_{\rm unq}}}
\newcommand{\NSR}{\ensuremath{N_{\rm unq}^{\rm SR}}}

\newcommand{\Nt}{\ensuremath{N_{\rm T}}}

\newcommand{\BitDepth}{\ensuremath{N_{\rm \frac{bits}{int}}}}
\newcommand{\IntPerVec}{\ensuremath{N_{\rm \frac{int}{vec}}}}

\newcommand{\Np}{\ensuremath{N_{\rm p}}}
\newcommand{\QuditDepth}{\ensuremath{N_{\rm \frac{qubit}{qudit}}}}

\newcommand{\Ne}{\ensuremath{n_{\rm e}}}

\newcommand{\EFCI}{\ensuremath{E_{\rm FCI}}}
\newcommand{\ChemAcc}{\ensuremath{\Delta E_{\rm chem.acc.}}}
\newcommand{\IPR}{\ensuremath{\mathrm{IPR}}}
\newcommand{\HilbSize}{\ensuremath{\left|\mathcal{H}\right|}}

  	\newcommand{\WeightOne}{0.9}
\newcommand{\WeightTwo}{0.1}
  	\newcommand{\WeightThree}{-0.2}
  	\newcommand{\WeightFour}{-0.2}
	\newcommand{\WeightFive}{0.3}
	
	\newcommand{\PhatOne}{\PauliI}
\newcommand{\PhatTwo}{\ensuremath{\PauliZ_1 \PauliZ_2}}
\newcommand{\PhatThree}{\ensuremath{\PauliX_0 \PauliX_2}}
\newcommand{\PhatFour}{\ensuremath{\PauliX_1 \PauliX_3}}
\newcommand{\PhatFive}{\ensuremath{\PauliY_1 \PauliY_2}}

\newcommand{\topk}{top-$K$}

\renewcommand{\Re}{\operatorname{Re}}

\newcommand{\mathexp}{\mathop{\mathbb{E}}}

\newcommand{\imagi}{\mathrm{i}}
\newcommand{\expe}{\mathrm{e}}

\newcommand{\etal}{\textit{et al.}}
\newcommand{\abinit}{\emph{ab initio}}
\newcommand{\Abinit}{\emph{Ab initio}}

\begin{document}
	\title{Neural quantum states and peaked molecular wave functions: curse or blessing?}
	\author{Aleksei Malyshev}	
	\email{aleksei.malyshev@physics.ox.ac.uk}
        \affiliation{University of Oxford, Clarendon Laboratory, Parks Road, Oxford,  OX1 3PU, UK}
        
    \author{Markus Schmitt}	
	\affiliation{Forschungszentrum J\"ulich GmbH, Peter Gr\"unberg Institute,
Quantum Control (PGI-8), 52425 J\"ulich, Germany}
    \affiliation{University of Regensburg, 93053 Regensburg, Germany}

	\author{A.\ I.\ Lvovsky}
	\affiliation{University of Oxford, Clarendon Laboratory, Parks Road, Oxford,  OX1 3PU, UK}	

\begin{abstract}
The field of neural quantum states has recently experienced a tremendous progress, making them a competitive tool of computational quantum many-body physics. 
However, their largest achievements to date mostly concern interacting spin systems, while their utility for quantum chemistry remains yet to be demonstrated.
Two main complications are the peaked structure of the molecular wave functions, which 
impedes sampling, and large number of terms in second quantised Hamiltonians, which 
hinders scaling to larger molecule sizes.
In this paper we address these issues jointly and argue that the peaked structure might actually be key to drastically more efficient calculations.
Specifically, we introduce a novel algorithm for autoregressive sampling without replacement and a procedure to calculate a computationally cheaper surrogate for the local energy.
We complement them with a custom modification of the stochastic reconfiguration optimisation technique and a highly optimised GPU implementation.
As a result, our calculations require substantially less resources and exhibit more than order of magnitude speedup compared to the previous works.
On a single GPU we study molecules comprising up to 118 qubits and outperform the ``golden standard'' CCSD(T) benchmark in Hilbert spaces of $\sim 10^{15}$ Slater determinants, which is orders of magnitude larger than what was previously achieved.
We believe that our work underscores the prospect of NQS for challenging quantum chemistry calculations and serves as a favourable ground for the future method development.
 
\end{abstract}
	
\maketitle

\section{Introduction}\label{sec:intro}
The past decade witnessed an explosive growth of deep learning applications stipulated by a remarkable ability of neural networks to efficiently represent complicated multidimensional probability distributions.
To exploit these capabilities in the context of computational quantum many-body physics, Carleo and Troyer pioneered the idea of \emph{neural quantum states (NQS)}, i.e., an efficient representation of a quantum state with a generative neural network~\cite{carleo_troyer_rbm}.
Since its conception the field flourished considerably \cite{medvidovic_nqs_review,lange_nqs_review,Hermann2023} and now neural quantum states are capable to tackle various problems, including low energy states~\cite{carleo_symmetries, carleo_convolutional, hibbat_allah_recurrent, transformer_j1_j2, bohrdt_doped_antiferromagnets,nomura_rbm_j1_j2, roth_group_convolutional,pei_bose_hubbard,nqs_nuclei_1, nqs_nuclei_2, nqs_nuclei_3, fermi_net, pauli_net, nekluydov_wasserstein, choo_nqs_j1_j2,  barrett_autoregressive_qchem, zhao_scalable_qchem, transformer_nqs_for_qchem,  malyshev_anqs_with_quantum_numbers, yoshioka_crystalline_rbm, liu_backflow_quantum_chemistry}, unitary time evolution~\cite{schmitt_dynamics_in_2d,Schmitt2022,zachary_anqs_dynamics,sinibaldi_unbiasing,MendesSantos2023,MendesSantos2024,nnqs_for_quantum_computing, matija_nqs_for_qc}, open systems~\cite{neural_density_operators,carleo_open_systems, Reh2021,vicentini_open_systems, autoregressive_open_systems} and quantum state tomography~\cite{nnqs_pure_tomography, tiunov_tomography, murali_tomography, fedotova_tomography}, placing them among state-of-the-art methods of computational quantum many-body physics.

A key advantage of NQS in comparison to alternative methods is the fact that the wave function representation can in principle be efficient irrespective of the interaction graph of the underlying physical system. Therefore, NQS are a promising candidate to tackle the paradigmatic problem of \abinit{} quantum chemistry to calculate the ground states of molecules in second quantisation.
Finding an algorithmic solution to this problem is of outstanding interest, since having access to the ground state of a molecule allows one to deduce most of the physical and chemical properties of the latter.
As for any quantum many-body problem, computational approaches are hindered by exponential growth of the Hilbert space with respect to the molecule size, and thus a trade off between the solution accuracy and its efficiency is inevitable.
A rich computational toolbox has been elaborated in quantum chemistry, with different methods relying on different approximations, and no method performing universally well across a multitude of molecular systems.
Thus, the development of versatile and widely applicable techniques remains a pivotal quest in the field. 

However, ever since the first work by Choo~\etal{}~\cite{carleo_quantum_chemistry} on NQS application to \emph{ab initio} quantum chemistry two outstanding challenges were recognised.
First, the molecular ground state wave functions are often of peaked structure, i.e.~dominated by few high amplitude components in the computational basis, which constitutes a major obstacle for sampling the Born distribution.
Second, molecular Hamiltonians have excessive number of terms, which renders the stochastic estimation of energy highly expensive and poses a formidable challenge of computational nature.
A range of subsequent works focused on gradually lifting these limitations by employing better sampling techniques~\cite{barrett_autoregressive_qchem}, streamlining the energy calculations~\cite{transformer_nqs_for_qchem}, parallelising the computational load~\cite{transformer_nqs_for_qchem, zhao_scalable_qchem}, or giving up sampling altogether and resorting to deterministic optimisation~\cite{li_deterministic_nqs_qchem}.
While such collective effort pushed the boundary of what is in principle achievable with NQS to system sizes of several dozens to hundred qubits, competitive energies were so far reported only for systems still amenable to exact diagonalisation.
Thus, at its current state the field lacks a compelling evidence suggesting that NQS can \emph{both} scale up to challenging system sizes and achieve state-of-the-art energies there.

In this work, we supplement NQS quantum chemistry calculations with a set of advances of both conceptual and practical nature aimed at improving its computational performance. As a result, we obtain an optimisation procedure which provides orders of magnitude speedup as compared to the conventional NQS optimisation and yet retains the same accuracy level.
Using only a single GPU, we surpass  conventional quantum chemistry methods such as CISD, CCSD and CCSD(T) for system sizes that have been previously   inaccessible to NQS (118 qubits, $10^{15}$ Slater determinants, $~3\cdot 10^6$ Hamiltonian terms).

At a high level, we leverage the previously unwelcome peakedness of the wave function. Our contributions include (i) a previously unexplored algorithm for autoregressive sampling without replacement; (ii) two novel approaches to energy evaluation with improved asymptotic complexity; (iii) a modification of the stochastic reconfiguration (SR) optimisation technique tailored to ANQS; (iv) compressed data representation enabling memory and compute efficient GPU implementation.

The remainder of this paper is organised as follows.
We provide a relevant background on variational Monte Carlo, quantum chemistry in second quantised formalism and autoregressive neural quantum states in Section~\ref{sec:background}.
In Section~\ref{sec:high_level_picture} we overview at a high-level our contributions, while their technical details are left for Methods.
In Section~\ref{sec:results} we present results of NQS quantum chemistry calculations for a range of molecules beyond exact diagonalisation, showcasing an improved computational performance of the method.
Finally, in Section~\ref{sec:conclusion} we conclude our work and indicate directions for future research.

\section{Background}\label{sec:background}

\subsection{\Abinit{} quantum chemistry}\label{sec:quatum_chemistry}
In second quantisation molecular Hamiltonians take the following general form:
\begin{align}
    \hat H_{\rm SQ}=\sum_{i,j}t_{ij}(\hat c_i^\dagger\hat c_j+h.c.)+\sum_{i,j,k,l}V_{i,j,k,l}\hat c_i^\dagger\hat c_j^\dagger\hat c_k\hat c_l\,
    \label{eq:fermionic_qc}
\end{align}
where fermionic creation and annihilation operators $\hat c_i^\dagger, \hat c_i$ act on a set of $N$ molecular orbitals.
In the field of NQS one usually applies the Jordan-Wigner transformation and maps orbitals to a system of $N$ qubits~\cite{carleo_quantum_chemistry, barrett_autoregressive_qchem, zhao_scalable_qchem}.
This produces a Hamiltonian of the following form:
\begin{equation}\label{equ:jw_hamiltonian}
    \Hhat_{\rm JW} =  \sum_{l=0}^{\Nt-1} h_l \bigotimes_{i=0}^{N-1} \hat{P}_i^{(l)}, 
\end{equation}
where \Nt{} is the number of terms and $\hat{P}_i^{(l)} \in \left\lbrace \PauliI_i, \PauliX_i, \PauliY_i, \PauliZ_i\right\rbrace $ are Pauli operators acting on the $i$-th qubit.
For \Ne{} electrons, this Hamiltonian acts on the Hilbert space spanned with $\binom{N}{\Ne}$ Slater determinants.
Each determinant corresponds to \Ne{} occupied orbitals and is represented with a basis vector $\ket{\xvec} = \ket{x_0 x_1 \ldots x_{N-1}},\ x_i \in \left\lbrace0, 1\right\rbrace$. 
Here the $i$-th orbital is occupied if $x_i = 1$. 
The molecular orbitals are usually obtained via the Hartree-Fock procedure.
The basis vector $\ket{\xvec_{\rm HF}} \coloneqq \ket{\underbrace{11\ldots1}_{\Ne}\underbrace{0\ldots0}_{N - \Ne}}$ represents the Hartree-Fock Slater determinant, which is a mean-field approximation of the true ground state.

The exact diagonalisation of the Hamiltonian, referred to as \emph{full configuration interaction (FCI)} in quantum chemistry, is clearly restricted by the factorial complexity scaling of Hilbert space size.
The traditional quantum chemistry methods trade off accuracy for polynomial complexity and approximate the ground state by choosing a class of trial states.
Two most prominent are families of \emph{configuration interaction (CI)} and \emph{coupled cluster (CC)} methods, which consider computational spaces spanned by excitations over $\ket{\xvec_{\rm HF}}$ up to a certain level.
Among these methods, CISD, CCSD and CCSD(T) are commonly used, with the latter considered as the \emph{de facto} ``golden standard''.

\subsection{Variational Monte Carlo}\label{sec:vmc}
To approximate the ground state of $\hat H$ we use a variational ansatz for the wave function, $\ket{\psi}=\sum_\xvec \psi(\xvec)\ket{\xvec}$. 
Here, $\ket{\xvec{}}$ is a computational basis vector in the Hilbert space of $N$ qubits and $\psi(\mathbf x)$ is a ``black box'' function which depends on a set of parameters $\theta$ and returns the quantum state amplitude for $\ket{\xvec{}}$.
We represent $\ket{\psi}$ with a neural network, and refer to the former as \emph{neural quantum state (NQS)}.
We numerically optimise the network parameters to minimise the energy expectation value $\Estate \coloneqq \frac{\braket{\psi|\Hhat|\psi}}{\braket{\psi|\psi}}$, and thus find the best ground state approximation.

To that end we employ the framework of iterative stochastic optimisation known as \emph{variational Monte Carlo (VMC)}.
This approach relies on the fact that quantum expectation values, such as the mean energy \Estate{}, can be cast in the form of mean values with respect to the Born distribution $p(\xvec) \coloneqq \frac{\left|\psi(\xvec)\right|^2}{\sum_{\xvec} \left|\psi(\xvec)\right|^2}$.
Hence, they can be estimated efficiently by obtaining a batch \Batch{} of \Ns{} samples from the distribution $p(\xvec)$.
Let $\Unique \coloneqq \textrm{Unique}(\Batch)$ be the set of unique samples in \Batch{} and let $n(\xvec)$ be the occurrence number for the sample \xvec{} so that $\sum_{\xvec \in \Unique} n(\xvec) = \Ns$.
We call the set of pairs $\Set{\left( \xvec, n(\xvec)\right)| \xvec \in \Unique }$ \emph{the sampling statistics}.
Using the sampling statistics, one estimates the energy of the state as a Monte Carlo expectation value as follows:
\begin{equation}\label{equ:energy_estimator_via_unique}
\Estate  = \mathbb{E} (\Eloc(\xvec) ) \approx  \sum_{\xvec \in \Unique} \Eloc(\xvec) \cdot \frac{n(\xvec)}{\Ns},
\end{equation}
where \Eloc(\xvec) is the so-called \emph{local energy} estimator:
\begin{equation}\label{equ:local_energy}
\Eloc(\xvec) \coloneqq\sum_{\xvecp: \Helem \neq 0} \frac{\Helem \psi(\xvecp)}{\psi(\xvec)}. %
\end{equation}
Similarly, one estimates the gradient of the energy according to Ref.~\cite{vicentini2022netket}:
\begin{equation}\label{equ:grad_def}
    \begin{gathered}
    \nabla_{\theta} \Estate = 2 \Re \left\lbrace \mathbb{E} \left[  \Eloc (\xvec) \cdot  \nabla_{\theta} \ln \psit^*(\xvec) \right] \right.  \\ 
     \qquad\quad \qquad\qquad-  \left. \mathbb{E} \left[  \Eloc (\xvec) \right] \cdot \mathbb{E} \left[\nabla_{\theta} \ln \psit^*(\xvec) \right] \right\rbrace.
    \end{gathered}
\end{equation}
This estimate is employed to update the ansatz parameters following the gradient descent rule or any advanced modification thereof~\cite{vicentini2022netket}.

\subsection{Autoregressive neural quantum states}\label{sec:anqs}
The quantum chemistry VMC optimisation is hindered by the fact that the ground state wave functions often have a distinct peaked structure.
The dominant contribution comes from the Hartree-Fock Slater determinant, i.e. $p(\xvec_{\rm HF})$ can be as high as 0.99 or more.
As a result, after the initial iterations $\xvec_{\rm HF}$ constitutes the majority of samples in the batch \Batch{}.
Correspondingly, one needs to resort to large batch sizes \Ns{} to explore a non-trivial portion of the Hilbert space and obtain meaningful gradient estimates.
This results in a large computational overhead if one optimises an NQS using  conventional sampling methods such as  Metropolis-Hastings~\cite{carleo_troyer_rbm, carleo_quantum_chemistry}.

To mitigate this issue we follow Ref.~\cite{barrett_autoregressive_qchem} and employ \emph{autoregressive} neural quantum states (ANQS).
The wave function $\psi(\xvec)$ of an ANQS is a product of single qubit conditional wave functions $\psi_i(x_i|\xvec_{<i})$ such that the conditional wave function for the $i$-th qubit depends only on the values of previous $i-1$ qubits:
\begin{equation}\label{equ:autoregr_psi_decomp}
    \psi(\xvec) = \prod_{i=0}^{N-1} \psi_i(x_i | \xvec_{<i}).
\end{equation}
Such a construction enables quick and unbiased sampling from the underlying Born distribution~\cite{autoregressive_originals, barrett_autoregressive_qchem}.
Importantly, Ref.~\cite{barrett_autoregressive_qchem} developed an algorithm which samples the \emph{statistics} corresponding to a batch of \Ns{} samples without producing each sample individually. 
The complexity of this algorithm is dominated by evaluating amplitudes of only \Nunq{} unique basis vectors, where $\Nunq \coloneqq  \left| \Unique \right|$. 
This  opened the avenue to batch sizes \Ns{} as big as $10^{12}$ (with the number of unique samples $\Nunq \sim 10^3\text{--}10^4$) and was adopted in subsequent research on ANQS quantum chemistry calculations~\cite{zhao_scalable_qchem, transformer_nqs_for_qchem, malyshev_anqs_with_quantum_numbers}.

\section{Method advancements}\label{sec:high_level_picture}
\subsection{Sampling without replacement}\label{sec:sampling_without_replacement}
Both Metropolis-Hastings and conventional autoregressive sampling are examples of  so-called sampling \emph{with} replacement. Such sampling is \emph{memoryless}, i.e.~if \xvec{} was obtained as $k$-th sample ($k < \Ns$), it can still be produced by the ansatz as $k^\prime$-th sample, where $k^\prime > k$. As a result, the number of obtained unique samples \Nunq{} cannot be set in advance; it emerges during sampling. 
One controls \Nunq{} only indirectly by changing the batch size \Ns{}, and the same value of \Ns{} can produce vastly different \Nunq{} depending on the probability distribution encoded by the ansatz. The autoregressive statistics sampling of Ref.~\cite{barrett_autoregressive_qchem}, while being more efficient, also emulates  sampling with replacement and hence suffers from  the same shortcoming.

It is however desirable to have direct control over \Nunq{}.
First, it defines the computational complexity of both autoregressive statistics sampling and local energy calculations,  while \Ns{} does not.
Second, \Nunq{} serves as a direct measure for the Hilbert space portion explored by an ANQS.

To address this issue, we resort to sampling from $p(\xvec)$ \emph{without} replacement.
Such sampling is \emph{memoryful}, i.e. if \xvec{} was obtained as the $k$-th sample, then (i) it can never be obtained in any of subsequent samplings; (ii) the remaining samples are produced from a renormalised probability distribution in which the samples that have already occurred are assigned zero probability.
As a result, we obtain a batch of \emph{precisely} \Nunq{} unique samples. 
Our sampling procedure, based on the algorithm by Kool {\it et al.}~\cite{ancestral_gumbel_top_k_sampling}, invokes the ancestral Gumbel \topk{} trick \cite{gumbel_gumbel, maddison_gumbel}, to autoregressively produce \Nunq{} samples without replacement in parallel. We describe further details in Methods and present  an ablation study showcasing the computational advantage
in the Supplementary Material.

\subsection{Streamlined local energy calculations}
The calculation of local energies constitutes the central computational bottleneck of NQS quantum chemistry calculations since the number of terms in molecular Hamiltonians grows quartically with the number of orbitals, $\Nt = \mathcal{O}(N^4)$~\cite{mcardle_review}.
According to Eq.~\eqref{equ:local_energy} one has to evaluate ansatz amplitudes for all \xvecp{} coupled via \Hhat{} to sampled $\xvec \in \Unique$, i.e.~such \xvecp{} that $\Helem \neq 0$.
This requires $\mathcal{O}(\Nunq \Nt)$ additional ansatz evaluations, which quickly becomes unwieldy as the system size grows. 

To mitigate this problem, we build upon the proposal of Wu {\it et al.}~\cite{transformer_nqs_for_qchem} and replace the local energy calculation \eqref{equ:local_energy}  with its computationally cheaper surrogate $\Elocvar(\xvec)$ which contains contributions only from those \xvecp{} which belong to the set of unique sampled basis vectors.
Hence, for the full energy calculation one requires only the amplitudes of basis vectors in \Unique{}, which can be computed and stored in memory with $\mathcal{O}\left(\Nunq\right)$ ansatz evaluations immediately after \Unique{} is obtained.

In addition, instead of evaluating the state energy as a Monte Carlo expectation \eqref{equ:energy_estimator_via_unique} of the local energy, we directly calculate the variational energy of the instantly sampled state:

\begin{equation}\label{equ:variational_energy}
    \Evar = \sum_{\xvec \in \Unique} \Elocvar(\xvec) \cdot \frac{p(\xvec)}{\mathcal{N}},
    \text{where } \mathcal{N} \coloneqq \sum_{\xvec \in \Unique} p(\xvec).
\end{equation}
Here we weigh local energy surrogate values by the renormalised probabilities obtained \emph{directly from ansatz}, as opposed to weighing them by empirical frequencies $\frac{n(\xvec)}{\Ns}$.
The reason for this is twofold.
First, sampling without replacement does not provide any empirical frequencies, but only the unique samples themselves, and thus one cannot use Eq.~\eqref{equ:energy_estimator_via_unique} directly.
Second, under such definition, the value calculated via Eq.~\eqref{equ:variational_energy} becomes variational, in that it is always an upper bound for the ground state energy since it corresponds to a physical state spanned by the vectors in \Unique.
Hence, in what follows we refer to the value given by Eq.~\eqref{equ:variational_energy} as the variational energy, and to the values of $\Elocvar(\xvec)$ as the variational (proxy of) local energy.

\subsubsection*{Improved asymptotic complexity}\label{sec:three_local_energy_approaches}
Although the approach of Ref.~\cite{transformer_nqs_for_qchem} reduces the number of \emph{ansatz evaluations} significantly below $\mathcal{O}(\Nunq \Nt)$, the number of \emph{computational operations} required to calculate \Evar{} still scales with \Nt{}.
This is because there are \Nunq{} values of $\Elocvar(\xvec)$ to be calculated; to obtain each of them, the authors iterate over \Nt{} Hamiltonian terms and calculate the corresponding coupled candidate \xvecp{}.
If  $\xvecp$ belongs to \Unique{}, they add the relevant matrix element to the final value.
In Ref.~\cite{transformer_nqs_for_qchem} checking whether \xvecp{} belongs to \Unique{} is implemented with binary search in $\mathcal{O}(\log \Nunq)$ operations, and thus in total one performs $\tilde{\mathcal{O}}(\Nunq \Nt)$ operations.
Here $\tilde{\mathcal{O}}$ indicates the complexity with omitted logarithmic scaling factors.

In this work we propose two novel methods to evaluate~\Evar, which have asymptotic complexity that does not depend on \Nt{}.
Specifically, we split the calculation of \Evar{} in two steps.
First, we obtain all pairs $(\xvec, \xvecp) \in \Unique \times \Unique$ which are coupled via the Hamiltonian.
We refer to this stage as \FindSampledAndCoupled{} procedure.
Second, we evaluate the matrix elements \Helem{} and add the corresponding factors to $\Elocvar(\xvec)$ and $\Elocvar(\xvecp)$.
Our methods focus on improving the asymptotic complexity of \FindSampledAndCoupled{}. The benchmark implementation we compare with is that of Ref.~\cite{transformer_nqs_for_qchem}, to which we refer as \CoupleViaUniqueXY{}.

When $\Nunq \ll \Nt$ the majority of candidate \xvecp{} will not belong to \Unique{}.
To take advantage of this, our first approach, to which we refer as \CoupleAllToAll{}, iterates over each pair $(\xvec, \xvecp) \in \Unique \times \Unique$ and checks whether there exists a Hamiltonian term coupling \xvec{} to \xvecp{}, 
see Methods for details. 
This approach brings the complexity of \FindSampledAndCoupled{} down to $\tilde{\mathcal{O}}(\Nunq^2)$.

Our second approach employs an empirical observation that in practice very few pairs $(\xvec, \xvecp)$ are actually coupled via the Hamiltonian. 
Thus, iterating over each element of $\Unique \times \Unique$ results in unnecessary computations too.
We address this issue by preprocessing $\Unique$ and reorganising it into a data structure known as \emph{prefix tree} or simply \emph{trie}, which we describe in more details in Methods.
This data structure allows one to exploit the sparsity of couplings within $\Unique \times \Unique$ and substantially speed up \FindSampledAndCoupled{}.
We refer to this implementation of \FindSampledAndCoupled{} as \CoupleViaPrefixTree{}.%

\subsection{ANQS-tailored stochastic reconfiguration}
Standard gradient descent performs poorly in curved parameter spaces, where the learning rate has to be selected carefully so that optimisation does not overshoot narrow ravines.
\emph{Stochastic reconfiguration} (SR) is a gradient postprocessing technique that modifies the energy gradient at every iteration to account for the curvature of underlying variational manifold.
To that end, for an ansatz with \Np{} parameters $\{\theta_p\}$ one evaluates the so-called $\Np\times\Np$ \emph{quantum geometric tensor (QGT)}~\cite{meyer_quantum_fisher_information, stokes_quantum_natural_gradient}:
\begin{equation}\label{equ:qgt_definition}
    S_{pq} = \braket{\partial_{p}\psi|\partial_{q} \psi} - \braket{\partial_{p}\psi|\psi}\braket{\psi|\partial_{q} \psi};\ p, q \in 1..\Np,
\end{equation}
where $\ket{\partial_{p} \psi}\coloneqq \frac{\partial \ket{\psi}}{\partial \theta_p}$. 
QGT captures the local curvature of the parameter space and allows one to ``flatten'' it by multiplying the energy gradient with an inverse of QGT: $\nabla E \rightarrow S^{-1}\nabla E$.

In VMC applications the overlaps in Eq.~\eqref{equ:qgt_definition} are estimated via sampling similarly to Eqs.~\eqref{equ:energy_estimator_via_unique} and \eqref{equ:grad_def}~\cite{vicentini2022netket}. 
We introduce two modifications to their calculation.
First, in the vein of Eq.~\eqref{equ:variational_energy} we evaluate stochastic averages using renormalised probabilities $\frac{p(\xvec)}{\mathcal{N}}$ of unique samples, as opposed to their occurrence numbers.
Second, evaluating $\frac{\partial \psi(\xvec)}{\partial \theta_p}$ for each \xvec{} in \Unique{} is computationally heavy, and thus we restrict the set of unique samples used to evaluate $S$.
Specifically, we take only first $\NSR \ll \Nunq$ unique samples corresponding to the highest probabilities.
In all experiments we use $\NSR = 100$ chosen after a study on how \NSR{} impacts the achieved variational energies (see Supplementary Material).
Apart from a computational benefit, this also seems to numerically stabilise the inversion of $S$, which is performed  following the recipes of  Refs.~\cite{chen_min_sr} and \cite{rende_reg_sr}.

We use the SR-transformed gradients jointly with the Adam optimiser.%
We observe that SR substantially boosts the optimisation convergence, i.e. allows achieving lower energies earlier in the optimisation.
This observation contrasts with previous reports that cast doubt on the merit of SR for ANQS optimisation~\cite{zachary_anqs_dynamics,bohrdt_doped_antiferromagnets}.

\subsection{GPU implementation}\label{sec:gpu_implementation}
Apart from improving the asymptotic complexity of optimisation, we also focus on the practical aspects of its implementation, namely on fully leveraging the inherent massive parallelism of GPUs.
We adapt the approach of Wu {\it et al.}~\cite{transformer_nqs_for_qchem} to store  each basis vector \xvec{} in a compressed form as a tuple of integers. This leads to a factor of eight memory requirement reduction compared to previous implementations, in which each bit was stored as a one-byte integer. This is critical given our algorithms require simultaneous handling of data structures containing $\mathcal{O}(\Nunq{}^2)\gtrsim10^9$ bit vectors of length $N$ (see Supplementary Material). %

We advance this method further by implementing \emph{all} steps of local energy calculation as a sequence of \emph{bitwise} operations on \xvec{} stored in such compressed format. This allows us to replace looping over qubits with vectorised bitwise processor operations, leading to highly accelerated calculations. Specifically, encoding the basis vectors in 64-bit integers speeds up the calculation by a factor of up to 64. As a result, experiments for systems with up to $3\cdot 10^6$ Hamiltonian terms required only a single GPU with 24GB of RAM and barely over a second per iteration.
This is a substantial reduction in comparison to Refs.~\cite{zhao_scalable_qchem} and~\cite{transformer_nqs_for_qchem} which relied on several GPUs and took dozens of seconds per iteration.
In addition, our implementation employs standard tensor algebra routines of PyTorch software library~\cite{pytorch} and may be run without any custom CUDA code.
We refer the reader to Supplementary Material for an overview of the key aspects of our code.

\section{Results}\label{sec:results}
The main implicit assumption behind our approach is that a large enough \Nunq{} allows sampling the most important amplitudes as long as the true ground state wave function is peaked enough.
In this section we empirically demonstrate the validity of this assumption. We obtain energies that are better than those provided by state-of-the art quantum chemistry methods and/or below chemical accuracy.
Equally as importantly, our approach provides an order of magnitude computational speedup compared to existing works. 

\subsection{\ce{Li2O} and \ce{BeF2} dissociation curves}\label{sec:li2o_bef2_diss_curves}
\begin{figure}
	\centering
	\includegraphics[width=\linewidth]{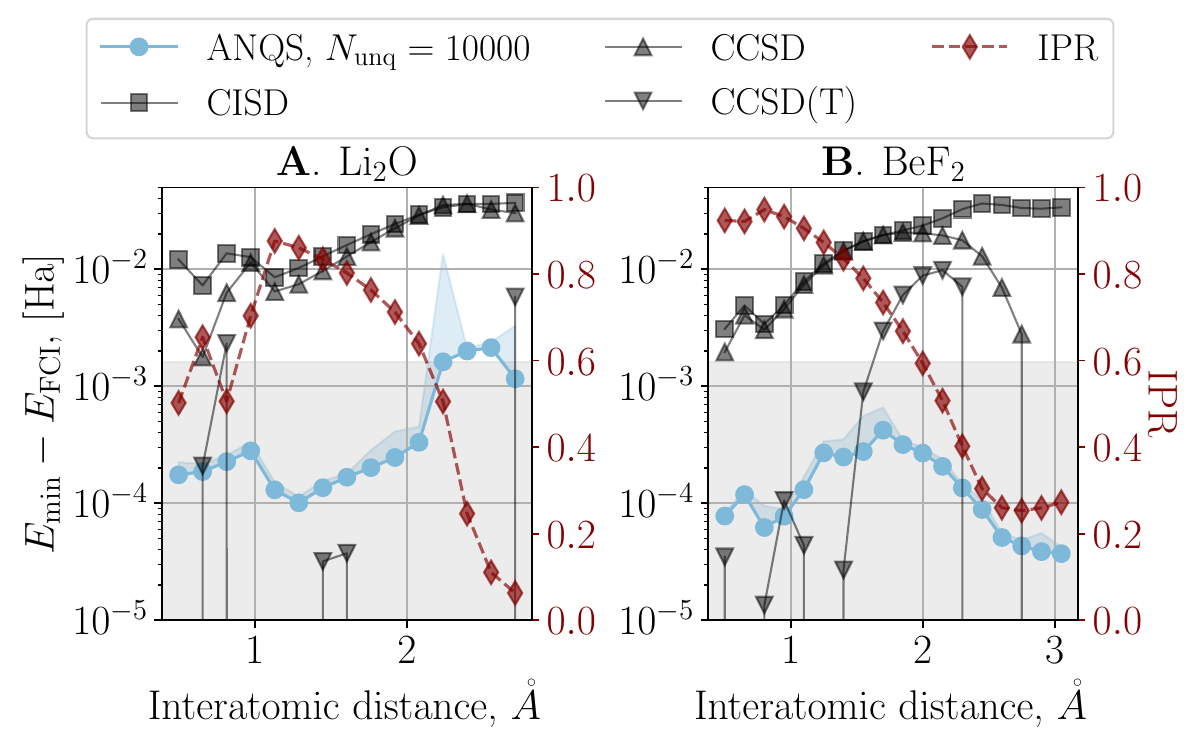}
	\caption{Dissociation curves for \ce{Li2O} and \ce{BeF2} molecules. The grey shaded areas correspond to energy differences within chemical accuracy. The blue shaded areas depict the spread of values from minimum to maximum across seeds. We plot energy differences to the exact diagonalisation (FCI) result ((left y-axis) and ground state IPR (right y-axis) as a function of interatomic distance.}
	\label{fig:li2o_bef2_diss_curves}
\end{figure}

In the first set of experiments we study \ce{Li2O} and \ce{BeF2} molecules both requiring $N=30$ qubits in the minimum basis set STO-3G.
These are amongst the largest molecules still amenable to exact diagonalisation.
Thus, it is possible to check whether ANQS can achieve \emph{chemical accuracy}, i.e.~whether it can converge to energies within $\ChemAcc \coloneqq 1.6$ mHa of the true ground state energy \EFCI{}.
In addition, the unfavourable scaling of local energy calculations is already apparent for these molecules, and in the previous works an optimisation based on the full \Eloc{} required an excessive compute time ranging from hours~\cite{malyshev_anqs_with_quantum_numbers} to days~\cite{barrett_autoregressive_qchem}.
As a result, little progress on extending the ANQS calculations beyond these molecules has been reported so far~\cite{zhao_scalable_qchem}.

We quantify the peakedness of the ground states using the \emph{inverse participation ratio (IPR)} defined as follows:
\begin{equation}
    \IPR \coloneqq \mathexp\left[{p(\xvec)}\right] = \sum_{\xvec} p(\xvec) \cdot p(\xvec).
\end{equation}
The maximum value of \IPR{} is 1 and it is achieved when only one basis vector contributes to a wave function, and its minimum is $\frac{1}{2^N}$, reached when $\forall \xvec\ |\psi(\xvec)| = \frac{1}{\sqrt{2^N}}$.

For each molecule we explore a range of interatomic distances and for each distance we optimise an ANQS with $\Nunq = 10^4$.
The minimum achieved variational energies depicted in Fig~\ref{fig:li2o_bef2_diss_curves} consistently surpass those obtained by the conventional methods like CISD and CCSD.
While ANQS does not always outperform CCSD(T), it is important to remember that the latter can yield unphysical energies below \EFCI{}, thus warranting cautious interpretation of such comparison.
Crucially, ANQS reliably achieves energies within chemical accuracy for nearly all points on dissociation curves, with a notable exception around 2.4--2.6 Å for \ce{Li2O}, a region where traditional methods deliver poorer energies too.
\begin{figure*}
	\centering
	\includegraphics[width=\linewidth]{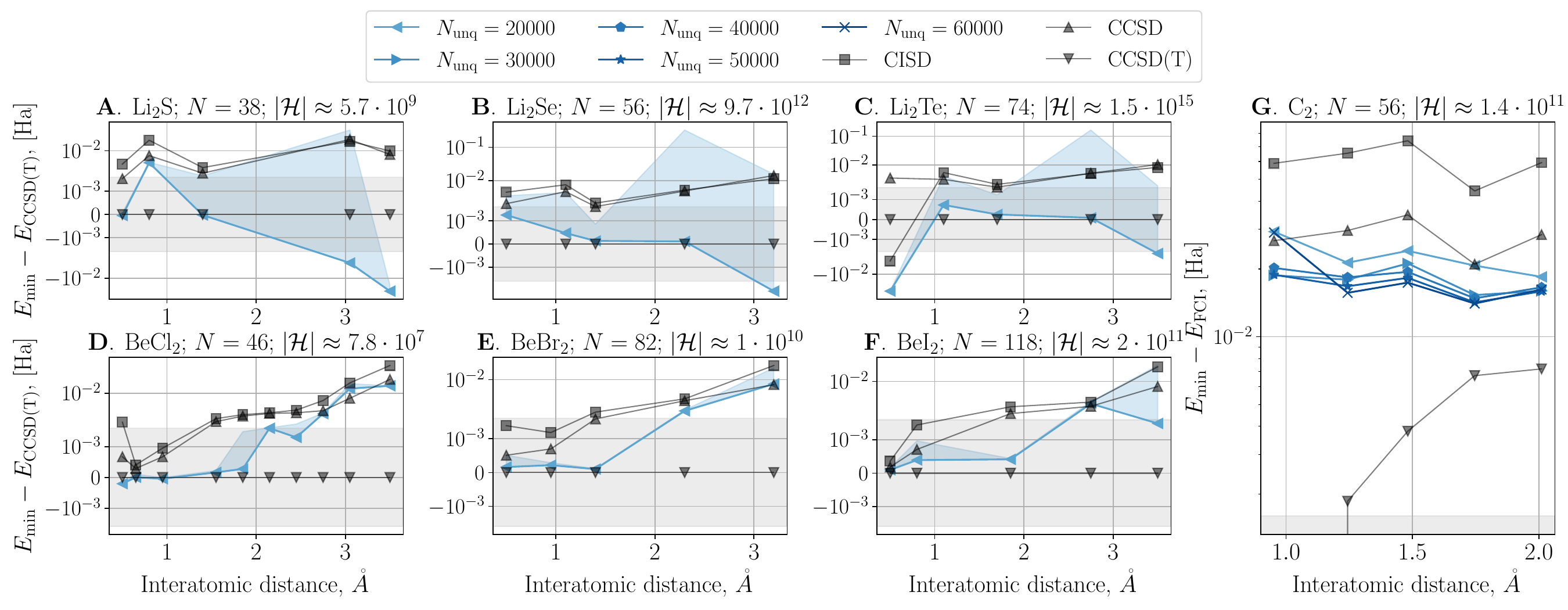}
	\caption{Dissociation curves for bigger molecules as described in the main text. For each molecule we show the 
 number of qubits in the Hamiltonian $N$ as well as the size of the physical Hilbert space. The \ce{C2} molecule is considered in cc-pvdz basis set. We plot energy differences to a reference energy as a function of interatomic distance. \textbf{A--F.} The reference energy is obtained with CCSD(T). \textbf{G.}  The reference energy is obtained with FCI as reported in~\cite{c2_diss_curve}. 
	The grey shaded areas correspond to the energy differences within chemical accuracy. The blue shaded areas depict the spread of values from minimum to maximum across seeds.}
	\label{fig:all_diss_curves}
\end{figure*}
We also plot the IPR corresponding to each interatomic distance.
For \ce{Li2O}, a higher peakedness generally correlates with the improved optimisation accuracy. 
However, the accuracy for \ce{BeF2} shows a more complicated pattern: it drops with an initial decrease in IPR from approximately 0.9 to 0.75, before improving again as the IPR decreases further to around 0.25. 
Thus, while the data are compatible with the hypothesis that high peakedness is sufficient for successful optimisation, the overall success may also be influenced by other factors, such as the phase structure of the ground state. 
Identifying these factors is an important direction for future research.

Finally, let us note that for \ce{Li2O} molecule with the geometry taken from PubChem database~\cite{pubchem} we achieve chemical accuracy on average in 750 seconds.
This is 25 times faster compared to Ref.~\cite{malyshev_anqs_with_quantum_numbers}, which to the best of our knowledge reported the fastest optimisation for the same molecule so far. The per-iteration time also improves to 0.18 seconds as compared to 8.3 seconds reported in the same work.

\subsection{Further \ce{Li} and \ce{Be} compounds}\label{sec:li_be_diss_curves}

In the next set of experiments we expand the ANQS quantum chemistry calculations to previously challenging system sizes. 
We obtain dissociation curves for two groups of molecules of increasing size; the first group includes \ce{Li2S}, \ce{Li2Se} and \ce{Li2Te} molecules, while the second consists of \ce{BeCl2}, \ce{BeBr2} and \ce{BeI2}.
We base our selection on their chemical similarity with the already studied \ce{Li2O} and \ce{BeF2}: the anion is replaced by a heavier element in the same group 6 (7) of the periodic table.
Thus we can systematically scale the number of qubits $N$, the Hilbert space size and the number of terms in Hamiltonian \Nt{}.
We note, however, that since the Hilbert space dimension is $\HilbSize{}\coloneqq\binom{N}{\Ne}$, the largest values of $N$ do not necessarily amount to the largest values of \HilbSize{}.
For example, the largest studied Hilbert space is that of \ce{Li2Te} molecule with 74 qubits and $\sim 10^{15}$ Slater determinants, which is three orders of magnitude more than the largest FCI study performed so far~\cite{largest_fci}.
At the same time, the Hilbert space size of \ce{BeI2} molecule with 118 qubits is on the order of ``mere'' $10^{11}$ Slater determinants.

We plot the results of this set of experiments in Fig.~\ref{fig:all_diss_curves}A--F.
We do not have access to FCI energies for these molecules, and thus we plot the difference between the minimum variational energies achieved by ANQS and those obtained with CCSD(T).
It can be seen that in the dominant majorities of experiments ANQS outperforms  traditional quantum chemistry methods such as CISD and CCSD.
When compared to CCSD(T), ANQS results mostly come close to within chemical accuracy, and in some cases surpass them.
We consider this as an evidence in favour of the remarkable prospect of ANQS.

\subsection{\ce{C2} molecule in cc-pvdz basis set}\label{sec:c2_diss_curve}
We also apply ANQS to a paradigmatic strongly correlated system: \ce{C2} molecule in the cc-pvdz basis~\cite{c2_diss_curve}.
The results are presented in Fig~\ref{fig:all_diss_curves}G, where we plot the difference between the lowest energy achieved by ANQS and \EFCI{} as given in Ref.~\cite{c2_diss_curve}.
While we find that increasing $\Nunq{}$ systematically improves the result, none of \Nunq{} values allowed reaching chemical accuracy or surpassing the CCSD(T) energies, even though \ce{C2} has a moderate number of qubits (56) and Hilbert space size ($\sim 10^{11}$) as compared to \ce{Li} and \ce{Be} compounds studied in the previous section. 
Nevertheless, an increase in \Nunq{} does result in better accuracy, and, starting from $\Nunq = 40000$, ANQS energies surpass those of CISD and CCSD methods.
We leave it for future work to explore whether increasing $\Nunq{}$ further would continue to improve the convergence, or whether the limiting factor is rather the expressivity of the chosen ansatz.

\subsection{Comparison of \FindSampledAndCoupled{} implementations}\label{sec:findsc_timings}
\begin{figure}
	\centering
	\includegraphics[width=\linewidth]{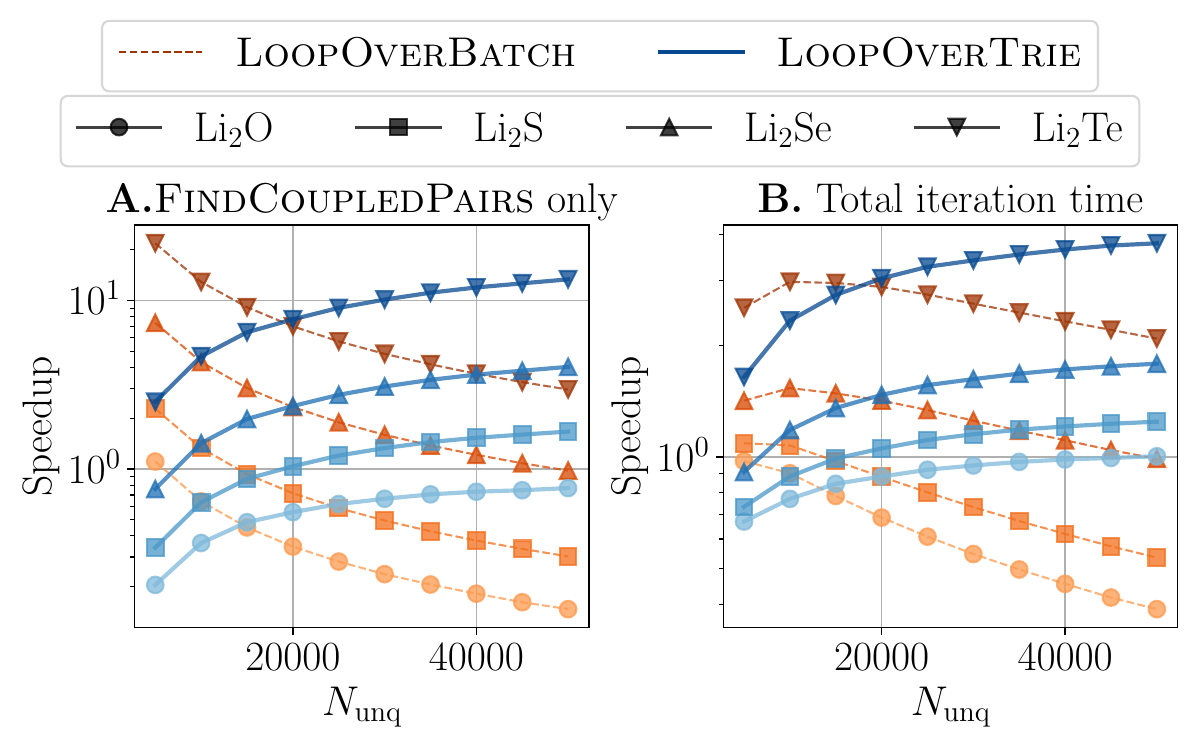}
	\caption{The speedup of \CoupleAllToAll{} and \CoupleViaPrefixTree{} routines with respect to baseline \CoupleViaUniqueXY{}. The speedups are calculated for: \textbf{A.} Only \FindSampledAndCoupled{} procedure; \textbf{B.} Total iteration time.}
	\label{fig:findsc_speedups}
\end{figure}

In the final set of experiments we benchmark three implementations of the \FindSampledAndCoupled{} procedure.
In Fig.~\ref{fig:findsc_speedups}A we plot the speedups achieved by \CoupleAllToAll{} and \CoupleViaPrefixTree{} algorithms compared to the baseline \CoupleViaUniqueXY{} implementation. 
For small values of \Nunq{} the \CoupleAllToAll{} implementation performs the best, offering up to 10x reduction in time.
However, as \Nunq{} grows, the advantage of \CoupleAllToAll{} becomes less and less pronounced, until it vanishes.
For example, for \ce{Li2O} molecule and $\Nunq = 5 \cdot 10^4$ it is an order of magnitude slower than the other approaches.

The opposite holds for \CoupleViaPrefixTree{}: at low \Nunq{} it struggles to compete with \CoupleAllToAll{}.
However, as \emph{both} \Nunq{} and \Nt{} grow, \CoupleViaPrefixTree{} becomes more and more advantageous: for example, for \ce{Li2Se} molecule and $\Nunq = 5\cdot10^4$ it is five times faster than either \CoupleViaUniqueXY{} or \CoupleAllToAll{}.
For \ce{Li2Te} molecule and the same \Nunq{} it demonstrates a tenfold speedup compared to \CoupleViaUniqueXY{}.

To accurately estimate the speedup associated with these new techniques, it is important to keep in mind that \FindSampledAndCoupled{} takes only a part of the compute time in each iteration. The local energy calculation involves three more important subroutines: sampling of \xvec{}, amplitude evaluation and computing matrix elements \Helem{}. In Fig.~\ref{fig:findsc_speedups}B, we show the gains associated with the new \FindSampledAndCoupled{} algorithms with respect to the total iteration time.
In the case of \ce{Li2Te} molecule and $\Nunq{} = 5\cdot10^4$, the tenfold improvement of \CoupleViaPrefixTree{} shrinks down to a factor of four.

\begin{figure*}
	\centering
	\includegraphics[width=\linewidth]{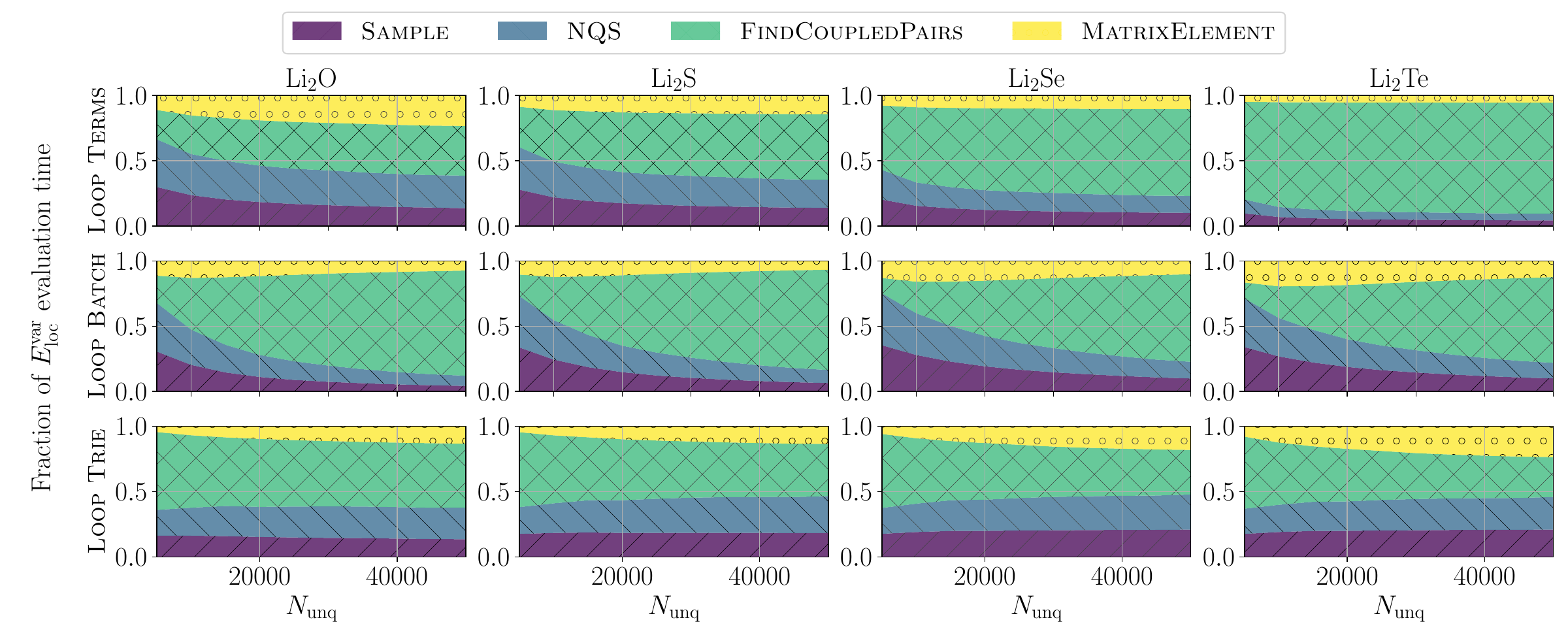}
	\caption{The fraction of the energy evaluation time spent on each of four crucial iteration parts: (i) sampling; (ii) evaluation of amplitudes for basis vectors in \Unique{}; (iii) finding the coupled pairs $(\xvec, \xvecp)$; (iv) calculating the matrix elements between the coupled pairs. Top to bottom rows correspond to different implementations of \FindSampledAndCoupled{} procedure: \CoupleViaUniqueXY{}, \CoupleAllToAll{}, \CoupleViaPrefixTree{}.}
	\label{fig:local_energy_fractions}
\end{figure*}

In Fig.~\ref{fig:local_energy_fractions} we plot the fraction of the time spent on each of the four subroutines during the energy calculation.
Both for \CoupleViaUniqueXY{} and \CoupleAllToAll{} the fraction of \FindSampledAndCoupled{} grows as \Nunq{} increases.
For the largest molecule (\ce{Li2Te}) at $\Nunq = 5\cdot 10^4$, it amounts for the majority of energy calculation time, while the remaining subroutines take at most 10\% of the calculations each.
At the same time, \FindSampledAndCoupled{} implemented as \CoupleViaPrefixTree{} constitutes only a third of the energy calculation time, and, more generally, decreases as both \Nunq{} and \Nt{} grow.
Thus, computations related directly to the local energy calculation (as opposed to sampling and amplitude evaluation) are no longer a bottleneck of NQS quantum chemistry.

\section{Discussion}\label{sec:conclusion}

Our results suggest that the peaked ground state wave functions typical for molecules should be not only mitigated in ANQS quantum chemistry calculations, but rather celebrated in certain cases.
In particular, it allowed us to leverage an ANQS-specific sampling algorithm which produces the desired number of unique samples.
Thus, we were able to focus on a small physically relevant portion of the Hilbert space and greatly accelerate the stochastic estimation of the variational energy by ignoring the contributions from the unsampled part of the Hilbert space.
Consequently, we were able to achieve energies that compete with those obtained with the traditional quantum chemistry methods, for computational spaces several orders of magnitude larger than those previously attempted with ANQS.

Remarkably, we even came across certain geometries where the peakedness was less pronounced, and yet focusing only on the sampled subspace still yielded energies within chemical accuracy to the ground state --- this was the case, for example, with  \ce{BeF2}.
At the same time, we identified cases when our method struggled to reach competitive energies, such as \ce{C2}. However, it remains to be explored, whether this indicates a general limitation of our method or whether it is due to unrelated issues like bottlenecks in the optimisation~\cite{bukov_non_stoquastic}.

Our results point out further directions for algorithmic improvements and theoretical analysis that would scale up our method. The first observation is that our method is readily amenable to  parallelisation across several GPUs. We expect this to enable addressing even larger problem sizes~\cite{zhao_scalable_qchem, transformer_nqs_for_qchem}, potentially reaching complex organic compounds.

Additionally, the impact of the underlying neural network architecture on the accuracy of optimisation remains a largely unexplored topic.
Recently, neural network architectures that are specifically tailored to fermionic systems showed significant promise %
in application to quantum many-body problems~\cite{luo_backflow, moreno_hfds}, including quantum chemistry ~\cite{liu_backflow_quantum_chemistry, Nys2024}. Combining the enhanced expressivity exhibited by these models with the sampling efficiency of ANQS can potentially give rise to a powerful new approach~\cite{humeniuk_slater_jastrow}.

We lack understanding on how the choice of \Nunq{} influences the learning ability of ANQS.
However daunting this question might seem, first steps in similar direction were already made by Astrakhantsev {\it et al.}~\cite{astrakhantsev_algorithmic_phase_transition}.
This work analysed an algorithmic phase transition in the learning dynamics of a variational quantum circuit related to the number of samples produced at each gradient descent step.
It is natural to extend these techniques  to ANQS optimisation.

Finally, as discussed, the variational energy \eqref{equ:variational_energy} is an upper bound to the true energy of the NQS, but this upper bound may not be sufficiently tight. %
The so-called \emph{projected} energies provide a more accurate estimate~\cite{cleland_poor_fciqmc_energy}. It is to be explored whether ANQS optimisation can be synthesised with this approach to energy calculation for improved optimisation.

Regardless of the actual research path taken, we believe that our work showcases the high promise of ANQS \abinit{} quantum chemistry calculations beyond the FCI limit. %
We are optimistic that the ideas presented in this paper, particularly aimed at reducing time and memory requirements of the method, will democratise and facilitate further experiments, thus serving as a fruitful basis for the rapid future progress.

\section*{Methods}\label{sec:technical details}

\subsection{Autoregressive sampling without replacement}\label{sec:gumbel_sampling}
For sampling without replacement we employ an algorithm developed in Ref.~\cite{ancestral_gumbel_top_k_sampling} which extends the Gumbel \topk{} trick~\cite{gumbel_gumbel, maddison_gumbel} to autoregressive probability distributions. 
This algorithm is not our contribution \emph{per se}, yet we overview it since it touches upon the ideas previously unexplored within the field of NQS optimisation.

\subsubsection{Conventional sampling without replacement}\label{sec:Cswr}
A conventional way to obtain $K$ samples without replacement from a probability distribution $p(\xvec)$ is to  produce samples one-by-one and adaptively change the sampled distribution in the following way. 
Suppose one has sampled without replacement $k-1$ unique samples constituting the set $\Unique^{(k-1)}$. 
Then, one obtains the $k$-th unique sample by sampling from the following probability distribution:
\begin{equation}\label{equ:renorm_prob}
    p^{(k)}(\xvec) = \begin{cases}
        0, \text{ if } \xvec \in \Unique^{(k-1)}, \\ 
        \frac{p(\xvec)}{\sum_{\xvecp \not \in \Unique^{(k-1)}} p(\xvec)} \equiv  \frac{p(\xvec)}{1 - \sum_{\xvecp \in \Unique^{(k-1)}} p(\xvec)}  \text{ otherwise}.
    \end{cases}
\end{equation}
In other words, one manually removes the probability mass associated with already produced samples and then renormalises the remaining probabilities.

\subsubsection{Gumbel \topk{} trick}
The Gumbel \topk{} trick replaces sequential sampling from $p(\xvec)$ with multiple samplings of another random variable, referred to as the \emph{Gumbel noise}. The cumulative distribution function of Gumbel noise is given by $F_{\textsc{G}}(g) = e^{-e^{-g}}$. The advantage of this approach is that Gumbel noise sampling can be  performed in a parallel fashion.
The algorithm proceeds  as follows~\cite{ancestral_gumbel_top_k_sampling}:

\begin{enumerate}
    \item We obtain a set of i.i.d. samples of Gumbel noise $G_{\xvec}$, one per each \xvec{}.
 This can be achieved by inverse transform sampling as $G = F_{\textsc{G}}^{-1}(U) = -\log \left(-\log U \right)$, where $U \sim \mathrm{Uniform}(0, 1)$.  
    \item We evaluate \emph{log}-probabilities corresponding to each possible outcome $l_\xvec \coloneqq \log p(\xvec)$, which we further refer to as \emph{unperturbed} log-probabilities.
    \item We calculate the so-called \emph{perturbed} log-probabilities by adding an Gumbel noise sample to each unperturbed log-probability: $L_{\xvec} \coloneqq l_\xvec + G_\xvec$. 
    \item We sort the perturbed log-probabilities in the descending order and take first $K$ \xvec's corresponding to the highest perturbed log-probabilities as an output of the sampling without replacement algorithm.
\end{enumerate}

To intuitively understand the Gumbel \topk{} trick, %
let us suppose we have already selected $k <K$ basis vectors %
and are now seeking to add the $(k+1)$-st.
Let $\xvec$ and \xvecp{} be two  basis vectors that are not among the $k$ already selected.
Let us evaluate the probability of the event that $L_{\xvec} < g_{{\xvec^\prime}}$, which equals to the probability that \xvecp{} is  selected in this step \emph{conditioned} on the event that either \xvec{} or \xvecp{} is  selected.

This probability is $\Pr(L_{\xvec} < L_{\xvec^\prime}) =\quad \Pr(l_\xvec +  G_\xvec < l_{\xvec^\prime} + G_{\xvec^\prime}) = \Pr(G_\xvec - G_{\xvec^{\prime}} < l_{\xvec^\prime} - l_\xvec)$.
Thus, one seeks the probability for the difference $\Delta G \coloneqq G_\xvec - G_\xvecp$ of two independent Gumbel variables to be less than the  given value $l_{\xvec^\prime} - l_\xvec$.
This probability is given by the value of the CDF $F_{\Delta G}(l_{\xvec^\prime} - l_\xvec)$ of the random variable $\Delta G$.
It can be calculated to be $F_{\Delta G}(l_{\xvec^\prime} - l_\xvec) = \frac{1}{1 + e^{-(l_{\xvec^\prime} - l_\xvec)}}$.
Since $e^{-(l_{\xvec^\prime} - l_\xvec)} = \frac{p(\xvec)}{p(\xvecp)}$, we obtain that $\Pr({L_\xvec} < {L_{\xvec^\prime}}) = \frac{p(\xvecp)}{p(\xvec) + p(\xvecp)}$.
Thus, conditioned on the assumption that either \xvec{} or \xvecp{} is  selected, $\xvecp$ is selected with the probability $\frac{p(\xvecp)}{p(\xvec) + p(\xvecp)}$, whereas $\xvec$ is selected with the probability $\frac{p(\xvec)}{p(\xvec) + p(\xvecp)}$, which is precisely what one expects from sampling without replacement as per Eq.~\eqref{equ:renorm_prob}.

\subsubsection{Autoregressive Gumbel \topk{} trick}
To explain the autoregressive Gumbel \topk{} sampling let us focus on $K=1$ for simplicity.
In this case the ordinary Gumbel \topk{} trick prescribes evaluating perturbed log-probabilities \emph{for all} \xvec{} and finding the single maximum among them.
For $N$ qubits this amounts to $2^N$ perturbed log-probabilities and is clearly unfeasible.
Hence, the approach of Ref.~\cite{ancestral_gumbel_top_k_sampling} effectively resorts to a binary search to find the \xvec{} corresponding to the maximum of the perturbed log-probabilities.

The authors show that given a vector $\xvec_{<i}$ the maximum of perturbed log-probabilities of its ``children'', i.e. $\max_{\xvec_{\geq i}} \left(\log p(\xvec_{<i} \xvec_{\geq i}) + G_{\xvec_{\geq i}}\right)$ is distributed as $ \log p(\xvec_{<i}) + G_{\xvec_{<i}} \eqqcolon L_{\xvec_{<i}}$.
In other words, the \emph{perturbed} log-probability of $\xvec_{<i}$ can serve as an upper bound for the perturbed log-probabilities of its children.
Consequently, given $\xvec_{<i}$ one might estimate the perturbed log-probabilities of $\xvec_{<i}0$ and $\xvec_{<i}1$ and keep only that partially sampled vector which has the larger perturbed log-probability.
By starting this process with an empty vector $\xvec_{<0} \equiv \varnothing$ and repeating it iteratively, one obtains the full \xvec{} corresponding to the highest perturbed log-probability without exponential growth of complexity.

There is, however, a subtlety.
Since the perturbed log-probability of $\xvec_{<i}$ bounds the perturbed log-probabilities of $\xvec_{<i}0$ and $\xvec_{<i}1$ from above, the latter, being sampled at a \emph{later} stage, should not exceed the value $L_{\xvec_{<i}}$ obtained at an \emph{earlier} stage.
In other words, sampling of $L_{\xvec_{<i}0}$ and $L_{\xvec_{<i}1}$ should be \emph{conditioned} on the value of $L_{\xvec_{<i}}$.
The authors of Ref.~\cite{ancestral_gumbel_top_k_sampling} show that the correct conditioned values denoted as $\tilde{L}_{\xvec_{<i}0}$ and $\tilde{L}_{\xvec_{<i}1}$ can be obtained by starting with the unconditioned values, finding their maximum $Z \coloneqq \max\left\lbrace L_{\xvec_{<i}0}, L_{\xvec_{<i}1}\right\rbrace$ and calculating the conditioned ones as $\tilde{L}_{\xvec_{<i}x_i} = -\log\left(\expe^{-\tilde{L}_{\xvec_{<i}}} - \expe^{-Z} + \expe^{-L_{\xvec_{<i}x_i}}\right)$.
As a result, the autoregressive sampling without replacement is described with the following pseudocode:
\nextalgo
\begin{algorithm}[H]
   \caption{Autoregressive Gumbel \topk{} sampling}\label{algo:ar_gumbel_top_k}
    \begin{algorithmic}[1]
        \Function{ARGumbelTopK}{$p_i(x_i|\xvec_{<i});\ K$}
                \Statex \hspace{\algorithmicindent}\texttt{\# Initialise the set of partially sampled vectors and their  (perturbed) log-probabilities.}
                \State  $\UniqueGumbel \coloneqq \left[ \left( \varnothing, 0, 0 \right) \right]$
                \For{$i$ \PFrom{} 0 \PTo{} $N-1$}
                    \State $\UniqueGumbelp \coloneqq \PEmptyList$
                    \For{$(\xvec_{<i}, l_{\xvec_{<i}}, \tilde{L}_{\xvec_{<i}})$ \PIn{} \UniqueGumbel{}}
                        \Statex \hspace{\algorithmicindent}\hspace{\algorithmicindent}\hspace{\algorithmicindent}{\texttt{\# Evaluate the unconditioned log-probs} }
                        \For{$x_{i}$ \PIn{} $\left\lbrace0, 1\right\rbrace$}
                            \State $l_{\xvec_{<i}x_i} \coloneqq l_{\xvec_{<i}} + \log p_i(x_i|\xvec_{<i})$
                            \State $L_{\xvec_{<i}x_i} \coloneqq l_{\xvec_{<i}x_i} + \textsc{Gumbel}(0)$
                        \EndFor
                        \State $Z \coloneqq \max\left\lbrace L_{\xvec_{<i}0}, L_{\xvec_{<i}1}\right\rbrace$
                        \Statex \hspace{\algorithmicindent}\hspace{\algorithmicindent}\hspace{\algorithmicindent}{\texttt{\# Evaluate the conditioned log-probs} }
                        \For{$x_{i}$ \PIn{} $\left\lbrace0, 1\right\rbrace$}
                            \State $\tilde{L}_{\xvec_{<i}x_i}~\coloneqq~-~\log\left(\expe^{-\tilde{L}_{\xvec_{<i}}} - \expe^{-Z} + \expe^{-L_{\xvec_{<i}x_i}}\right)$
                            \State $\UniqueGumbelp\PAppend{(\xvec_{<i}x_i, l_{\xvec_{<i}x_i}, \tilde{L}_{\xvec_{<i}x_i})}$
                        \EndFor
                    \EndFor                        
                    \State Sort \UniqueGumbelp{} in the descending order of $\tilde{L}_{\xvec_{<i+1}}$.
                    \State $\UniqueGumbel \coloneqq \UniqueGumbelp{}[1:K]$.
                \EndFor
                \State $\Unique \coloneqq$ every \xvec{} in \UniqueGumbel{}.

            \State \textbf{return} \Unique{}
        \EndFunction
    \end{algorithmic}
\end{algorithm}
A generalisation to $K > 1$ is achieved by retaining \topk{} (perturbed) log-probabilities during sampling of each 
qubit, see Ref.~\cite{ancestral_gumbel_top_k_sampling} for more details.
The complexity of Algorithm~\ref{algo:ar_gumbel_top_k} is $\mathcal{O}\left(\Nunq \AnsatzComplexity \right)$, where \AnsatzComplexity{} is the complexity of one ansatz evaluation. 
Importantly, the ansatz evaluations can be parallelised over all $\xvec_{<i}$ for a given $i$ using batched GPU operations, and the only required loop is over the qubits.
As a result, the total complexity of producing \Nunq{} samples is comparable to that of evaluating their amplitudes, which is a hallmark of autoregressive ansatzes.

\subsection{\FindSampledAndCoupled{} implementations}
\subsubsection{Hamiltonian arithmetic}
We start our technical consideration of \FindSampledAndCoupled{} implementation by encoding each Hamiltonian term as follows.
We define the bit vector \Xvecl{} encoding the information about the positions of \PauliX{} operators in the Hamiltonian term $\Phatl{}\equiv \bigotimes_{i=0}^{N-1} \hat{P}_i^{(l)}$ as follows:
\begin{equation*}
        \Xvecli \coloneqq  \begin{cases}
            1, \text{ if } \Phatli = \PauliX;\\
            0, \text{ otherwise. }
        \end{cases}
\end{equation*}
The bit vectors \Yvecl{} and \Zvecl{} are defined in analogous way. 
Thus, one can store the information about each term as a tuple of four items $\left(h_l, \Xvecl, \Yvecl, \Zvecl \right)$.

Such representation gives rise to \emph{Hamiltonian arithmetic}: a set of rules to evaluate $\braket{\xvec|\Phatl|\xvecp}$ as a sequence of bitwise operations on $N$-bit vectors representing the term and basis vectors:
\begin{gather}\label{equ:pauli_melem}
    \braket{\xvec|\Phatl|\xvecp} = \begin{cases}
        0 \text{, if } \xvec \XORop \xvecp \neq \XYvecl;\\
        \expe^{i\phi\left(\xvecp, \Phatl \right)} \text{, otherwise, }
    \end{cases} \\
    \text{where } \phi\left(\xvecp, \Phatl \right) \coloneqq \frac{\pi}{2}\left|\Yvecl \right| + \pi \left|\xvecp \ANDop \YZvecl \right|,\nonumber
\end{gather}
where $\ORop$, $\ANDop$ and $\XORop$ denote the bitwise OR, AND and XOR operations, respectively, $\left|\xvec\right| \coloneqq \sum_{i=0}^{N-1} x_i$  is the Hamming weight of a \xvec, $\XYvecl \coloneqq \Xvecl \ORop \Yvecl$ and $\YZvecl \coloneqq \Yvecl \ORop \Zvecl$.

We see that each \Phatl{} couples \xvec{} to a single \xvecp{} which can be calculated as $\xvecp = \xvec \XORop \XYvecl$. In other words, a pair (\xvec{}, \xvecp{}) is coupled by a given Hamiltonian term \Phatl{} if $\XYvecl$ equals $\xvec \XORop \xvecp$. The matrix element itself is just a complex exponent with the phase $\phi\left(\xvecp, \Phatl \right)$.

\subsubsection{Unique \XYvec}
Note, however, that in realistic molecular Hamiltonians \emph{several} \Phatl{} might have the same \XYvec{} representation, i.e. the number of \emph{unique} \XYvecl{} is less than \Nt{}.
Let us denote the set of all unique \XYvecl{} as $\UniqueXYSet$; the size $\left| \UniqueXYSet \right|$ of this set is still $\mathcal{O}(\Nt) =  \mathcal{O}(N^4)$.
For a given \xvec{}, every implementation of \FindSampledAndCoupled{} seeks to find a set of \xvecp{} such that $\xvec \XORop \xvecp \in \UniqueXYSet$.
We denote such set as $\SampledAndCoupledx \coloneqq \Set{\xvecp| \xvec \XORop \xvecp \in \UniqueXYSet}$. 

\subsubsection{\CoupleViaUniqueXY}\label{sec:coupling_to_unique_xy}
The \CoupleViaUniqueXY{} implementation can be described with the following pseudocode:
\algosuffix{1}
\setcounter{algomain}{2}
\begin{algorithm}[H]
    \caption{Find \SampledAndCoupledx{} via looping over unique \XYvecl{}.}\label{algo:couple_to_unique_xy}
    \begin{algorithmic}[1]
        \Function{\CoupleViaUniqueXY}{\xvec}
            \State $\SampledAndCoupledx \coloneqq \PEmptyList$
            \For{\XYvec{} \PIn{} \UniqueXYSet}
                \State $\xvecp \coloneqq \xvec \XORop \XYvec$
                \If{\xvecp{} \PIn{} \Unique}
                    \State $\SampledAndCoupledx\PAppend{\xvecp}$
                \EndIf
            \EndFor
        \State \textbf{return} \SampledAndCoupledx
        \EndFunction
    \end{algorithmic}
\end{algorithm}
Invoking \CoupleViaUniqueXY{} for all \xvec{} in \Unique{} results in the time complexity $\mathcal{O}\left(\Nunq \cdot \Nt \cdot (\BitwiseComplexity + \FindInUComplexity )\right)$.
Here \BitwiseComplexity{} is the complexity of a bitwise operation on a pair of basis vectors, which, in principle, is $\mathcal{O}\left(N\right)$.
However, with the compressed storing of $N$-bit vectors as tuples of integers discussed in Section~\ref{sec:gpu_implementation} it is possible to reduce the scaling constant by the bit depth of those integers (64 in our case).
At the same time, \FindInUComplexity{} is the complexity of checking whether a candidate vector \xvecp{} belongs to the batch of sampled unique vectors.
In Ref.~\cite{transformer_nqs_for_qchem} \Unique{} is stored as an ordered table and the checking is performed via binary search; hence $\FindInUComplexity = \mathcal{O}\left(\log \Nunq \right)$. 
However, implementing binary search posed a significant challenge for our GPU-based implementation.
Hence, in Supplementary Material we propose an alternative approach with similar asymptotic complexity.
\subsubsection{\CoupleAllToAll{}}\label{sec:all_to_all_coupling}
The pseudocode for \CoupleAllToAll{} is as follows:
\setcounter{algomain}{2}
\algosuffix{2}
\begin{algorithm}[H]
    \caption{Find \SampledAndCoupledx{} via looping over \Unique.}\label{algo:couple_all_to_all}
    \begin{algorithmic}[1]
        \Function{\CoupleAllToAll}{\xvec}
            \State $\SampledAndCoupledx \coloneqq \PEmptyList$
            \For{\xvecp{} \PIn{} \Unique}
                \State $\XYvec \coloneqq \xvec \XORop \xvecp$
                \If{\XYvec{} \PIn{} \UniqueXYSet{}}
                    \State $\SampledAndCoupledx\PAppend{\xvecp}$
                \EndIf
            \EndFor
            \State \textbf{return} \SampledAndCoupledx
        \EndFunction
    \end{algorithmic}
\end{algorithm}

Invoking \CoupleViaUniqueXY{} for all \xvec{} in \Unique{} results in the time complexity $\mathcal{O}\left(\Nunq^2 (\BitwiseComplexity + \FindInXYComplexity) \right)$. 
Similarly to \CoupleViaUniqueXY{} $\FindInXYComplexity{}=\mathcal{O}\left(\log \Nt \right)$ is the complexity of checking whether each calculated \XYvec{} belongs to \UniqueXYSet{}. To implement this check in $\ll \mathcal{O}(\Nt)$ operations for each pair, we create an ordered data structure that lists both the $(\xvec, \xvecp)$ pairs and the Hamiltonian terms  encoded as bit vectors in a way enabling rapid lookup, as further explained in the Supplementary. %

\subsubsection{Prefix tree}
\begin{figure}[b]
	\centering
	\includegraphics[width=\linewidth]{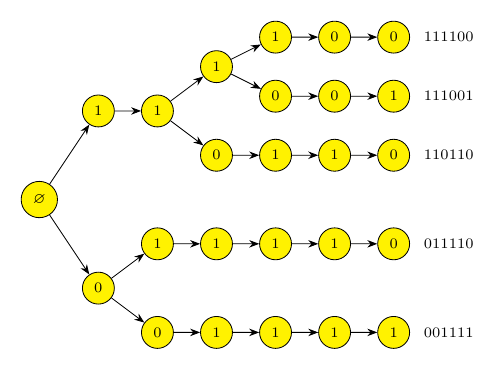}
	\caption{Prefix tree for a set of five bit vectors $\left\lbrace001111, 011110, 110110, 111001, 111100\right\rbrace$. One can see that, for example, the strings $110110$ and $111001$ are encoded by the same first two levels of the prefix tree, while the remaining levels are different for them.}
	\label{fig:prefix_tree}
\end{figure}

Prefix tree (also known as \emph{trie}) is a data structure used to store a set of bit vectors (or more generally strings over some alphabet) in a compressed way.
Suppose one has to store two strings $abcd$ and $abce$.
Both strings share the same prefix $abc$; to avoid keeping redundant information, one might just store the prefix $abc$, two possible endings $d$ and $e$ and the way the endings are connected to the prefix.
The prefix tree develops this idea further and applies it to \emph{all} possible prefixes in a set of strings as illustrated in Fig.~\ref{fig:prefix_tree}.
From a more formal perspective, prefix tree is a directed tree, where the root node corresponds to the ``start of a string'' symbol, ordinary nodes contain string symbols, and every path in the tree from the root to a leaf represents a particular string in the set.

\subsubsection{Prefix tree construction}\label{sec:construct_prefix_tree}
Algorithm~\ref{algo:construct_prefix_tree} presents a pseudocode to construct a prefix tree from a set of unique bit vectors. 
We store the tree \Tree{} as a list of nodes representing unique prefixes.
We presume that each node is a data structure with three fields: \Value{}, storing $x_i$; \Parent{}, storing the reference to a parent node at the previous level; and array of two references \Children{} to children nodes at the next level. 
By default, the references are set to \None{} indicating that the link between the node and its parent/children has not been established yet.
The prefix $\xvec_{<i}$ of a node can be reconstructed by traversing the prefix tree up via \texttt{parent} references.
\nextalgo
\begin{algorithm}[H]
    \caption{Constrution of prefix tree}\label{algo:construct_prefix_tree}
    \begin{algorithmic}[1]
        \Function{ConstructPrefixTree}{\Unique}
            \State $\Tree \coloneqq \left[\  \right]$
            \State $\Tree \PAppend{\Node(\texttt{value}=\varnothing)} $
            \For{\xvec{} \PIn{} \Unique}
                \State $\CurNode \coloneqq \Tree[0]$
                \For{$i$ \PFrom{} 0 \PTo{} $N-1$}
                    \If{$\CurNode.\Children[x_i]$ $=$ \None{}}
                        \State $\NewNode \coloneqq~\Node(\text{\Value=$x_i$})$
                        \State $\NewNode.\Parent \coloneqq \CurNode$
                        \State $\CurNode.\Children[x_i] \coloneqq \NewNode$
                        \State $\Tree \PAppend{\NewNode} $
                    \EndIf
                    \State $\CurNode \coloneqq \CurNode.\Children[x_i]$
                \EndFor
            \EndFor

            \State \textbf{return} \Tree
        \EndFunction
    \end{algorithmic}
\end{algorithm}

\subsubsection{\CoupleViaPrefixTree{}}
As discussed, Algorithm~\ref{algo:couple_all_to_all} evaluates $\xvec{}\XORop\xvecp{}$ for each pair of unique basis vectors and then checks whether the result belongs to \UniqueXYSet{}.
This requires looping over \emph{every} bit in \xvec{} and \xvecp{}.
Importantly, during this loop, Algorithm~\ref{algo:couple_all_to_all} does not try to check whether a \emph{partial} XOR of \xvec{} and \xvecp{} is valid, i.e.~whether it might be a prefix of any \XYvec{} in \UniqueXYSet{}.
Such approach results in unnecessary calculations, since it might become apparent early in the loop that \xvec{} and \xvecp{} cannot be coupled by any term in \Hhat{}.

In Algorithm~\ref{algo:couple_prefix_tree} we address this issue and remove the ``futil'' pairs $(\xvec_{<i}, \xvecp_{<i})$ from consideration as early as possible. 
To that end, we store \emph{both} \Unique{} and \UniqueXYSet{} as prefix trees, which we denote as $\Tree^\Unique$ and $\Tree^\UniqueXYSet$ correspondingly.
For a given \xvec{}, we traverse the prefix tree $\Tree^\Unique$ level-by-level and keep only those paths which correspond to a valid prefix of the same length in $\Tree^\UniqueXYSet$.
Note that checking whether $\XYvec_{<i}$ belongs to $\Tree^\UniqueXYSet$ does not require any search operations.
Instead, since each prefix $\XYvec_{<i}$ is built sequentially starting from the root, one only needs to examine whether the node corresponding to $\XYvec_{<i-1}$ has a child corresponding to $x_i \XORop x^{\prime}_i$.

\setcounter{algomain}{1}
\algosuffix{3}
\begin{algorithm}[H]
    \caption{Find \SampledAndCoupledx{} via prefix tree.}\label{algo:couple_prefix_tree}
    \begin{algorithmic}[1]
        \Function{\CoupleViaPrefixTree{}}{\xvec}
            \State $\UniqueTree \coloneqq \Call{ConstructPrefixTree}{\Unique}$
            \State $\UniqueXYTree \coloneqq \Call{ConstructPrefixTree}{\UniqueXYSet}$
            \Statex \hspace{\algorithmicindent}\texttt{\# Initialise a set of coupled $\xvecp_{<i}$ and corresponding $\XYvec_{<i}$.}
            \State $\SCXYx \coloneqq \left[(\UniqueTree[0], \UniqueXYTree[0]) \right]$
            \Statex            

            \For{$i$ \PFrom{} 0 \PTo{} $N-1$}
                \State $\SCXYx^\prime \coloneqq \left[\ \right]$
                \Statex
                \For{(\UnqNode, \UnqXYNode) \PIn{} \SCXYx}
                    \For{$x_{i}^\prime$ \PIn\ $\left\lbrace0, 1\right\rbrace$}
                        \State $\NewUnqNode \coloneqq \UnqNode.\Children[x_{i}^\prime]$
                        \State $\NewUnqXYNode \coloneqq \UnqXYNode.\Children[x_i \XORop x_i^\prime]$
                        \Statex 
                        \If {\NewUnqNode{}~$\neq$~\None{}~\textbf{and}~\NewUnqXYNode{}~$\neq$~\None{}}
                            \State $\SCXYx^\prime\PAppend{(\NewUnqNode,\ \NewUnqXYNode{})}$
                        \Else
                            \State \textbf{continue}
                        \EndIf
                    \EndFor
                \EndFor
                \State $\SCXYx \coloneqq \SCXYx^\prime$
            \EndFor
            \State Reconstruct \SampledAndCoupledx{} from every \UnqNode{} in \SCXYx{}.
            \State \textbf{return} \SampledAndCoupledx
        \EndFunction
    \end{algorithmic}
\end{algorithm}

\subsubsection{\CoupleViaPrefixTree{} complexity}
To conclude the discussion of \CoupleViaPrefixTree{}, let us consider its worst case computational complexity.
The worst case corresponds to the situation when \emph{all} \xvec{} in \Unique{} are coupled to each other and \emph{no} paths are dropped during the traversal of \UniqueTree{}.
Hence, starting from some level we have $\Nunq{}$ paths.
Consequently, the number of operations required to traverse through $N$ levels of \UniqueTree{} is bounded from above by $\mathcal{O}(N \cdot \Nunq)$.
Since we invoke \CoupleViaPrefixTree{} for each \xvec{} in \Unique{}, the total complexity of \FindSampledAndCoupled{} becomes $\mathcal{O}(N\cdot\Nunq^2)$.
Thus, we recover the complexity of all-to-all coupling Algorithm~\ref{algo:couple_all_to_all}, assuming that $\BitwiseComplexity = \mathcal{O}(N)$.

The above discussion leads us to expect \CoupleViaPrefixTree{} to be at least as fast as \CoupleAllToAll{}.
However, this is not always the case in practice, as evident from Fig.~\ref{fig:findsc_speedups}. 
This is because \CoupleViaPrefixTree{} requires explicit looping over $N$ levels of prefix trees, and thus loses the benefits of the compressed basis vector storing. 
Nevertheless, for large systems efficient exploitation of coupling sparsity outweighs this disadvantage, and \CoupleViaPrefixTree{} does demonstrate practical speedups.

\subsection{Experimental particulars}
\subsubsection{General setup}
All experiments were performed on a single NVIDIA RTX A5000 GPU with 24 GB of RAM.
For a given molecule and its geometry, we obtain the qubit Hamiltonian with OpenFermion software library~\cite{open_fermion}, which uses PySCF~\cite{pyscf2} and Psi4~\cite{psi4} quantum chemistry packages as underlying backends to calculate one- and two-body integrals.
We study all molecules in the minimal basis set STO-3G, unless specified otherwise.
For each molecule we obtain the maximum set of symmetries and ensure symmetry-aware sampling as described in Ref.~\cite{malyshev_anqs_with_quantum_numbers}.
For each geometry we run calculations using three  different seeds of the underlying pseudorandom number generator.
When discussing achieved energies we report the \emph{minimum} value obtained across seeds unless specified otherwise.
For other metrics we report the \emph{mean} value.
We invite the reader interested in specific values of the hyperapameters and/or reproducing the experimental results to explore the supplementary GitHub repository~\cite{anqs_qchem_github_repo}.

\subsubsection{Ansatz architecture}\label{sec:ansatz_architecture}
We employ an ANQS architecture which represents each conditional wave function with \emph{two} real-valued subnetworks, separately encoding absolute value $\log\left|\psi \right|$ and phase $\varphi$ of an amplitude $\psi = \expe^{\log\left|\psi \right| + \imagi \varphi}$. 
This allows us to perform SR gradient postprocessing without worrying about numerical instability issues arising with complex-valued subnetworks.

Each ANQS subnetwork is a multi-layer perceptron with two hidden layers of width 64.
There is a residual connection adding the output of the first hidden layer to the output of the second \emph{before} the second layer is activated.
The hidden layers are activated with the hyperbolic tangent function, while no elementwise activation is applied to the last layer.
Instead, we apply a \emph{global} activation to the last layer which we describe below.

\subsubsection{Local pruning strategy}
At each level of the autoregressive tree, we mask unphysical nodes (set their probabilities to zero) as described in Barrett {\it at al.} \cite{barrett_autoregressive_qchem}. We do not use more sophisticated masking strategies proposed in 
 our earlier work \cite{malyshev_anqs_with_quantum_numbers}. This is due to our lack of understanding how to reconcile discarding the samples required by these strategies with autoregressive sampling without replacement.

\subsubsection{Grouping qubits into qudits}
We alleviate the expressivity restriction brought about by our masking strategy by grouping multiple qubits into qudits and sampling the latter. The number \QuditDepth{} of qubits per qudit is a hyperparameter, which we chose to equal 6 as a result of a study described in Supplementary Material . 
If $N$ is not a multiple of \QuditDepth{}, the last subnetwork groups the remaining $\left(N \mod \QuditDepth\right)$ qubits into a last qudit.
A further advantage of grouping qubits into qudits is faster sampling and amplitude evaluation, which we also cover in Supplementary Material.

\subsubsection{Global activation}
Once the log-amplitude subnetwork produces unnormalised values for $\log \left|\psi \left(x_i |\xvec_{<i}\right)\right|$, we apply a global activation function to them.
Specifically, we shift them by their average, i.e. $\log \left|\psi \left(x_i |\xvec_{<i}\right)\right|  \rightarrow \log \left|\psi \left(x_i |\xvec_{<i}\right)\right|  - \sum_{x_i}\log \left|\psi \left(x_i |\xvec_{<i}\right)\right|$, where the summation over $x_i$ includes all possible $2^{\QuditDepth}$ values of $i$-th qudit.
In our preliminary experiments we observed empirically that such global activation improves the optimisation convergence.

\section*{Acknowledgments} 
MS acknowledges support from the Helmholtz Initiative and Networking Fund, grant no.~VH-NG-1711.

\bibliography{bibliography}

\begin{thebibliography}{73}%
\makeatletter
\providecommand \@ifxundefined [1]{%
 \@ifx{#1\undefined}
}%
\providecommand \@ifnum [1]{%
 \ifnum #1\expandafter \@firstoftwo
 \else \expandafter \@secondoftwo
 \fi
}%
\providecommand \@ifx [1]{%
 \ifx #1\expandafter \@firstoftwo
 \else \expandafter \@secondoftwo
 \fi
}%
\providecommand \natexlab [1]{#1}%
\providecommand \enquote  [1]{``#1''}%
\providecommand \bibnamefont  [1]{#1}%
\providecommand \bibfnamefont [1]{#1}%
\providecommand \citenamefont [1]{#1}%
\providecommand \href@noop [0]{\@secondoftwo}%
\providecommand \href [0]{\begingroup \@sanitize@url \@href}%
\providecommand \@href[1]{\@@startlink{#1}\@@href}%
\providecommand \@@href[1]{\endgroup#1\@@endlink}%
\providecommand \@sanitize@url [0]{\catcode `\\12\catcode `\$12\catcode
  `\&12\catcode `\#12\catcode `\^12\catcode `\_12\catcode `\%12\relax}%
\providecommand \@@startlink[1]{}%
\providecommand \@@endlink[0]{}%
\providecommand \url  [0]{\begingroup\@sanitize@url \@url }%
\providecommand \@url [1]{\endgroup\@href {#1}{\urlprefix }}%
\providecommand \urlprefix  [0]{URL }%
\providecommand \Eprint [0]{\href }%
\providecommand \doibase [0]{https://doi.org/}%
\providecommand \selectlanguage [0]{\@gobble}%
\providecommand \bibinfo  [0]{\@secondoftwo}%
\providecommand \bibfield  [0]{\@secondoftwo}%
\providecommand \translation [1]{[#1]}%
\providecommand \BibitemOpen [0]{}%
\providecommand \bibitemStop [0]{}%
\providecommand \bibitemNoStop [0]{.\EOS\space}%
\providecommand \EOS [0]{\spacefactor3000\relax}%
\providecommand \BibitemShut  [1]{\csname bibitem#1\endcsname}%
\let\auto@bib@innerbib\@empty
%</preamble>
\bibitem [{\citenamefont {Carleo}\ and\ \citenamefont
  {Troyer}(2017)}]{carleo_troyer_rbm}%
  \BibitemOpen
  \bibfield  {author} {\bibinfo {author} {\bibfnamefont {G.}~\bibnamefont
  {Carleo}}\ and\ \bibinfo {author} {\bibfnamefont {M.}~\bibnamefont
  {Troyer}},\ }\bibfield  {title} {\bibinfo {title} {Solving the quantum
  many-body problem with artificial neural networks},\ }\href
  {https://doi.org/10.1126/science.aag2302} {\bibfield  {journal} {\bibinfo
  {journal} {Science}\ }\textbf {\bibinfo {volume} {355}},\ \bibinfo {pages}
  {602} (\bibinfo {year} {2017})}\BibitemShut {NoStop}%
\bibitem [{\citenamefont {Medvidović}\ and\ \citenamefont
  {Moreno}(2024)}]{medvidovic_nqs_review}%
  \BibitemOpen
  \bibfield  {author} {\bibinfo {author} {\bibfnamefont {M.}~\bibnamefont
  {Medvidović}}\ and\ \bibinfo {author} {\bibfnamefont {J.~R.}\ \bibnamefont
  {Moreno}},\ }\href@noop {} {\bibinfo {title} {Neural-network quantum states
  for many-body physics}} (\bibinfo {year} {2024}),\ \Eprint
  {https://arxiv.org/abs/2402.11014} {arXiv:2402.11014 [cond-mat.dis-nn]}
  \BibitemShut {NoStop}%
\bibitem [{\citenamefont {Lange}\ \emph {et~al.}(2024)\citenamefont {Lange},
  \citenamefont {de~Walle}, \citenamefont {Abedinnia},\ and\ \citenamefont
  {Bohrdt}}]{lange_nqs_review}%
  \BibitemOpen
  \bibfield  {author} {\bibinfo {author} {\bibfnamefont {H.}~\bibnamefont
  {Lange}}, \bibinfo {author} {\bibfnamefont {A.~V.}\ \bibnamefont {de~Walle}},
  \bibinfo {author} {\bibfnamefont {A.}~\bibnamefont {Abedinnia}},\ and\
  \bibinfo {author} {\bibfnamefont {A.}~\bibnamefont {Bohrdt}},\ }\href@noop {}
  {\bibinfo {title} {From architectures to applications: A review of neural
  quantum states}} (\bibinfo {year} {2024}),\ \Eprint
  {https://arxiv.org/abs/2402.09402} {arXiv:2402.09402 [cond-mat.dis-nn]}
  \BibitemShut {NoStop}%
\bibitem [{\citenamefont {Hermann}\ \emph {et~al.}(2023)\citenamefont
  {Hermann}, \citenamefont {Spencer}, \citenamefont {Choo}, \citenamefont
  {Mezzacapo}, \citenamefont {Foulkes}, \citenamefont {Pfau}, \citenamefont
  {Carleo},\ and\ \citenamefont {Noé}}]{Hermann2023}%
  \BibitemOpen
  \bibfield  {author} {\bibinfo {author} {\bibfnamefont {J.}~\bibnamefont
  {Hermann}}, \bibinfo {author} {\bibfnamefont {J.}~\bibnamefont {Spencer}},
  \bibinfo {author} {\bibfnamefont {K.}~\bibnamefont {Choo}}, \bibinfo {author}
  {\bibfnamefont {A.}~\bibnamefont {Mezzacapo}}, \bibinfo {author}
  {\bibfnamefont {W.~M.~C.}\ \bibnamefont {Foulkes}}, \bibinfo {author}
  {\bibfnamefont {D.}~\bibnamefont {Pfau}}, \bibinfo {author} {\bibfnamefont
  {G.}~\bibnamefont {Carleo}},\ and\ \bibinfo {author} {\bibfnamefont
  {F.}~\bibnamefont {Noé}},\ }\bibfield  {title} {\bibinfo {title} {Ab initio
  quantum chemistry with neural-network wavefunctions},\ }\href
  {https://doi.org/10.1038/s41570-023-00516-8} {\bibfield  {journal} {\bibinfo
  {journal} {Nature Reviews Chemistry}\ }\textbf {\bibinfo {volume} {7}},\
  \bibinfo {pages} {692–709} (\bibinfo {year} {2023})}\BibitemShut {NoStop}%
\bibitem [{\citenamefont {Choo}\ \emph {et~al.}(2018)\citenamefont {Choo},
  \citenamefont {Carleo}, \citenamefont {Regnault},\ and\ \citenamefont
  {Neupert}}]{carleo_symmetries}%
  \BibitemOpen
  \bibfield  {author} {\bibinfo {author} {\bibfnamefont {K.}~\bibnamefont
  {Choo}}, \bibinfo {author} {\bibfnamefont {G.}~\bibnamefont {Carleo}},
  \bibinfo {author} {\bibfnamefont {N.}~\bibnamefont {Regnault}},\ and\
  \bibinfo {author} {\bibfnamefont {T.}~\bibnamefont {Neupert}},\ }\bibfield
  {title} {\bibinfo {title} {Symmetries and many-body excitations with
  neural-network quantum states},\ }\href
  {https://link.aps.org/doi/10.1103/PhysRevLett.121.167204} {\bibfield
  {journal} {\bibinfo  {journal} {Phys. Rev. Lett.}\ }\textbf {\bibinfo
  {volume} {121}},\ \bibinfo {pages} {167204} (\bibinfo {year}
  {2018})}\BibitemShut {NoStop}%
\bibitem [{\citenamefont {Choo}\ \emph
  {et~al.}(2019{\natexlab{a}})\citenamefont {Choo}, \citenamefont {Neupert},\
  and\ \citenamefont {Carleo}}]{carleo_convolutional}%
  \BibitemOpen
  \bibfield  {author} {\bibinfo {author} {\bibfnamefont {K.}~\bibnamefont
  {Choo}}, \bibinfo {author} {\bibfnamefont {T.}~\bibnamefont {Neupert}},\ and\
  \bibinfo {author} {\bibfnamefont {G.}~\bibnamefont {Carleo}},\ }\bibfield
  {title} {\bibinfo {title} {Two-dimensional frustrated
  ${J}_{1}\text{\ensuremath{-}}{J}_{2}$ model studied with neural network
  quantum states},\ }\href
  {https://link.aps.org/doi/10.1103/PhysRevB.100.125124} {\bibfield  {journal}
  {\bibinfo  {journal} {Phys. Rev. B}\ }\textbf {\bibinfo {volume} {100}},\
  \bibinfo {pages} {125124} (\bibinfo {year} {2019}{\natexlab{a}})}\BibitemShut
  {NoStop}%
\bibitem [{\citenamefont {Hibat-Allah}\ \emph {et~al.}(2020)\citenamefont
  {Hibat-Allah}, \citenamefont {Ganahl}, \citenamefont {Hayward}, \citenamefont
  {Melko},\ and\ \citenamefont {Carrasquilla}}]{hibbat_allah_recurrent}%
  \BibitemOpen
  \bibfield  {author} {\bibinfo {author} {\bibfnamefont {M.}~\bibnamefont
  {Hibat-Allah}}, \bibinfo {author} {\bibfnamefont {M.}~\bibnamefont {Ganahl}},
  \bibinfo {author} {\bibfnamefont {L.~E.}\ \bibnamefont {Hayward}}, \bibinfo
  {author} {\bibfnamefont {R.~G.}\ \bibnamefont {Melko}},\ and\ \bibinfo
  {author} {\bibfnamefont {J.}~\bibnamefont {Carrasquilla}},\ }\bibfield
  {title} {\bibinfo {title} {Recurrent neural network wave functions},\
  }\href@noop {} {\bibfield  {journal} {\bibinfo  {journal} {Physical Review
  Research}\ }\textbf {\bibinfo {volume} {2}},\ \bibinfo {pages} {023358}
  (\bibinfo {year} {2020})}\BibitemShut {NoStop}%
\bibitem [{\citenamefont {Viteritti}\ \emph {et~al.}(2022)\citenamefont
  {Viteritti}, \citenamefont {Rende},\ and\ \citenamefont
  {Becca}}]{transformer_j1_j2}%
  \BibitemOpen
  \bibfield  {author} {\bibinfo {author} {\bibfnamefont {L.~L.}\ \bibnamefont
  {Viteritti}}, \bibinfo {author} {\bibfnamefont {R.}~\bibnamefont {Rende}},\
  and\ \bibinfo {author} {\bibfnamefont {F.}~\bibnamefont {Becca}},\ }\bibfield
   {title} {\bibinfo {title} {Transformer variational wave functions for
  frustrated quantum spin systems},\ }\href@noop {} {\bibfield  {journal}
  {\bibinfo  {journal} {arXiv preprint arXiv:2211.05504}\ } (\bibinfo {year}
  {2022})}\BibitemShut {NoStop}%
\bibitem [{\citenamefont {Lange}\ \emph {et~al.}(2023)\citenamefont {Lange},
  \citenamefont {Döschl}, \citenamefont {Carrasquilla},\ and\ \citenamefont
  {Bohrdt}}]{bohrdt_doped_antiferromagnets}%
  \BibitemOpen
  \bibfield  {author} {\bibinfo {author} {\bibfnamefont {H.}~\bibnamefont
  {Lange}}, \bibinfo {author} {\bibfnamefont {F.}~\bibnamefont {Döschl}},
  \bibinfo {author} {\bibfnamefont {J.}~\bibnamefont {Carrasquilla}},\ and\
  \bibinfo {author} {\bibfnamefont {A.}~\bibnamefont {Bohrdt}},\ }\href@noop {}
  {\bibinfo {title} {Neural network approach to quasiparticle dispersions in
  doped antiferromagnets}} (\bibinfo {year} {2023}),\ \Eprint
  {https://arxiv.org/abs/2310.08578} {arXiv:2310.08578 [cond-mat.str-el]}
  \BibitemShut {NoStop}%
\bibitem [{\citenamefont {Nomura}(2021)}]{nomura_rbm_j1_j2}%
  \BibitemOpen
  \bibfield  {author} {\bibinfo {author} {\bibfnamefont {Y.}~\bibnamefont
  {Nomura}},\ }\bibfield  {title} {\bibinfo {title} {Helping restricted
  boltzmann machines with quantum-state representation by restoring symmetry},\
  }\href@noop {} {\bibfield  {journal} {\bibinfo  {journal} {Journal of
  Physics: Condensed Matter}\ }\textbf {\bibinfo {volume} {33}},\ \bibinfo
  {pages} {174003} (\bibinfo {year} {2021})}\BibitemShut {NoStop}%
\bibitem [{\citenamefont {Roth}\ and\ \citenamefont
  {MacDonald}(2021)}]{roth_group_convolutional}%
  \BibitemOpen
  \bibfield  {author} {\bibinfo {author} {\bibfnamefont {C.}~\bibnamefont
  {Roth}}\ and\ \bibinfo {author} {\bibfnamefont {A.~H.}\ \bibnamefont
  {MacDonald}},\ }\bibfield  {title} {\bibinfo {title} {Group convolutional
  neural networks improve quantum state accuracy},\ }\href@noop {} {\bibfield
  {journal} {\bibinfo  {journal} {arXiv preprint arXiv:2104.05085}\ } (\bibinfo
  {year} {2021})}\BibitemShut {NoStop}%
\bibitem [{\citenamefont {Pei}\ and\ \citenamefont
  {Clark}(2024)}]{pei_bose_hubbard}%
  \BibitemOpen
  \bibfield  {author} {\bibinfo {author} {\bibfnamefont {M.~Y.}\ \bibnamefont
  {Pei}}\ and\ \bibinfo {author} {\bibfnamefont {S.~R.}\ \bibnamefont
  {Clark}},\ }\href@noop {} {\bibinfo {title} {Specialising neural-network
  quantum states for the bose hubbard model}} (\bibinfo {year} {2024}),\
  \Eprint {https://arxiv.org/abs/2402.15424} {arXiv:2402.15424
  [cond-mat.quant-gas]} \BibitemShut {NoStop}%
\bibitem [{\citenamefont {Adams}\ \emph {et~al.}(2021)\citenamefont {Adams},
  \citenamefont {Carleo}, \citenamefont {Lovato},\ and\ \citenamefont
  {Rocco}}]{nqs_nuclei_1}%
  \BibitemOpen
  \bibfield  {author} {\bibinfo {author} {\bibfnamefont {C.}~\bibnamefont
  {Adams}}, \bibinfo {author} {\bibfnamefont {G.}~\bibnamefont {Carleo}},
  \bibinfo {author} {\bibfnamefont {A.}~\bibnamefont {Lovato}},\ and\ \bibinfo
  {author} {\bibfnamefont {N.}~\bibnamefont {Rocco}},\ }\bibfield  {title}
  {\bibinfo {title} {Variational monte carlo calculations of
  $a\ensuremath{\le}4$ nuclei with an artificial neural-network correlator
  ansatz},\ }\href {https://doi.org/10.1103/PhysRevLett.127.022502} {\bibfield
  {journal} {\bibinfo  {journal} {Phys. Rev. Lett.}\ }\textbf {\bibinfo
  {volume} {127}},\ \bibinfo {pages} {022502} (\bibinfo {year}
  {2021})}\BibitemShut {NoStop}%
\bibitem [{\citenamefont {Lovato}\ \emph {et~al.}(2022)\citenamefont {Lovato},
  \citenamefont {Adams}, \citenamefont {Carleo},\ and\ \citenamefont
  {Rocco}}]{nqs_nuclei_2}%
  \BibitemOpen
  \bibfield  {author} {\bibinfo {author} {\bibfnamefont {A.}~\bibnamefont
  {Lovato}}, \bibinfo {author} {\bibfnamefont {C.}~\bibnamefont {Adams}},
  \bibinfo {author} {\bibfnamefont {G.}~\bibnamefont {Carleo}},\ and\ \bibinfo
  {author} {\bibfnamefont {N.}~\bibnamefont {Rocco}},\ }\bibfield  {title}
  {\bibinfo {title} {Hidden-nucleons neural-network quantum states for the
  nuclear many-body problem},\ }\href
  {https://doi.org/10.1103/PhysRevResearch.4.043178} {\bibfield  {journal}
  {\bibinfo  {journal} {Phys. Rev. Res.}\ }\textbf {\bibinfo {volume} {4}},\
  \bibinfo {pages} {043178} (\bibinfo {year} {2022})}\BibitemShut {NoStop}%
\bibitem [{\citenamefont {Yang}\ and\ \citenamefont
  {Zhao}(2023)}]{nqs_nuclei_3}%
  \BibitemOpen
  \bibfield  {author} {\bibinfo {author} {\bibfnamefont {Y.~L.}\ \bibnamefont
  {Yang}}\ and\ \bibinfo {author} {\bibfnamefont {P.~W.}\ \bibnamefont
  {Zhao}},\ }\bibfield  {title} {\bibinfo {title} {Deep-neural-network approach
  to solving the ab initio nuclear structure problem},\ }\href
  {https://doi.org/10.1103/PhysRevC.107.034320} {\bibfield  {journal} {\bibinfo
   {journal} {Phys. Rev. C}\ }\textbf {\bibinfo {volume} {107}},\ \bibinfo
  {pages} {034320} (\bibinfo {year} {2023})}\BibitemShut {NoStop}%
\bibitem [{\citenamefont {Pfau}\ \emph {et~al.}(2020)\citenamefont {Pfau},
  \citenamefont {Spencer}, \citenamefont {Matthews},\ and\ \citenamefont
  {Foulkes}}]{fermi_net}%
  \BibitemOpen
  \bibfield  {author} {\bibinfo {author} {\bibfnamefont {D.}~\bibnamefont
  {Pfau}}, \bibinfo {author} {\bibfnamefont {J.~S.}\ \bibnamefont {Spencer}},
  \bibinfo {author} {\bibfnamefont {A.~G. D.~G.}\ \bibnamefont {Matthews}},\
  and\ \bibinfo {author} {\bibfnamefont {W.~M.~C.}\ \bibnamefont {Foulkes}},\
  }\bibfield  {title} {\bibinfo {title} {Ab initio solution of the
  many-electron schr\"odinger equation with deep neural networks},\ }\href
  {https://doi.org/10.1103/PhysRevResearch.2.033429} {\bibfield  {journal}
  {\bibinfo  {journal} {Phys. Rev. Res.}\ }\textbf {\bibinfo {volume} {2}},\
  \bibinfo {pages} {033429} (\bibinfo {year} {2020})}\BibitemShut {NoStop}%
\bibitem [{\citenamefont {Hermann}\ \emph {et~al.}(2020)\citenamefont
  {Hermann}, \citenamefont {Sch{\"a}tzle},\ and\ \citenamefont
  {No{\'e}}}]{pauli_net}%
  \BibitemOpen
  \bibfield  {author} {\bibinfo {author} {\bibfnamefont {J.}~\bibnamefont
  {Hermann}}, \bibinfo {author} {\bibfnamefont {Z.}~\bibnamefont
  {Sch{\"a}tzle}},\ and\ \bibinfo {author} {\bibfnamefont {F.}~\bibnamefont
  {No{\'e}}},\ }\bibfield  {title} {\bibinfo {title} {Deep-neural-network
  solution of the electronic schr{\"o}dinger equation},\ }\href
  {https://doi.org/https://doi.org/10.1038/s41557-020-0544-y} {\bibfield
  {journal} {\bibinfo  {journal} {Nature Chemistry}\ }\textbf {\bibinfo
  {volume} {12}},\ \bibinfo {pages} {891} (\bibinfo {year} {2020})}\BibitemShut
  {NoStop}%
\bibitem [{\citenamefont {Neklyudov}\ \emph {et~al.}(2023)\citenamefont
  {Neklyudov}, \citenamefont {Nys}, \citenamefont {Thiede}, \citenamefont
  {Carrasquilla}, \citenamefont {Liu}, \citenamefont {Welling},\ and\
  \citenamefont {Makhzani}}]{nekluydov_wasserstein}%
  \BibitemOpen
  \bibfield  {author} {\bibinfo {author} {\bibfnamefont {K.}~\bibnamefont
  {Neklyudov}}, \bibinfo {author} {\bibfnamefont {J.}~\bibnamefont {Nys}},
  \bibinfo {author} {\bibfnamefont {L.}~\bibnamefont {Thiede}}, \bibinfo
  {author} {\bibfnamefont {J.}~\bibnamefont {Carrasquilla}}, \bibinfo {author}
  {\bibfnamefont {Q.}~\bibnamefont {Liu}}, \bibinfo {author} {\bibfnamefont
  {M.}~\bibnamefont {Welling}},\ and\ \bibinfo {author} {\bibfnamefont
  {A.}~\bibnamefont {Makhzani}},\ }\bibfield  {title} {\bibinfo {title}
  {Wasserstein quantum monte carlo: A novel approach for solving the quantum
  many-body schr\"{o}dinger equation},\ }in\ \href
  {https://proceedings.neurips.cc/paper_files/paper/2023/file/c8450235f227f136242f774b2799581f-Paper-Conference.pdf}
  {\emph {\bibinfo {booktitle} {Advances in Neural Information Processing
  Systems}}},\ Vol.~\bibinfo {volume} {36},\ \bibinfo {editor} {edited by\
  \bibinfo {editor} {\bibfnamefont {A.}~\bibnamefont {Oh}}, \bibinfo {editor}
  {\bibfnamefont {T.}~\bibnamefont {Neumann}}, \bibinfo {editor} {\bibfnamefont
  {A.}~\bibnamefont {Globerson}}, \bibinfo {editor} {\bibfnamefont
  {K.}~\bibnamefont {Saenko}}, \bibinfo {editor} {\bibfnamefont
  {M.}~\bibnamefont {Hardt}},\ and\ \bibinfo {editor} {\bibfnamefont
  {S.}~\bibnamefont {Levine}}}\ (\bibinfo  {publisher} {Curran Associates,
  Inc.},\ \bibinfo {year} {2023})\ pp.\ \bibinfo {pages}
  {63461--63482}\BibitemShut {NoStop}%
\bibitem [{\citenamefont {Choo}\ \emph
  {et~al.}(2019{\natexlab{b}})\citenamefont {Choo}, \citenamefont {Neupert},\
  and\ \citenamefont {Carleo}}]{choo_nqs_j1_j2}%
  \BibitemOpen
  \bibfield  {author} {\bibinfo {author} {\bibfnamefont {K.}~\bibnamefont
  {Choo}}, \bibinfo {author} {\bibfnamefont {T.}~\bibnamefont {Neupert}},\ and\
  \bibinfo {author} {\bibfnamefont {G.}~\bibnamefont {Carleo}},\ }\bibfield
  {title} {\bibinfo {title} {Two-dimensional frustrated j 1- j 2 model studied
  with neural network quantum states},\ }\href@noop {} {\bibfield  {journal}
  {\bibinfo  {journal} {Physical Review B}\ }\textbf {\bibinfo {volume}
  {100}},\ \bibinfo {pages} {125124} (\bibinfo {year}
  {2019}{\natexlab{b}})}\BibitemShut {NoStop}%
\bibitem [{\citenamefont {Barrett}\ \emph {et~al.}(2022)\citenamefont
  {Barrett}, \citenamefont {Malyshev},\ and\ \citenamefont
  {Lvovsky}}]{barrett_autoregressive_qchem}%
  \BibitemOpen
  \bibfield  {author} {\bibinfo {author} {\bibfnamefont {T.~D.}\ \bibnamefont
  {Barrett}}, \bibinfo {author} {\bibfnamefont {A.}~\bibnamefont {Malyshev}},\
  and\ \bibinfo {author} {\bibfnamefont {A.~I.}\ \bibnamefont {Lvovsky}},\
  }\bibfield  {title} {\bibinfo {title} {Autoregressive neural-network
  wavefunctions for ab initio quantum chemistry},\ }\href
  {https://doi.org/10.1038/s42256-022-00461-z} {\bibfield  {journal} {\bibinfo
  {journal} {Nature Machine Intelligence}\ }\textbf {\bibinfo {volume} {4}},\
  \bibinfo {pages} {351} (\bibinfo {year} {2022})}\BibitemShut {NoStop}%
\bibitem [{\citenamefont {Zhao}\ \emph {et~al.}(2023)\citenamefont {Zhao},
  \citenamefont {Stokes},\ and\ \citenamefont
  {Veerapaneni}}]{zhao_scalable_qchem}%
  \BibitemOpen
  \bibfield  {author} {\bibinfo {author} {\bibfnamefont {T.}~\bibnamefont
  {Zhao}}, \bibinfo {author} {\bibfnamefont {J.}~\bibnamefont {Stokes}},\ and\
  \bibinfo {author} {\bibfnamefont {S.}~\bibnamefont {Veerapaneni}},\
  }\bibfield  {title} {\bibinfo {title} {Scalable neural quantum states
  architecture for quantum chemistry},\ }\href
  {https://doi.org/10.1088/2632-2153/acdb2f} {\bibfield  {journal} {\bibinfo
  {journal} {Machine Learning: Science and Technology}\ }\textbf {\bibinfo
  {volume} {4}},\ \bibinfo {pages} {025034} (\bibinfo {year}
  {2023})}\BibitemShut {NoStop}%
\bibitem [{\citenamefont {Wu}\ \emph {et~al.}(2023)\citenamefont {Wu},
  \citenamefont {Guo}, \citenamefont {Fan}, \citenamefont {Zhou},\ and\
  \citenamefont {Shang}}]{transformer_nqs_for_qchem}%
  \BibitemOpen
  \bibfield  {author} {\bibinfo {author} {\bibfnamefont {Y.}~\bibnamefont
  {Wu}}, \bibinfo {author} {\bibfnamefont {C.}~\bibnamefont {Guo}}, \bibinfo
  {author} {\bibfnamefont {Y.}~\bibnamefont {Fan}}, \bibinfo {author}
  {\bibfnamefont {P.}~\bibnamefont {Zhou}},\ and\ \bibinfo {author}
  {\bibfnamefont {H.}~\bibnamefont {Shang}},\ }\bibfield  {title} {\bibinfo
  {title} {Nnqs-transformer: an efficient and scalable neural network quantum
  states approach for ab initio quantum chemistry},\ }in\ \href
  {https://doi.org/10.1145/3581784.3607061} {\emph {\bibinfo {booktitle}
  {Proceedings of the International Conference for High Performance Computing,
  Networking, Storage and Analysis}}},\ \bibinfo {series and number} {SC '23}\
  (\bibinfo  {publisher} {Association for Computing Machinery},\ \bibinfo
  {address} {New York, NY, USA},\ \bibinfo {year} {2023})\BibitemShut {NoStop}%
\bibitem [{\citenamefont {Malyshev}\ \emph {et~al.}(2023)\citenamefont
  {Malyshev}, \citenamefont {Arrazola},\ and\ \citenamefont
  {Lvovsky}}]{malyshev_anqs_with_quantum_numbers}%
  \BibitemOpen
  \bibfield  {author} {\bibinfo {author} {\bibfnamefont {A.}~\bibnamefont
  {Malyshev}}, \bibinfo {author} {\bibfnamefont {J.~M.}\ \bibnamefont
  {Arrazola}},\ and\ \bibinfo {author} {\bibfnamefont {A.~I.}\ \bibnamefont
  {Lvovsky}},\ }\href@noop {} {\bibinfo {title} {Autoregressive neural quantum
  states with quantum number symmetries}} (\bibinfo {year} {2023}),\ \Eprint
  {https://arxiv.org/abs/2310.04166} {arXiv:2310.04166 [quant-ph]} \BibitemShut
  {NoStop}%
\bibitem [{\citenamefont {Yoshioka}\ \emph {et~al.}(2021)\citenamefont
  {Yoshioka}, \citenamefont {Mizukami},\ and\ \citenamefont
  {Nori}}]{yoshioka_crystalline_rbm}%
  \BibitemOpen
  \bibfield  {author} {\bibinfo {author} {\bibfnamefont {N.}~\bibnamefont
  {Yoshioka}}, \bibinfo {author} {\bibfnamefont {W.}~\bibnamefont {Mizukami}},\
  and\ \bibinfo {author} {\bibfnamefont {F.}~\bibnamefont {Nori}},\ }\bibfield
  {title} {\bibinfo {title} {Solving quasiparticle band spectra of real solids
  using neural-network quantum states},\ }\href@noop {} {\bibfield  {journal}
  {\bibinfo  {journal} {Communications Physics}\ }\textbf {\bibinfo {volume}
  {4}},\ \bibinfo {pages} {1} (\bibinfo {year} {2021})}\BibitemShut {NoStop}%
\bibitem [{\citenamefont {Liu}\ and\ \citenamefont
  {Clark}(2024)}]{liu_backflow_quantum_chemistry}%
  \BibitemOpen
  \bibfield  {author} {\bibinfo {author} {\bibfnamefont {A.-J.}\ \bibnamefont
  {Liu}}\ and\ \bibinfo {author} {\bibfnamefont {B.~K.}\ \bibnamefont
  {Clark}},\ }\href@noop {} {\bibinfo {title} {Neural network backflow for
  ab-initio quantum chemistry}} (\bibinfo {year} {2024}),\ \Eprint
  {https://arxiv.org/abs/2403.03286} {arXiv:2403.03286 [physics.chem-ph]}
  \BibitemShut {NoStop}%
\bibitem [{\citenamefont {Schmitt}\ and\ \citenamefont
  {Heyl}(2020)}]{schmitt_dynamics_in_2d}%
  \BibitemOpen
  \bibfield  {author} {\bibinfo {author} {\bibfnamefont {M.}~\bibnamefont
  {Schmitt}}\ and\ \bibinfo {author} {\bibfnamefont {M.}~\bibnamefont {Heyl}},\
  }\bibfield  {title} {\bibinfo {title} {Quantum many-body dynamics in two
  dimensions with artificial neural networks},\ }\href@noop {} {\bibfield
  {journal} {\bibinfo  {journal} {Physical Review Letters}\ }\textbf {\bibinfo
  {volume} {125}},\ \bibinfo {pages} {100503} (\bibinfo {year}
  {2020})}\BibitemShut {NoStop}%
\bibitem [{\citenamefont {Schmitt}\ \emph {et~al.}(2022)\citenamefont
  {Schmitt}, \citenamefont {Rams}, \citenamefont {Dziarmaga}, \citenamefont
  {Heyl},\ and\ \citenamefont {Zurek}}]{Schmitt2022}%
  \BibitemOpen
  \bibfield  {author} {\bibinfo {author} {\bibfnamefont {M.}~\bibnamefont
  {Schmitt}}, \bibinfo {author} {\bibfnamefont {M.~M.}\ \bibnamefont {Rams}},
  \bibinfo {author} {\bibfnamefont {J.}~\bibnamefont {Dziarmaga}}, \bibinfo
  {author} {\bibfnamefont {M.}~\bibnamefont {Heyl}},\ and\ \bibinfo {author}
  {\bibfnamefont {W.~H.}\ \bibnamefont {Zurek}},\ }\bibfield  {title} {\bibinfo
  {title} {Quantum phase transition dynamics in the two-dimensional
  transverse-field ising model},\ }\href
  {https://doi.org/10.1126/sciadv.abl6850} {\bibfield  {journal} {\bibinfo
  {journal} {Science Advances}\ }\textbf {\bibinfo {volume} {8}},\ \bibinfo
  {pages} {eabl6850} (\bibinfo {year} {2022})},\ \Eprint
  {https://arxiv.org/abs/https://www.science.org/doi/pdf/10.1126/sciadv.abl6850}
  {https://www.science.org/doi/pdf/10.1126/sciadv.abl6850} \BibitemShut
  {NoStop}%
\bibitem [{\citenamefont {Donatella}\ \emph {et~al.}(2023)\citenamefont
  {Donatella}, \citenamefont {Denis}, \citenamefont {Le~Boit\'e},\ and\
  \citenamefont {Ciuti}}]{zachary_anqs_dynamics}%
  \BibitemOpen
  \bibfield  {author} {\bibinfo {author} {\bibfnamefont {K.}~\bibnamefont
  {Donatella}}, \bibinfo {author} {\bibfnamefont {Z.}~\bibnamefont {Denis}},
  \bibinfo {author} {\bibfnamefont {A.}~\bibnamefont {Le~Boit\'e}},\ and\
  \bibinfo {author} {\bibfnamefont {C.}~\bibnamefont {Ciuti}},\ }\bibfield
  {title} {\bibinfo {title} {Dynamics with autoregressive neural quantum
  states: Application to critical quench dynamics},\ }\href
  {https://doi.org/10.1103/PhysRevA.108.022210} {\bibfield  {journal} {\bibinfo
   {journal} {Phys. Rev. A}\ }\textbf {\bibinfo {volume} {108}},\ \bibinfo
  {pages} {022210} (\bibinfo {year} {2023})}\BibitemShut {NoStop}%
\bibitem [{\citenamefont {Sinibaldi}\ \emph {et~al.}(2023)\citenamefont
  {Sinibaldi}, \citenamefont {Giuliani}, \citenamefont {Carleo},\ and\
  \citenamefont {Vicentini}}]{sinibaldi_unbiasing}%
  \BibitemOpen
  \bibfield  {author} {\bibinfo {author} {\bibfnamefont {A.}~\bibnamefont
  {Sinibaldi}}, \bibinfo {author} {\bibfnamefont {C.}~\bibnamefont {Giuliani}},
  \bibinfo {author} {\bibfnamefont {G.}~\bibnamefont {Carleo}},\ and\ \bibinfo
  {author} {\bibfnamefont {F.}~\bibnamefont {Vicentini}},\ }\bibfield  {title}
  {\bibinfo {title} {Unbiasing time-dependent {V}ariational {M}onte {C}arlo by
  projected quantum evolution},\ }\href
  {https://doi.org/10.22331/q-2023-10-10-1131} {\bibfield  {journal} {\bibinfo
  {journal} {{Quantum}}\ }\textbf {\bibinfo {volume} {7}},\ \bibinfo {pages}
  {1131} (\bibinfo {year} {2023})}\BibitemShut {NoStop}%
\bibitem [{\citenamefont {Mendes-Santos}\ \emph {et~al.}(2023)\citenamefont
  {Mendes-Santos}, \citenamefont {Schmitt},\ and\ \citenamefont
  {Heyl}}]{MendesSantos2023}%
  \BibitemOpen
  \bibfield  {author} {\bibinfo {author} {\bibfnamefont {T.}~\bibnamefont
  {Mendes-Santos}}, \bibinfo {author} {\bibfnamefont {M.}~\bibnamefont
  {Schmitt}},\ and\ \bibinfo {author} {\bibfnamefont {M.}~\bibnamefont
  {Heyl}},\ }\bibfield  {title} {\bibinfo {title} {Highly resolved spectral
  functions of two-dimensional systems with neural quantum states},\ }\href
  {https://doi.org/10.1103/PhysRevLett.131.046501} {\bibfield  {journal}
  {\bibinfo  {journal} {Phys. Rev. Lett.}\ }\textbf {\bibinfo {volume} {131}},\
  \bibinfo {pages} {046501} (\bibinfo {year} {2023})}\BibitemShut {NoStop}%
\bibitem [{\citenamefont {Mendes-Santos}\ \emph {et~al.}(2024)\citenamefont
  {Mendes-Santos}, \citenamefont {Schmitt}, \citenamefont {Angelone},
  \citenamefont {Rodriguez}, \citenamefont {Scholl}, \citenamefont {Williams},
  \citenamefont {Barredo}, \citenamefont {Lahaye}, \citenamefont {Browaeys},
  \citenamefont {Heyl},\ and\ \citenamefont {Dalmonte}}]{MendesSantos2024}%
  \BibitemOpen
  \bibfield  {author} {\bibinfo {author} {\bibfnamefont {T.}~\bibnamefont
  {Mendes-Santos}}, \bibinfo {author} {\bibfnamefont {M.}~\bibnamefont
  {Schmitt}}, \bibinfo {author} {\bibfnamefont {A.}~\bibnamefont {Angelone}},
  \bibinfo {author} {\bibfnamefont {A.}~\bibnamefont {Rodriguez}}, \bibinfo
  {author} {\bibfnamefont {P.}~\bibnamefont {Scholl}}, \bibinfo {author}
  {\bibfnamefont {H.~J.}\ \bibnamefont {Williams}}, \bibinfo {author}
  {\bibfnamefont {D.}~\bibnamefont {Barredo}}, \bibinfo {author} {\bibfnamefont
  {T.}~\bibnamefont {Lahaye}}, \bibinfo {author} {\bibfnamefont
  {A.}~\bibnamefont {Browaeys}}, \bibinfo {author} {\bibfnamefont
  {M.}~\bibnamefont {Heyl}},\ and\ \bibinfo {author} {\bibfnamefont
  {M.}~\bibnamefont {Dalmonte}},\ }\bibfield  {title} {\bibinfo {title}
  {Wave-function network description and kolmogorov complexity of quantum
  many-body systems},\ }\href {https://doi.org/10.1103/PhysRevX.14.021029}
  {\bibfield  {journal} {\bibinfo  {journal} {Phys. Rev. X}\ }\textbf {\bibinfo
  {volume} {14}},\ \bibinfo {pages} {021029} (\bibinfo {year}
  {2024})}\BibitemShut {NoStop}%
\bibitem [{\citenamefont {Jónsson}\ \emph {et~al.}(2018)\citenamefont
  {Jónsson}, \citenamefont {Bauer},\ and\ \citenamefont
  {Carleo}}]{nnqs_for_quantum_computing}%
  \BibitemOpen
  \bibfield  {author} {\bibinfo {author} {\bibfnamefont {B.}~\bibnamefont
  {Jónsson}}, \bibinfo {author} {\bibfnamefont {B.}~\bibnamefont {Bauer}},\
  and\ \bibinfo {author} {\bibfnamefont {G.}~\bibnamefont {Carleo}},\
  }\href@noop {} {\bibinfo {title} {Neural-network states for the classical
  simulation of quantum computing}} (\bibinfo {year} {2018}),\ \Eprint
  {https://arxiv.org/abs/1808.05232} {arXiv:1808.05232 [quant-ph]} \BibitemShut
  {NoStop}%
\bibitem [{\citenamefont {Medvidovi{\'c}}\ and\ \citenamefont
  {Carleo}(2021)}]{matija_nqs_for_qc}%
  \BibitemOpen
  \bibfield  {author} {\bibinfo {author} {\bibfnamefont {M.}~\bibnamefont
  {Medvidovi{\'c}}}\ and\ \bibinfo {author} {\bibfnamefont {G.}~\bibnamefont
  {Carleo}},\ }\bibfield  {title} {\bibinfo {title} {Classical variational
  simulation of the quantum approximate optimization algorithm},\ }\href@noop
  {} {\bibfield  {journal} {\bibinfo  {journal} {npj Quantum Information}\
  }\textbf {\bibinfo {volume} {7}},\ \bibinfo {pages} {1} (\bibinfo {year}
  {2021})}\BibitemShut {NoStop}%
\bibitem [{\citenamefont {Torlai}\ and\ \citenamefont
  {Melko}(2018)}]{neural_density_operators}%
  \BibitemOpen
  \bibfield  {author} {\bibinfo {author} {\bibfnamefont {G.}~\bibnamefont
  {Torlai}}\ and\ \bibinfo {author} {\bibfnamefont {R.~G.}\ \bibnamefont
  {Melko}},\ }\bibfield  {title} {\bibinfo {title} {Latent space purification
  via neural density operators},\ }\href
  {https://link.aps.org/doi/10.1103/PhysRevLett.120.240503} {\bibfield
  {journal} {\bibinfo  {journal} {Phys. Rev. Lett.}\ }\textbf {\bibinfo
  {volume} {120}},\ \bibinfo {pages} {240503} (\bibinfo {year}
  {2018})}\BibitemShut {NoStop}%
\bibitem [{\citenamefont {Hartmann}\ and\ \citenamefont
  {Carleo}(2019)}]{carleo_open_systems}%
  \BibitemOpen
  \bibfield  {author} {\bibinfo {author} {\bibfnamefont {M.~J.}\ \bibnamefont
  {Hartmann}}\ and\ \bibinfo {author} {\bibfnamefont {G.}~\bibnamefont
  {Carleo}},\ }\bibfield  {title} {\bibinfo {title} {Neural-network approach to
  dissipative quantum many-body dynamics},\ }\href
  {https://link.aps.org/doi/10.1103/PhysRevLett.122.250502} {\bibfield
  {journal} {\bibinfo  {journal} {Phys. Rev. Lett.}\ }\textbf {\bibinfo
  {volume} {122}},\ \bibinfo {pages} {250502} (\bibinfo {year}
  {2019})}\BibitemShut {NoStop}%
\bibitem [{\citenamefont {Reh}\ \emph {et~al.}(2021)\citenamefont {Reh},
  \citenamefont {Schmitt},\ and\ \citenamefont {G\"arttner}}]{Reh2021}%
  \BibitemOpen
  \bibfield  {author} {\bibinfo {author} {\bibfnamefont {M.}~\bibnamefont
  {Reh}}, \bibinfo {author} {\bibfnamefont {M.}~\bibnamefont {Schmitt}},\ and\
  \bibinfo {author} {\bibfnamefont {M.}~\bibnamefont {G\"arttner}},\ }\bibfield
   {title} {\bibinfo {title} {Time-dependent variational principle for open
  quantum systems with artificial neural networks},\ }\href
  {https://doi.org/10.1103/PhysRevLett.127.230501} {\bibfield  {journal}
  {\bibinfo  {journal} {Phys. Rev. Lett.}\ }\textbf {\bibinfo {volume} {127}},\
  \bibinfo {pages} {230501} (\bibinfo {year} {2021})}\BibitemShut {NoStop}%
\bibitem [{\citenamefont {Vicentini}\ \emph {et~al.}(2019)\citenamefont
  {Vicentini}, \citenamefont {Biella}, \citenamefont {Regnault},\ and\
  \citenamefont {Ciuti}}]{vicentini_open_systems}%
  \BibitemOpen
  \bibfield  {author} {\bibinfo {author} {\bibfnamefont {F.}~\bibnamefont
  {Vicentini}}, \bibinfo {author} {\bibfnamefont {A.}~\bibnamefont {Biella}},
  \bibinfo {author} {\bibfnamefont {N.}~\bibnamefont {Regnault}},\ and\
  \bibinfo {author} {\bibfnamefont {C.}~\bibnamefont {Ciuti}},\ }\bibfield
  {title} {\bibinfo {title} {Variational neural-network ansatz for steady
  states in open quantum systems},\ }\href
  {https://doi.org/10.1103/PhysRevLett.122.250503} {\bibfield  {journal}
  {\bibinfo  {journal} {Phys. Rev. Lett.}\ }\textbf {\bibinfo {volume} {122}},\
  \bibinfo {pages} {250503} (\bibinfo {year} {2019})}\BibitemShut {NoStop}%
\bibitem [{\citenamefont {Luo}\ \emph {et~al.}(2022)\citenamefont {Luo},
  \citenamefont {Chen}, \citenamefont {Carrasquilla},\ and\ \citenamefont
  {Clark}}]{autoregressive_open_systems}%
  \BibitemOpen
  \bibfield  {author} {\bibinfo {author} {\bibfnamefont {D.}~\bibnamefont
  {Luo}}, \bibinfo {author} {\bibfnamefont {Z.}~\bibnamefont {Chen}}, \bibinfo
  {author} {\bibfnamefont {J.}~\bibnamefont {Carrasquilla}},\ and\ \bibinfo
  {author} {\bibfnamefont {B.~K.}\ \bibnamefont {Clark}},\ }\bibfield  {title}
  {\bibinfo {title} {Autoregressive neural network for simulating open quantum
  systems via a probabilistic formulation},\ }\href
  {https://doi.org/10.1103/PhysRevLett.128.090501} {\bibfield  {journal}
  {\bibinfo  {journal} {Phys. Rev. Lett.}\ }\textbf {\bibinfo {volume} {128}},\
  \bibinfo {pages} {090501} (\bibinfo {year} {2022})}\BibitemShut {NoStop}%
\bibitem [{\citenamefont {Torlai}\ \emph {et~al.}(2018)\citenamefont {Torlai},
  \citenamefont {Mazzola}, \citenamefont {Carrasquilla}, \citenamefont
  {Troyer}, \citenamefont {Melko},\ and\ \citenamefont
  {Carleo}}]{nnqs_pure_tomography}%
  \BibitemOpen
  \bibfield  {author} {\bibinfo {author} {\bibfnamefont {G.}~\bibnamefont
  {Torlai}}, \bibinfo {author} {\bibfnamefont {G.}~\bibnamefont {Mazzola}},
  \bibinfo {author} {\bibfnamefont {J.}~\bibnamefont {Carrasquilla}}, \bibinfo
  {author} {\bibfnamefont {M.}~\bibnamefont {Troyer}}, \bibinfo {author}
  {\bibfnamefont {R.}~\bibnamefont {Melko}},\ and\ \bibinfo {author}
  {\bibfnamefont {G.}~\bibnamefont {Carleo}},\ }\bibfield  {title} {\bibinfo
  {title} {Neural-network quantum state tomography},\ }\href
  {https://doi.org/10.1038/s41567-018-0048-5} {\bibfield  {journal} {\bibinfo
  {journal} {Nature Physics}\ }\textbf {\bibinfo {volume} {14}},\ \bibinfo
  {pages} {447} (\bibinfo {year} {2018})}\BibitemShut {NoStop}%
\bibitem [{\citenamefont {Tiunov}\ \emph {et~al.}(2020)\citenamefont {Tiunov},
  \citenamefont {(Vyborova)}, \citenamefont {Ulanov}, \citenamefont {Lvovsky},\
  and\ \citenamefont {Fedorov}}]{tiunov_tomography}%
  \BibitemOpen
  \bibfield  {author} {\bibinfo {author} {\bibfnamefont {E.~S.}\ \bibnamefont
  {Tiunov}}, \bibinfo {author} {\bibfnamefont {V.~V.~T.}\ \bibnamefont
  {(Vyborova)}}, \bibinfo {author} {\bibfnamefont {A.~E.}\ \bibnamefont
  {Ulanov}}, \bibinfo {author} {\bibfnamefont {A.~I.}\ \bibnamefont
  {Lvovsky}},\ and\ \bibinfo {author} {\bibfnamefont {A.~K.}\ \bibnamefont
  {Fedorov}},\ }\bibfield  {title} {\bibinfo {title} {Experimental quantum
  homodyne tomography via machine learning},\ }\href
  {https://doi.org/10.1364/OPTICA.389482} {\bibfield  {journal} {\bibinfo
  {journal} {Optica}\ }\textbf {\bibinfo {volume} {7}},\ \bibinfo {pages} {448}
  (\bibinfo {year} {2020})}\BibitemShut {NoStop}%
\bibitem [{\citenamefont {Kurmapu}\ \emph {et~al.}(2023)\citenamefont
  {Kurmapu}, \citenamefont {Tiunova}, \citenamefont {Tiunov}, \citenamefont
  {Ringbauer}, \citenamefont {Maier}, \citenamefont {Blatt}, \citenamefont
  {Monz}, \citenamefont {Fedorov},\ and\ \citenamefont
  {Lvovsky}}]{murali_tomography}%
  \BibitemOpen
  \bibfield  {author} {\bibinfo {author} {\bibfnamefont {M.~K.}\ \bibnamefont
  {Kurmapu}}, \bibinfo {author} {\bibfnamefont {V.}~\bibnamefont {Tiunova}},
  \bibinfo {author} {\bibfnamefont {E.}~\bibnamefont {Tiunov}}, \bibinfo
  {author} {\bibfnamefont {M.}~\bibnamefont {Ringbauer}}, \bibinfo {author}
  {\bibfnamefont {C.}~\bibnamefont {Maier}}, \bibinfo {author} {\bibfnamefont
  {R.}~\bibnamefont {Blatt}}, \bibinfo {author} {\bibfnamefont
  {T.}~\bibnamefont {Monz}}, \bibinfo {author} {\bibfnamefont {A.~K.}\
  \bibnamefont {Fedorov}},\ and\ \bibinfo {author} {\bibfnamefont
  {A.}~\bibnamefont {Lvovsky}},\ }\bibfield  {title} {\bibinfo {title}
  {Reconstructing complex states of a $20$-qubit quantum simulator},\ }\href
  {https://doi.org/10.1103/PRXQuantum.4.040345} {\bibfield  {journal} {\bibinfo
   {journal} {PRX Quantum}\ }\textbf {\bibinfo {volume} {4}},\ \bibinfo {pages}
  {040345} (\bibinfo {year} {2023})}\BibitemShut {NoStop}%
\bibitem [{\citenamefont {Fedotova}\ \emph {et~al.}(2023)\citenamefont
  {Fedotova}, \citenamefont {Kuznetsov}, \citenamefont {Tiunov}, \citenamefont
  {Ulanov},\ and\ \citenamefont {Lvovsky}}]{fedotova_tomography}%
  \BibitemOpen
  \bibfield  {author} {\bibinfo {author} {\bibfnamefont {E.}~\bibnamefont
  {Fedotova}}, \bibinfo {author} {\bibfnamefont {N.}~\bibnamefont {Kuznetsov}},
  \bibinfo {author} {\bibfnamefont {E.}~\bibnamefont {Tiunov}}, \bibinfo
  {author} {\bibfnamefont {A.~E.}\ \bibnamefont {Ulanov}},\ and\ \bibinfo
  {author} {\bibfnamefont {A.~I.}\ \bibnamefont {Lvovsky}},\ }\bibfield
  {title} {\bibinfo {title} {Continuous-variable quantum tomography of
  high-amplitude states},\ }\href {https://doi.org/10.1103/PhysRevA.108.042430}
  {\bibfield  {journal} {\bibinfo  {journal} {Phys. Rev. A}\ }\textbf {\bibinfo
  {volume} {108}},\ \bibinfo {pages} {042430} (\bibinfo {year}
  {2023})}\BibitemShut {NoStop}%
\bibitem [{\citenamefont {Choo}\ \emph {et~al.}(2020)\citenamefont {Choo},
  \citenamefont {Mezzacapo},\ and\ \citenamefont
  {Carleo}}]{carleo_quantum_chemistry}%
  \BibitemOpen
  \bibfield  {author} {\bibinfo {author} {\bibfnamefont {K.}~\bibnamefont
  {Choo}}, \bibinfo {author} {\bibfnamefont {A.}~\bibnamefont {Mezzacapo}},\
  and\ \bibinfo {author} {\bibfnamefont {G.}~\bibnamefont {Carleo}},\
  }\bibfield  {title} {\bibinfo {title} {Fermionic neural-network states for
  ab-initio electronic structure},\ }\href
  {https://doi.org/10.1038/s41467-020-15724-9} {\bibfield  {journal} {\bibinfo
  {journal} {Nature Communications}\ }\textbf {\bibinfo {volume} {11}},\
  \bibinfo {pages} {2368} (\bibinfo {year} {2020})}\BibitemShut {NoStop}%
\bibitem [{\citenamefont {Li}\ \emph {et~al.}(2023)\citenamefont {Li},
  \citenamefont {Huang}, \citenamefont {Zhang}, \citenamefont {Li},
  \citenamefont {Cao}, \citenamefont {Lv},\ and\ \citenamefont
  {Hu}}]{li_deterministic_nqs_qchem}%
  \BibitemOpen
  \bibfield  {author} {\bibinfo {author} {\bibfnamefont {X.}~\bibnamefont
  {Li}}, \bibinfo {author} {\bibfnamefont {J.-C.}\ \bibnamefont {Huang}},
  \bibinfo {author} {\bibfnamefont {G.-Z.}\ \bibnamefont {Zhang}}, \bibinfo
  {author} {\bibfnamefont {H.-E.}\ \bibnamefont {Li}}, \bibinfo {author}
  {\bibfnamefont {C.-S.}\ \bibnamefont {Cao}}, \bibinfo {author} {\bibfnamefont
  {D.}~\bibnamefont {Lv}},\ and\ \bibinfo {author} {\bibfnamefont {H.-S.}\
  \bibnamefont {Hu}},\ }\bibfield  {title} {\bibinfo {title} {A nonstochastic
  optimization algorithm for neural-network quantum states},\ }\href
  {https://doi.org/10.1021/acs.jctc.3c00831} {\bibfield  {journal} {\bibinfo
  {journal} {Journal of Chemical Theory and Computation}\ }\textbf {\bibinfo
  {volume} {19}},\ \bibinfo {pages} {8156} (\bibinfo {year} {2023})},\ \bibinfo
  {note} {pMID: 37962975},\ \Eprint
  {https://arxiv.org/abs/https://doi.org/10.1021/acs.jctc.3c00831}
  {https://doi.org/10.1021/acs.jctc.3c00831} \BibitemShut {NoStop}%
\bibitem [{\citenamefont {Vicentini}\ \emph {et~al.}(2022)\citenamefont
  {Vicentini}, \citenamefont {Hofmann}, \citenamefont {Szabó}, \citenamefont
  {Wu}, \citenamefont {Roth}, \citenamefont {Giuliani}, \citenamefont {Pescia},
  \citenamefont {Nys}, \citenamefont {Vargas-Calderón}, \citenamefont
  {Astrakhantsev},\ and\ \citenamefont {Carleo}}]{vicentini2022netket}%
  \BibitemOpen
  \bibfield  {author} {\bibinfo {author} {\bibfnamefont {F.}~\bibnamefont
  {Vicentini}}, \bibinfo {author} {\bibfnamefont {D.}~\bibnamefont {Hofmann}},
  \bibinfo {author} {\bibfnamefont {A.}~\bibnamefont {Szabó}}, \bibinfo
  {author} {\bibfnamefont {D.}~\bibnamefont {Wu}}, \bibinfo {author}
  {\bibfnamefont {C.}~\bibnamefont {Roth}}, \bibinfo {author} {\bibfnamefont
  {C.}~\bibnamefont {Giuliani}}, \bibinfo {author} {\bibfnamefont
  {G.}~\bibnamefont {Pescia}}, \bibinfo {author} {\bibfnamefont
  {J.}~\bibnamefont {Nys}}, \bibinfo {author} {\bibfnamefont {V.}~\bibnamefont
  {Vargas-Calderón}}, \bibinfo {author} {\bibfnamefont {N.}~\bibnamefont
  {Astrakhantsev}},\ and\ \bibinfo {author} {\bibfnamefont {G.}~\bibnamefont
  {Carleo}},\ }\bibfield  {title} {\bibinfo {title} {{NetKet 3: Machine
  Learning Toolbox for Many-Body Quantum Systems}},\ }\href
  {https://doi.org/10.21468/SciPostPhysCodeb.7} {\bibfield  {journal} {\bibinfo
   {journal} {SciPost Phys. Codebases}\ ,\ \bibinfo {pages} {7}} (\bibinfo
  {year} {2022})}\BibitemShut {NoStop}%
\bibitem [{\citenamefont {Sharir}\ \emph {et~al.}(2020)\citenamefont {Sharir},
  \citenamefont {Levine}, \citenamefont {Wies}, \citenamefont {Carleo},\ and\
  \citenamefont {Shashua}}]{autoregressive_originals}%
  \BibitemOpen
  \bibfield  {author} {\bibinfo {author} {\bibfnamefont {O.}~\bibnamefont
  {Sharir}}, \bibinfo {author} {\bibfnamefont {Y.}~\bibnamefont {Levine}},
  \bibinfo {author} {\bibfnamefont {N.}~\bibnamefont {Wies}}, \bibinfo {author}
  {\bibfnamefont {G.}~\bibnamefont {Carleo}},\ and\ \bibinfo {author}
  {\bibfnamefont {A.}~\bibnamefont {Shashua}},\ }\bibfield  {title} {\bibinfo
  {title} {Deep autoregressive models for the efficient variational simulation
  of many-body quantum systems},\ }\href
  {https://link.aps.org/doi/10.1103/PhysRevLett.124.020503} {\bibfield
  {journal} {\bibinfo  {journal} {Phys. Rev. Lett.}\ }\textbf {\bibinfo
  {volume} {124}},\ \bibinfo {pages} {020503} (\bibinfo {year}
  {2020})}\BibitemShut {NoStop}%
\bibitem [{\citenamefont {Kool}\ \emph {et~al.}(2020)\citenamefont {Kool},
  \citenamefont {Van~Hoof},\ and\ \citenamefont
  {Welling}}]{ancestral_gumbel_top_k_sampling}%
  \BibitemOpen
  \bibfield  {author} {\bibinfo {author} {\bibfnamefont {W.}~\bibnamefont
  {Kool}}, \bibinfo {author} {\bibfnamefont {H.}~\bibnamefont {Van~Hoof}},\
  and\ \bibinfo {author} {\bibfnamefont {M.}~\bibnamefont {Welling}},\
  }\bibfield  {title} {\bibinfo {title} {Ancestral gumbel-top-k sampling for
  sampling without replacement},\ }\href@noop {} {\bibfield  {journal}
  {\bibinfo  {journal} {The Journal of Machine Learning Research}\ }\textbf
  {\bibinfo {volume} {21}},\ \bibinfo {pages} {1726} (\bibinfo {year}
  {2020})}\BibitemShut {NoStop}%
\bibitem [{\citenamefont {Gumbel}(1954)}]{gumbel_gumbel}%
  \BibitemOpen
  \bibfield  {author} {\bibinfo {author} {\bibfnamefont {E.~J.}\ \bibnamefont
  {Gumbel}},\ }\href@noop {} {\emph {\bibinfo {title} {Statistical theory of
  extreme values and some practical applications: a series of lectures}}},\
  Vol.~\bibinfo {volume} {33}\ (\bibinfo  {publisher} {US Government Printing
  Office},\ \bibinfo {year} {1954})\BibitemShut {NoStop}%
\bibitem [{\citenamefont {Maddison}\ \emph {et~al.}(2014)\citenamefont
  {Maddison}, \citenamefont {Tarlow},\ and\ \citenamefont
  {Minka}}]{maddison_gumbel}%
  \BibitemOpen
  \bibfield  {author} {\bibinfo {author} {\bibfnamefont {C.~J.}\ \bibnamefont
  {Maddison}}, \bibinfo {author} {\bibfnamefont {D.}~\bibnamefont {Tarlow}},\
  and\ \bibinfo {author} {\bibfnamefont {T.}~\bibnamefont {Minka}},\ }\bibfield
   {title} {\bibinfo {title} {{$A^\ast$ Sampling}},\ }in\ \href
  {https://proceedings.neurips.cc/paper_files/paper/2014/file/309fee4e541e51de2e41f21bebb342aa-Paper.pdf}
  {\emph {\bibinfo {booktitle} {Advances in Neural Information Processing
  Systems}}},\ Vol.~\bibinfo {volume} {27},\ \bibinfo {editor} {edited by\
  \bibinfo {editor} {\bibfnamefont {Z.}~\bibnamefont {Ghahramani}}, \bibinfo
  {editor} {\bibfnamefont {M.}~\bibnamefont {Welling}}, \bibinfo {editor}
  {\bibfnamefont {C.}~\bibnamefont {Cortes}}, \bibinfo {editor} {\bibfnamefont
  {N.}~\bibnamefont {Lawrence}},\ and\ \bibinfo {editor} {\bibfnamefont
  {K.}~\bibnamefont {Weinberger}}}\ (\bibinfo  {publisher} {Curran Associates,
  Inc.},\ \bibinfo {year} {2014})\BibitemShut {NoStop}%
\bibitem [{\citenamefont {McArdle}\ \emph {et~al.}(2020)\citenamefont
  {McArdle}, \citenamefont {Endo}, \citenamefont {Aspuru-Guzik}, \citenamefont
  {Benjamin},\ and\ \citenamefont {Yuan}}]{mcardle_review}%
  \BibitemOpen
  \bibfield  {author} {\bibinfo {author} {\bibfnamefont {S.}~\bibnamefont
  {McArdle}}, \bibinfo {author} {\bibfnamefont {S.}~\bibnamefont {Endo}},
  \bibinfo {author} {\bibfnamefont {A.}~\bibnamefont {Aspuru-Guzik}}, \bibinfo
  {author} {\bibfnamefont {S.~C.}\ \bibnamefont {Benjamin}},\ and\ \bibinfo
  {author} {\bibfnamefont {X.}~\bibnamefont {Yuan}},\ }\bibfield  {title}
  {\bibinfo {title} {Quantum computational chemistry},\ }\href
  {https://doi.org/10.1103/RevModPhys.92.015003} {\bibfield  {journal}
  {\bibinfo  {journal} {Rev. Mod. Phys.}\ }\textbf {\bibinfo {volume} {92}},\
  \bibinfo {pages} {015003} (\bibinfo {year} {2020})}\BibitemShut {NoStop}%
\bibitem [{\citenamefont {Meyer}(2021)}]{meyer_quantum_fisher_information}%
  \BibitemOpen
  \bibfield  {author} {\bibinfo {author} {\bibfnamefont {J.~J.}\ \bibnamefont
  {Meyer}},\ }\bibfield  {title} {\bibinfo {title} {Fisher {I}nformation in
  {N}oisy {I}ntermediate-{S}cale {Q}uantum {A}pplications},\ }\href
  {https://doi.org/10.22331/q-2021-09-09-539} {\bibfield  {journal} {\bibinfo
  {journal} {{Quantum}}\ }\textbf {\bibinfo {volume} {5}},\ \bibinfo {pages}
  {539} (\bibinfo {year} {2021})}\BibitemShut {NoStop}%
\bibitem [{\citenamefont {Stokes}\ \emph {et~al.}(2020)\citenamefont {Stokes},
  \citenamefont {Izaac}, \citenamefont {Killoran},\ and\ \citenamefont
  {Carleo}}]{stokes_quantum_natural_gradient}%
  \BibitemOpen
  \bibfield  {author} {\bibinfo {author} {\bibfnamefont {J.}~\bibnamefont
  {Stokes}}, \bibinfo {author} {\bibfnamefont {J.}~\bibnamefont {Izaac}},
  \bibinfo {author} {\bibfnamefont {N.}~\bibnamefont {Killoran}},\ and\
  \bibinfo {author} {\bibfnamefont {G.}~\bibnamefont {Carleo}},\ }\bibfield
  {title} {\bibinfo {title} {Quantum {N}atural {G}radient},\ }\href
  {https://doi.org/10.22331/q-2020-05-25-269} {\bibfield  {journal} {\bibinfo
  {journal} {{Quantum}}\ }\textbf {\bibinfo {volume} {4}},\ \bibinfo {pages}
  {269} (\bibinfo {year} {2020})}\BibitemShut {NoStop}%
\bibitem [{\citenamefont {Chen}\ and\ \citenamefont
  {Heyl}(2023)}]{chen_min_sr}%
  \BibitemOpen
  \bibfield  {author} {\bibinfo {author} {\bibfnamefont {A.}~\bibnamefont
  {Chen}}\ and\ \bibinfo {author} {\bibfnamefont {M.}~\bibnamefont {Heyl}},\
  }\href@noop {} {\bibinfo {title} {Efficient optimization of deep neural
  quantum states toward machine precision}} (\bibinfo {year} {2023}),\ \Eprint
  {https://arxiv.org/abs/2302.01941} {arXiv:2302.01941 [cond-mat.dis-nn]}
  \BibitemShut {NoStop}%
\bibitem [{\citenamefont {Rende}\ \emph {et~al.}(2023)\citenamefont {Rende},
  \citenamefont {Viteritti}, \citenamefont {Bardone}, \citenamefont {Becca},\
  and\ \citenamefont {Goldt}}]{rende_reg_sr}%
  \BibitemOpen
  \bibfield  {author} {\bibinfo {author} {\bibfnamefont {R.}~\bibnamefont
  {Rende}}, \bibinfo {author} {\bibfnamefont {L.~L.}\ \bibnamefont
  {Viteritti}}, \bibinfo {author} {\bibfnamefont {L.}~\bibnamefont {Bardone}},
  \bibinfo {author} {\bibfnamefont {F.}~\bibnamefont {Becca}},\ and\ \bibinfo
  {author} {\bibfnamefont {S.}~\bibnamefont {Goldt}},\ }\href@noop {} {\bibinfo
  {title} {A simple linear algebra identity to optimize large-scale neural
  network quantum states}} (\bibinfo {year} {2023}),\ \Eprint
  {https://arxiv.org/abs/2310.05715} {arXiv:2310.05715 [cond-mat.str-el]}
  \BibitemShut {NoStop}%
\bibitem [{\citenamefont {Ansel}\ \emph {et~al.}(2024)\citenamefont {Ansel},
  \citenamefont {Yang}, \citenamefont {He}, \citenamefont {Gimelshein},
  \citenamefont {Jain}, \citenamefont {Voznesensky}, \citenamefont {Bao},
  \citenamefont {Bell}, \citenamefont {Berard}, \citenamefont {Burovski},
  \citenamefont {Chauhan}, \citenamefont {Chourdia}, \citenamefont {Constable},
  \citenamefont {Desmaison}, \citenamefont {DeVito}, \citenamefont {Ellison},
  \citenamefont {Feng}, \citenamefont {Gong}, \citenamefont {Gschwind},
  \citenamefont {Hirsh}, \citenamefont {Huang}, \citenamefont {Kalambarkar},
  \citenamefont {Kirsch}, \citenamefont {Lazos}, \citenamefont {Lezcano},
  \citenamefont {Liang}, \citenamefont {Liang}, \citenamefont {Lu},
  \citenamefont {Luk}, \citenamefont {Maher}, \citenamefont {Pan},
  \citenamefont {Puhrsch}, \citenamefont {Reso}, \citenamefont {Saroufim},
  \citenamefont {Siraichi}, \citenamefont {Suk}, \citenamefont {Suo},
  \citenamefont {Tillet}, \citenamefont {Wang}, \citenamefont {Wang},
  \citenamefont {Wen}, \citenamefont {Zhang}, \citenamefont {Zhao},
  \citenamefont {Zhou}, \citenamefont {Zou}, \citenamefont {Mathews},
  \citenamefont {Chanan}, \citenamefont {Wu},\ and\ \citenamefont
  {Chintala}}]{pytorch}%
  \BibitemOpen
  \bibfield  {author} {\bibinfo {author} {\bibfnamefont {J.}~\bibnamefont
  {Ansel}}, \bibinfo {author} {\bibfnamefont {E.}~\bibnamefont {Yang}},
  \bibinfo {author} {\bibfnamefont {H.}~\bibnamefont {He}}, \bibinfo {author}
  {\bibfnamefont {N.}~\bibnamefont {Gimelshein}}, \bibinfo {author}
  {\bibfnamefont {A.}~\bibnamefont {Jain}}, \bibinfo {author} {\bibfnamefont
  {M.}~\bibnamefont {Voznesensky}}, \bibinfo {author} {\bibfnamefont
  {B.}~\bibnamefont {Bao}}, \bibinfo {author} {\bibfnamefont {P.}~\bibnamefont
  {Bell}}, \bibinfo {author} {\bibfnamefont {D.}~\bibnamefont {Berard}},
  \bibinfo {author} {\bibfnamefont {E.}~\bibnamefont {Burovski}}, \bibinfo
  {author} {\bibfnamefont {G.}~\bibnamefont {Chauhan}}, \bibinfo {author}
  {\bibfnamefont {A.}~\bibnamefont {Chourdia}}, \bibinfo {author}
  {\bibfnamefont {W.}~\bibnamefont {Constable}}, \bibinfo {author}
  {\bibfnamefont {A.}~\bibnamefont {Desmaison}}, \bibinfo {author}
  {\bibfnamefont {Z.}~\bibnamefont {DeVito}}, \bibinfo {author} {\bibfnamefont
  {E.}~\bibnamefont {Ellison}}, \bibinfo {author} {\bibfnamefont
  {W.}~\bibnamefont {Feng}}, \bibinfo {author} {\bibfnamefont {J.}~\bibnamefont
  {Gong}}, \bibinfo {author} {\bibfnamefont {M.}~\bibnamefont {Gschwind}},
  \bibinfo {author} {\bibfnamefont {B.}~\bibnamefont {Hirsh}}, \bibinfo
  {author} {\bibfnamefont {S.}~\bibnamefont {Huang}}, \bibinfo {author}
  {\bibfnamefont {K.}~\bibnamefont {Kalambarkar}}, \bibinfo {author}
  {\bibfnamefont {L.}~\bibnamefont {Kirsch}}, \bibinfo {author} {\bibfnamefont
  {M.}~\bibnamefont {Lazos}}, \bibinfo {author} {\bibfnamefont
  {M.}~\bibnamefont {Lezcano}}, \bibinfo {author} {\bibfnamefont
  {Y.}~\bibnamefont {Liang}}, \bibinfo {author} {\bibfnamefont
  {J.}~\bibnamefont {Liang}}, \bibinfo {author} {\bibfnamefont
  {Y.}~\bibnamefont {Lu}}, \bibinfo {author} {\bibfnamefont {C.}~\bibnamefont
  {Luk}}, \bibinfo {author} {\bibfnamefont {B.}~\bibnamefont {Maher}}, \bibinfo
  {author} {\bibfnamefont {Y.}~\bibnamefont {Pan}}, \bibinfo {author}
  {\bibfnamefont {C.}~\bibnamefont {Puhrsch}}, \bibinfo {author} {\bibfnamefont
  {M.}~\bibnamefont {Reso}}, \bibinfo {author} {\bibfnamefont {M.}~\bibnamefont
  {Saroufim}}, \bibinfo {author} {\bibfnamefont {M.~Y.}\ \bibnamefont
  {Siraichi}}, \bibinfo {author} {\bibfnamefont {H.}~\bibnamefont {Suk}},
  \bibinfo {author} {\bibfnamefont {M.}~\bibnamefont {Suo}}, \bibinfo {author}
  {\bibfnamefont {P.}~\bibnamefont {Tillet}}, \bibinfo {author} {\bibfnamefont
  {E.}~\bibnamefont {Wang}}, \bibinfo {author} {\bibfnamefont {X.}~\bibnamefont
  {Wang}}, \bibinfo {author} {\bibfnamefont {W.}~\bibnamefont {Wen}}, \bibinfo
  {author} {\bibfnamefont {S.}~\bibnamefont {Zhang}}, \bibinfo {author}
  {\bibfnamefont {X.}~\bibnamefont {Zhao}}, \bibinfo {author} {\bibfnamefont
  {K.}~\bibnamefont {Zhou}}, \bibinfo {author} {\bibfnamefont {R.}~\bibnamefont
  {Zou}}, \bibinfo {author} {\bibfnamefont {A.}~\bibnamefont {Mathews}},
  \bibinfo {author} {\bibfnamefont {G.}~\bibnamefont {Chanan}}, \bibinfo
  {author} {\bibfnamefont {P.}~\bibnamefont {Wu}},\ and\ \bibinfo {author}
  {\bibfnamefont {S.}~\bibnamefont {Chintala}},\ }\bibfield  {title} {\bibinfo
  {title} {Pytorch 2: Faster machine learning through dynamic python bytecode
  transformation and graph compilation},\ }in\ \href
  {https://doi.org/10.1145/3620665.3640366} {\emph {\bibinfo {booktitle}
  {Proceedings of the 29th ACM International Conference on Architectural
  Support for Programming Languages and Operating Systems, Volume 2 (ASPLOS
  '24)}}}\ (\bibinfo  {publisher} {ACM},\ \bibinfo {year} {2024})\BibitemShut
  {NoStop}%
\bibitem [{\citenamefont {Sharma}\ and\ \citenamefont
  {Alavi}(2015)}]{c2_diss_curve}%
  \BibitemOpen
  \bibfield  {author} {\bibinfo {author} {\bibfnamefont {S.}~\bibnamefont
  {Sharma}}\ and\ \bibinfo {author} {\bibfnamefont {A.}~\bibnamefont {Alavi}},\
  }\bibfield  {title} {\bibinfo {title} {{Multireference linearized coupled
  cluster theory for strongly correlated systems using matrix product
  states}},\ }\href {https://doi.org/10.1063/1.4928643} {\bibfield  {journal}
  {\bibinfo  {journal} {The Journal of Chemical Physics}\ }\textbf {\bibinfo
  {volume} {143}},\ \bibinfo {pages} {102815} (\bibinfo {year} {2015})},\
  \Eprint
  {https://arxiv.org/abs/https://pubs.aip.org/aip/jcp/article-pdf/doi/10.1063/1.4928643/15502606/102815\_1\_online.pdf}
  {https://pubs.aip.org/aip/jcp/article-pdf/doi/10.1063/1.4928643/15502606/102815\_1\_online.pdf}
  \BibitemShut {NoStop}%
\bibitem [{\citenamefont {Kim}\ \emph {et~al.}(2016)\citenamefont {Kim},
  \citenamefont {Thiessen}, \citenamefont {Bolton}, \citenamefont {Chen},
  \citenamefont {Fu}, \citenamefont {Gindulyte}, \citenamefont {Han},
  \citenamefont {He}, \citenamefont {He}, \citenamefont {Shoemaker} \emph
  {et~al.}}]{pubchem}%
  \BibitemOpen
  \bibfield  {author} {\bibinfo {author} {\bibfnamefont {S.}~\bibnamefont
  {Kim}}, \bibinfo {author} {\bibfnamefont {P.~A.}\ \bibnamefont {Thiessen}},
  \bibinfo {author} {\bibfnamefont {E.~E.}\ \bibnamefont {Bolton}}, \bibinfo
  {author} {\bibfnamefont {J.}~\bibnamefont {Chen}}, \bibinfo {author}
  {\bibfnamefont {G.}~\bibnamefont {Fu}}, \bibinfo {author} {\bibfnamefont
  {A.}~\bibnamefont {Gindulyte}}, \bibinfo {author} {\bibfnamefont
  {L.}~\bibnamefont {Han}}, \bibinfo {author} {\bibfnamefont {J.}~\bibnamefont
  {He}}, \bibinfo {author} {\bibfnamefont {S.}~\bibnamefont {He}}, \bibinfo
  {author} {\bibfnamefont {B.~A.}\ \bibnamefont {Shoemaker}}, \emph {et~al.},\
  }\bibfield  {title} {\bibinfo {title} {Pubchem substance and compound
  databases},\ }\href@noop {} {\bibfield  {journal} {\bibinfo  {journal}
  {Nucleic Acids Res.}\ }\textbf {\bibinfo {volume} {44}},\ \bibinfo {pages}
  {D1202} (\bibinfo {year} {2016})}\BibitemShut {NoStop}%
\bibitem [{\citenamefont {Gao}\ \emph {et~al.}(2024)\citenamefont {Gao},
  \citenamefont {Imamura}, \citenamefont {Kasagi},\ and\ \citenamefont
  {Yoshida}}]{largest_fci}%
  \BibitemOpen
  \bibfield  {author} {\bibinfo {author} {\bibfnamefont {H.}~\bibnamefont
  {Gao}}, \bibinfo {author} {\bibfnamefont {S.}~\bibnamefont {Imamura}},
  \bibinfo {author} {\bibfnamefont {A.}~\bibnamefont {Kasagi}},\ and\ \bibinfo
  {author} {\bibfnamefont {E.}~\bibnamefont {Yoshida}},\ }\bibfield  {title}
  {\bibinfo {title} {Distributed implementation of full configuration
  interaction for one trillion determinants},\ }\href
  {https://doi.org/10.1021/acs.jctc.3c01190} {\bibfield  {journal} {\bibinfo
  {journal} {Journal of Chemical Theory and Computation}\ }\textbf {\bibinfo
  {volume} {20}},\ \bibinfo {pages} {1185} (\bibinfo {year} {2024})},\ \bibinfo
  {note} {pMID: 38314701},\ \Eprint
  {https://arxiv.org/abs/https://doi.org/10.1021/acs.jctc.3c01190}
  {https://doi.org/10.1021/acs.jctc.3c01190} \BibitemShut {NoStop}%
\bibitem [{\citenamefont {Bukov}\ \emph {et~al.}(2021)\citenamefont {Bukov},
  \citenamefont {Schmitt},\ and\ \citenamefont
  {Dupont}}]{bukov_non_stoquastic}%
  \BibitemOpen
  \bibfield  {author} {\bibinfo {author} {\bibfnamefont {M.}~\bibnamefont
  {Bukov}}, \bibinfo {author} {\bibfnamefont {M.}~\bibnamefont {Schmitt}},\
  and\ \bibinfo {author} {\bibfnamefont {M.}~\bibnamefont {Dupont}},\
  }\bibfield  {title} {\bibinfo {title} {{Learning the ground state of a
  non-stoquastic quantum Hamiltonian in a rugged neural network landscape}},\
  }\href {https://doi.org/10.21468/SciPostPhys.10.6.147} {\bibfield  {journal}
  {\bibinfo  {journal} {SciPost Phys.}\ }\textbf {\bibinfo {volume} {10}},\
  \bibinfo {pages} {147} (\bibinfo {year} {2021})}\BibitemShut {NoStop}%
\bibitem [{\citenamefont {Luo}\ and\ \citenamefont
  {Clark}(2019)}]{luo_backflow}%
  \BibitemOpen
  \bibfield  {author} {\bibinfo {author} {\bibfnamefont {D.}~\bibnamefont
  {Luo}}\ and\ \bibinfo {author} {\bibfnamefont {B.~K.}\ \bibnamefont
  {Clark}},\ }\bibfield  {title} {\bibinfo {title} {Backflow transformations
  via neural networks for quantum many-body wave functions},\ }\href
  {https://doi.org/10.1103/PhysRevLett.122.226401} {\bibfield  {journal}
  {\bibinfo  {journal} {Phys. Rev. Lett.}\ }\textbf {\bibinfo {volume} {122}},\
  \bibinfo {pages} {226401} (\bibinfo {year} {2019})}\BibitemShut {NoStop}%
\bibitem [{\citenamefont {Robledo~Moreno}\ \emph {et~al.}(2022)\citenamefont
  {Robledo~Moreno}, \citenamefont {Carleo}, \citenamefont {Georges},\ and\
  \citenamefont {Stokes}}]{moreno_hfds}%
  \BibitemOpen
  \bibfield  {author} {\bibinfo {author} {\bibfnamefont {J.}~\bibnamefont
  {Robledo~Moreno}}, \bibinfo {author} {\bibfnamefont {G.}~\bibnamefont
  {Carleo}}, \bibinfo {author} {\bibfnamefont {A.}~\bibnamefont {Georges}},\
  and\ \bibinfo {author} {\bibfnamefont {J.}~\bibnamefont {Stokes}},\
  }\bibfield  {title} {\bibinfo {title} {Fermionic wave functions from
  neural-network constrained hidden states},\ }\href@noop {} {\bibfield
  {journal} {\bibinfo  {journal} {Proceedings of the National Academy of
  Sciences}\ }\textbf {\bibinfo {volume} {119}},\ \bibinfo {pages}
  {e2122059119} (\bibinfo {year} {2022})}\BibitemShut {NoStop}%
\bibitem [{\citenamefont {Nys}\ \emph {et~al.}(2024)\citenamefont {Nys},
  \citenamefont {Pescia},\ and\ \citenamefont {Carleo}}]{Nys2024}%
  \BibitemOpen
  \bibfield  {author} {\bibinfo {author} {\bibfnamefont {J.}~\bibnamefont
  {Nys}}, \bibinfo {author} {\bibfnamefont {G.}~\bibnamefont {Pescia}},\ and\
  \bibinfo {author} {\bibfnamefont {G.}~\bibnamefont {Carleo}},\ }\href@noop {}
  {\bibinfo {title} {Ab-initio variational wave functions for the
  time-dependent many-electron schrödinger equation}} (\bibinfo {year}
  {2024}),\ \Eprint {https://arxiv.org/abs/2403.07447} {arXiv:2403.07447
  [cond-mat.str-el]} \BibitemShut {NoStop}%
\bibitem [{\citenamefont {Humeniuk}\ \emph {et~al.}(2023)\citenamefont
  {Humeniuk}, \citenamefont {Wan},\ and\ \citenamefont
  {Wang}}]{humeniuk_slater_jastrow}%
  \BibitemOpen
  \bibfield  {author} {\bibinfo {author} {\bibfnamefont {S.}~\bibnamefont
  {Humeniuk}}, \bibinfo {author} {\bibfnamefont {Y.}~\bibnamefont {Wan}},\ and\
  \bibinfo {author} {\bibfnamefont {L.}~\bibnamefont {Wang}},\ }\bibfield
  {title} {\bibinfo {title} {{Autoregressive neural Slater-Jastrow ansatz for
  variational Monte Carlo simulation}},\ }\href
  {https://doi.org/10.21468/SciPostPhys.14.6.171} {\bibfield  {journal}
  {\bibinfo  {journal} {SciPost Phys.}\ }\textbf {\bibinfo {volume} {14}},\
  \bibinfo {pages} {171} (\bibinfo {year} {2023})}\BibitemShut {NoStop}%
\bibitem [{\citenamefont {Astrakhantsev}\ \emph {et~al.}(2023)\citenamefont
  {Astrakhantsev}, \citenamefont {Mazzola}, \citenamefont {Tavernelli},\ and\
  \citenamefont {Carleo}}]{astrakhantsev_algorithmic_phase_transition}%
  \BibitemOpen
  \bibfield  {author} {\bibinfo {author} {\bibfnamefont {N.}~\bibnamefont
  {Astrakhantsev}}, \bibinfo {author} {\bibfnamefont {G.}~\bibnamefont
  {Mazzola}}, \bibinfo {author} {\bibfnamefont {I.}~\bibnamefont
  {Tavernelli}},\ and\ \bibinfo {author} {\bibfnamefont {G.}~\bibnamefont
  {Carleo}},\ }\bibfield  {title} {\bibinfo {title} {Phenomenological theory of
  variational quantum ground-state preparation},\ }\href
  {https://doi.org/10.1103/PhysRevResearch.5.033225} {\bibfield  {journal}
  {\bibinfo  {journal} {Phys. Rev. Res.}\ }\textbf {\bibinfo {volume} {5}},\
  \bibinfo {pages} {033225} (\bibinfo {year} {2023})}\BibitemShut {NoStop}%
\bibitem [{\citenamefont {Cleland}\ \emph {et~al.}(2012)\citenamefont
  {Cleland}, \citenamefont {Booth}, \citenamefont {Overy},\ and\ \citenamefont
  {Alavi}}]{cleland_poor_fciqmc_energy}%
  \BibitemOpen
  \bibfield  {author} {\bibinfo {author} {\bibfnamefont {D.}~\bibnamefont
  {Cleland}}, \bibinfo {author} {\bibfnamefont {G.~H.}\ \bibnamefont {Booth}},
  \bibinfo {author} {\bibfnamefont {C.}~\bibnamefont {Overy}},\ and\ \bibinfo
  {author} {\bibfnamefont {A.}~\bibnamefont {Alavi}},\ }\bibfield  {title}
  {\bibinfo {title} {Taming the first-row diatomics: A full configuration
  interaction quantum monte carlo study},\ }\href
  {https://doi.org/10.1021/ct300504f} {\bibfield  {journal} {\bibinfo
  {journal} {Journal of Chemical Theory and Computation}\ }\textbf {\bibinfo
  {volume} {8}},\ \bibinfo {pages} {4138} (\bibinfo {year} {2012})},\ \bibinfo
  {note} {pMID: 26605580},\ \Eprint
  {https://arxiv.org/abs/https://doi.org/10.1021/ct300504f}
  {https://doi.org/10.1021/ct300504f} \BibitemShut {NoStop}%
\bibitem [{\citenamefont {McClean}\ \emph {et~al.}(2020)\citenamefont
  {McClean}, \citenamefont {Rubin}, \citenamefont {Sung}, \citenamefont
  {Kivlichan}, \citenamefont {Bonet-Monroig}, \citenamefont {Cao},
  \citenamefont {Dai}, \citenamefont {Fried}, \citenamefont {Gidney},
  \citenamefont {Gimby}, \citenamefont {Gokhale}, \citenamefont {Häner},
  \citenamefont {Hardikar}, \citenamefont {Havlíček}, \citenamefont
  {Higgott}, \citenamefont {Huang}, \citenamefont {Izaac}, \citenamefont
  {Jiang}, \citenamefont {Liu}, \citenamefont {McArdle}, \citenamefont
  {Neeley}, \citenamefont {O’Brien}, \citenamefont {O’Gorman},
  \citenamefont {Ozfidan}, \citenamefont {Radin}, \citenamefont {Romero},
  \citenamefont {Sawaya}, \citenamefont {Senjean}, \citenamefont {Setia},
  \citenamefont {Sim}, \citenamefont {Steiger}, \citenamefont {Steudtner},
  \citenamefont {Sun}, \citenamefont {Sun}, \citenamefont {Wang}, \citenamefont
  {Zhang},\ and\ \citenamefont {Babbush}}]{open_fermion}%
  \BibitemOpen
  \bibfield  {author} {\bibinfo {author} {\bibfnamefont {J.~R.}\ \bibnamefont
  {McClean}}, \bibinfo {author} {\bibfnamefont {N.~C.}\ \bibnamefont {Rubin}},
  \bibinfo {author} {\bibfnamefont {K.~J.}\ \bibnamefont {Sung}}, \bibinfo
  {author} {\bibfnamefont {I.~D.}\ \bibnamefont {Kivlichan}}, \bibinfo {author}
  {\bibfnamefont {X.}~\bibnamefont {Bonet-Monroig}}, \bibinfo {author}
  {\bibfnamefont {Y.}~\bibnamefont {Cao}}, \bibinfo {author} {\bibfnamefont
  {C.}~\bibnamefont {Dai}}, \bibinfo {author} {\bibfnamefont {E.~S.}\
  \bibnamefont {Fried}}, \bibinfo {author} {\bibfnamefont {C.}~\bibnamefont
  {Gidney}}, \bibinfo {author} {\bibfnamefont {B.}~\bibnamefont {Gimby}},
  \bibinfo {author} {\bibfnamefont {P.}~\bibnamefont {Gokhale}}, \bibinfo
  {author} {\bibfnamefont {T.}~\bibnamefont {Häner}}, \bibinfo {author}
  {\bibfnamefont {T.}~\bibnamefont {Hardikar}}, \bibinfo {author}
  {\bibfnamefont {V.}~\bibnamefont {Havlíček}}, \bibinfo {author}
  {\bibfnamefont {O.}~\bibnamefont {Higgott}}, \bibinfo {author} {\bibfnamefont
  {C.}~\bibnamefont {Huang}}, \bibinfo {author} {\bibfnamefont
  {J.}~\bibnamefont {Izaac}}, \bibinfo {author} {\bibfnamefont
  {Z.}~\bibnamefont {Jiang}}, \bibinfo {author} {\bibfnamefont
  {X.}~\bibnamefont {Liu}}, \bibinfo {author} {\bibfnamefont {S.}~\bibnamefont
  {McArdle}}, \bibinfo {author} {\bibfnamefont {M.}~\bibnamefont {Neeley}},
  \bibinfo {author} {\bibfnamefont {T.}~\bibnamefont {O’Brien}}, \bibinfo
  {author} {\bibfnamefont {B.}~\bibnamefont {O’Gorman}}, \bibinfo {author}
  {\bibfnamefont {I.}~\bibnamefont {Ozfidan}}, \bibinfo {author} {\bibfnamefont
  {M.~D.}\ \bibnamefont {Radin}}, \bibinfo {author} {\bibfnamefont
  {J.}~\bibnamefont {Romero}}, \bibinfo {author} {\bibfnamefont {N.~P.~D.}\
  \bibnamefont {Sawaya}}, \bibinfo {author} {\bibfnamefont {B.}~\bibnamefont
  {Senjean}}, \bibinfo {author} {\bibfnamefont {K.}~\bibnamefont {Setia}},
  \bibinfo {author} {\bibfnamefont {S.}~\bibnamefont {Sim}}, \bibinfo {author}
  {\bibfnamefont {D.~S.}\ \bibnamefont {Steiger}}, \bibinfo {author}
  {\bibfnamefont {M.}~\bibnamefont {Steudtner}}, \bibinfo {author}
  {\bibfnamefont {Q.}~\bibnamefont {Sun}}, \bibinfo {author} {\bibfnamefont
  {W.}~\bibnamefont {Sun}}, \bibinfo {author} {\bibfnamefont {D.}~\bibnamefont
  {Wang}}, \bibinfo {author} {\bibfnamefont {F.}~\bibnamefont {Zhang}},\ and\
  \bibinfo {author} {\bibfnamefont {R.}~\bibnamefont {Babbush}},\ }\bibfield
  {title} {\bibinfo {title} {Openfermion: the electronic structure package for
  quantum computers},\ }\href {https://doi.org/10.1088/2058-9565/ab8ebc}
  {\bibfield  {journal} {\bibinfo  {journal} {Quantum Science and Technology}\
  }\textbf {\bibinfo {volume} {5}},\ \bibinfo {pages} {034014} (\bibinfo {year}
  {2020})}\BibitemShut {NoStop}%
\bibitem [{\citenamefont {Sun}\ \emph {et~al.}(2020)\citenamefont {Sun},
  \citenamefont {Zhang}, \citenamefont {Banerjee}, \citenamefont {Bao},
  \citenamefont {Barbry}, \citenamefont {Blunt}, \citenamefont {Bogdanov},
  \citenamefont {Booth}, \citenamefont {Chen}, \citenamefont {Cui},
  \citenamefont {Eriksen}, \citenamefont {Gao}, \citenamefont {Guo},
  \citenamefont {Hermann}, \citenamefont {Hermes}, \citenamefont {Koh},
  \citenamefont {Koval}, \citenamefont {Lehtola}, \citenamefont {Li},
  \citenamefont {Liu}, \citenamefont {Mardirossian}, \citenamefont {McClain},
  \citenamefont {Motta}, \citenamefont {Mussard}, \citenamefont {Pham},
  \citenamefont {Pulkin}, \citenamefont {Purwanto}, \citenamefont {Robinson},
  \citenamefont {Ronca}, \citenamefont {Sayfutyarova}, \citenamefont
  {Scheurer}, \citenamefont {Schurkus}, \citenamefont {Smith}, \citenamefont
  {Sun}, \citenamefont {Sun}, \citenamefont {Upadhyay}, \citenamefont {Wagner},
  \citenamefont {Wang}, \citenamefont {White}, \citenamefont {Whitfield},
  \citenamefont {Williamson}, \citenamefont {Wouters}, \citenamefont {Yang},
  \citenamefont {Yu}, \citenamefont {Zhu}, \citenamefont {Berkelbach},
  \citenamefont {Sharma}, \citenamefont {Sokolov},\ and\ \citenamefont
  {Chan}}]{pyscf2}%
  \BibitemOpen
  \bibfield  {author} {\bibinfo {author} {\bibfnamefont {Q.}~\bibnamefont
  {Sun}}, \bibinfo {author} {\bibfnamefont {X.}~\bibnamefont {Zhang}}, \bibinfo
  {author} {\bibfnamefont {S.}~\bibnamefont {Banerjee}}, \bibinfo {author}
  {\bibfnamefont {P.}~\bibnamefont {Bao}}, \bibinfo {author} {\bibfnamefont
  {M.}~\bibnamefont {Barbry}}, \bibinfo {author} {\bibfnamefont {N.~S.}\
  \bibnamefont {Blunt}}, \bibinfo {author} {\bibfnamefont {N.~A.}\ \bibnamefont
  {Bogdanov}}, \bibinfo {author} {\bibfnamefont {G.~H.}\ \bibnamefont {Booth}},
  \bibinfo {author} {\bibfnamefont {J.}~\bibnamefont {Chen}}, \bibinfo {author}
  {\bibfnamefont {Z.-H.}\ \bibnamefont {Cui}}, \bibinfo {author} {\bibfnamefont
  {J.~J.}\ \bibnamefont {Eriksen}}, \bibinfo {author} {\bibfnamefont
  {Y.}~\bibnamefont {Gao}}, \bibinfo {author} {\bibfnamefont {S.}~\bibnamefont
  {Guo}}, \bibinfo {author} {\bibfnamefont {J.}~\bibnamefont {Hermann}},
  \bibinfo {author} {\bibfnamefont {M.~R.}\ \bibnamefont {Hermes}}, \bibinfo
  {author} {\bibfnamefont {K.}~\bibnamefont {Koh}}, \bibinfo {author}
  {\bibfnamefont {P.}~\bibnamefont {Koval}}, \bibinfo {author} {\bibfnamefont
  {S.}~\bibnamefont {Lehtola}}, \bibinfo {author} {\bibfnamefont
  {Z.}~\bibnamefont {Li}}, \bibinfo {author} {\bibfnamefont {J.}~\bibnamefont
  {Liu}}, \bibinfo {author} {\bibfnamefont {N.}~\bibnamefont {Mardirossian}},
  \bibinfo {author} {\bibfnamefont {J.~D.}\ \bibnamefont {McClain}}, \bibinfo
  {author} {\bibfnamefont {M.}~\bibnamefont {Motta}}, \bibinfo {author}
  {\bibfnamefont {B.}~\bibnamefont {Mussard}}, \bibinfo {author} {\bibfnamefont
  {H.~Q.}\ \bibnamefont {Pham}}, \bibinfo {author} {\bibfnamefont
  {A.}~\bibnamefont {Pulkin}}, \bibinfo {author} {\bibfnamefont
  {W.}~\bibnamefont {Purwanto}}, \bibinfo {author} {\bibfnamefont {P.~J.}\
  \bibnamefont {Robinson}}, \bibinfo {author} {\bibfnamefont {E.}~\bibnamefont
  {Ronca}}, \bibinfo {author} {\bibfnamefont {E.~R.}\ \bibnamefont
  {Sayfutyarova}}, \bibinfo {author} {\bibfnamefont {M.}~\bibnamefont
  {Scheurer}}, \bibinfo {author} {\bibfnamefont {H.~F.}\ \bibnamefont
  {Schurkus}}, \bibinfo {author} {\bibfnamefont {J.~E.~T.}\ \bibnamefont
  {Smith}}, \bibinfo {author} {\bibfnamefont {C.}~\bibnamefont {Sun}}, \bibinfo
  {author} {\bibfnamefont {S.-N.}\ \bibnamefont {Sun}}, \bibinfo {author}
  {\bibfnamefont {S.}~\bibnamefont {Upadhyay}}, \bibinfo {author}
  {\bibfnamefont {L.~K.}\ \bibnamefont {Wagner}}, \bibinfo {author}
  {\bibfnamefont {X.}~\bibnamefont {Wang}}, \bibinfo {author} {\bibfnamefont
  {A.}~\bibnamefont {White}}, \bibinfo {author} {\bibfnamefont {J.~D.}\
  \bibnamefont {Whitfield}}, \bibinfo {author} {\bibfnamefont {M.~J.}\
  \bibnamefont {Williamson}}, \bibinfo {author} {\bibfnamefont
  {S.}~\bibnamefont {Wouters}}, \bibinfo {author} {\bibfnamefont
  {J.}~\bibnamefont {Yang}}, \bibinfo {author} {\bibfnamefont {J.~M.}\
  \bibnamefont {Yu}}, \bibinfo {author} {\bibfnamefont {T.}~\bibnamefont
  {Zhu}}, \bibinfo {author} {\bibfnamefont {T.~C.}\ \bibnamefont {Berkelbach}},
  \bibinfo {author} {\bibfnamefont {S.}~\bibnamefont {Sharma}}, \bibinfo
  {author} {\bibfnamefont {A.~Y.}\ \bibnamefont {Sokolov}},\ and\ \bibinfo
  {author} {\bibfnamefont {G.~K.-L.}\ \bibnamefont {Chan}},\ }\bibfield
  {title} {\bibinfo {title} {{Recent developments in the PySCF program
  package}},\ }\href {https://doi.org/10.1063/5.0006074} {\bibfield  {journal}
  {\bibinfo  {journal} {The Journal of Chemical Physics}\ }\textbf {\bibinfo
  {volume} {153}},\ \bibinfo {pages} {024109} (\bibinfo {year}
  {2020})}\BibitemShut {NoStop}%
\bibitem [{\citenamefont {Smith}\ \emph {et~al.}(2020)\citenamefont {Smith},
  \citenamefont {Burns}, \citenamefont {Simmonett}, \citenamefont {Parrish},
  \citenamefont {Schieber}, \citenamefont {Galvelis}, \citenamefont {Kraus},
  \citenamefont {Kruse}, \citenamefont {Di~Remigio}, \citenamefont {Alenaizan},
  \citenamefont {James}, \citenamefont {Lehtola}, \citenamefont {Misiewicz},
  \citenamefont {Scheurer}, \citenamefont {Shaw}, \citenamefont {Schriber},
  \citenamefont {Xie}, \citenamefont {Glick}, \citenamefont {Sirianni},
  \citenamefont {O’Brien}, \citenamefont {Waldrop}, \citenamefont {Kumar},
  \citenamefont {Hohenstein}, \citenamefont {Pritchard}, \citenamefont
  {Brooks}, \citenamefont {Schaefer}, \citenamefont {Sokolov}, \citenamefont
  {Patkowski}, \citenamefont {DePrince}, \citenamefont {Bozkaya}, \citenamefont
  {King}, \citenamefont {Evangelista}, \citenamefont {Turney}, \citenamefont
  {Crawford},\ and\ \citenamefont {Sherrill}}]{psi4}%
  \BibitemOpen
  \bibfield  {author} {\bibinfo {author} {\bibfnamefont {D.~G.~A.}\
  \bibnamefont {Smith}}, \bibinfo {author} {\bibfnamefont {L.~A.}\ \bibnamefont
  {Burns}}, \bibinfo {author} {\bibfnamefont {A.~C.}\ \bibnamefont
  {Simmonett}}, \bibinfo {author} {\bibfnamefont {R.~M.}\ \bibnamefont
  {Parrish}}, \bibinfo {author} {\bibfnamefont {M.~C.}\ \bibnamefont
  {Schieber}}, \bibinfo {author} {\bibfnamefont {R.}~\bibnamefont {Galvelis}},
  \bibinfo {author} {\bibfnamefont {P.}~\bibnamefont {Kraus}}, \bibinfo
  {author} {\bibfnamefont {H.}~\bibnamefont {Kruse}}, \bibinfo {author}
  {\bibfnamefont {R.}~\bibnamefont {Di~Remigio}}, \bibinfo {author}
  {\bibfnamefont {A.}~\bibnamefont {Alenaizan}}, \bibinfo {author}
  {\bibfnamefont {A.~M.}\ \bibnamefont {James}}, \bibinfo {author}
  {\bibfnamefont {S.}~\bibnamefont {Lehtola}}, \bibinfo {author} {\bibfnamefont
  {J.~P.}\ \bibnamefont {Misiewicz}}, \bibinfo {author} {\bibfnamefont
  {M.}~\bibnamefont {Scheurer}}, \bibinfo {author} {\bibfnamefont {R.~A.}\
  \bibnamefont {Shaw}}, \bibinfo {author} {\bibfnamefont {J.~B.}\ \bibnamefont
  {Schriber}}, \bibinfo {author} {\bibfnamefont {Y.}~\bibnamefont {Xie}},
  \bibinfo {author} {\bibfnamefont {Z.~L.}\ \bibnamefont {Glick}}, \bibinfo
  {author} {\bibfnamefont {D.~A.}\ \bibnamefont {Sirianni}}, \bibinfo {author}
  {\bibfnamefont {J.~S.}\ \bibnamefont {O’Brien}}, \bibinfo {author}
  {\bibfnamefont {J.~M.}\ \bibnamefont {Waldrop}}, \bibinfo {author}
  {\bibfnamefont {A.}~\bibnamefont {Kumar}}, \bibinfo {author} {\bibfnamefont
  {E.~G.}\ \bibnamefont {Hohenstein}}, \bibinfo {author} {\bibfnamefont
  {B.~P.}\ \bibnamefont {Pritchard}}, \bibinfo {author} {\bibfnamefont {B.~R.}\
  \bibnamefont {Brooks}}, \bibinfo {author} {\bibfnamefont {I.}~\bibnamefont
  {Schaefer}, \bibfnamefont {Henry~F.}}, \bibinfo {author} {\bibfnamefont
  {A.~Y.}\ \bibnamefont {Sokolov}}, \bibinfo {author} {\bibfnamefont
  {K.}~\bibnamefont {Patkowski}}, \bibinfo {author} {\bibfnamefont
  {I.}~\bibnamefont {DePrince}, \bibfnamefont {A.~Eugene}}, \bibinfo {author}
  {\bibfnamefont {U.}~\bibnamefont {Bozkaya}}, \bibinfo {author} {\bibfnamefont
  {R.~A.}\ \bibnamefont {King}}, \bibinfo {author} {\bibfnamefont {F.~A.}\
  \bibnamefont {Evangelista}}, \bibinfo {author} {\bibfnamefont {J.~M.}\
  \bibnamefont {Turney}}, \bibinfo {author} {\bibfnamefont {T.~D.}\
  \bibnamefont {Crawford}},\ and\ \bibinfo {author} {\bibfnamefont {C.~D.}\
  \bibnamefont {Sherrill}},\ }\bibfield  {title} {\bibinfo {title} {{PSI4 1.4:
  Open-source software for high-throughput quantum chemistry}},\ }\href
  {https://doi.org/10.1063/5.0006002} {\bibfield  {journal} {\bibinfo
  {journal} {The Journal of Chemical Physics}\ }\textbf {\bibinfo {volume}
  {152}},\ \bibinfo {pages} {184108} (\bibinfo {year} {2020})}\BibitemShut
  {NoStop}%
\bibitem [{\citenamefont {Malyshev}(2024)}]{anqs_qchem_github_repo}%
  \BibitemOpen
  \bibfield  {author} {\bibinfo {author} {\bibfnamefont {A.}~\bibnamefont
  {Malyshev}},\ }\href@noop {} {\bibinfo {title} {Autoregressive neural quantum
  states for quantum chemistry}},\ \bibinfo {howpublished}
  {\url{https://github.com/Exferro/anqs_quantum_chemistry}} (\bibinfo {year}
  {2024})\BibitemShut {NoStop}%
\bibitem [{\citenamefont {Reh}\ \emph {et~al.}(2023)\citenamefont {Reh},
  \citenamefont {Schmitt},\ and\ \citenamefont {G{\"a}rttner}}]{markus_review}%
  \BibitemOpen
  \bibfield  {author} {\bibinfo {author} {\bibfnamefont {M.}~\bibnamefont
  {Reh}}, \bibinfo {author} {\bibfnamefont {M.}~\bibnamefont {Schmitt}},\ and\
  \bibinfo {author} {\bibfnamefont {M.}~\bibnamefont {G{\"a}rttner}},\
  }\bibfield  {title} {\bibinfo {title} {Optimizing design choices for neural
  quantum states},\ }\href@noop {} {\bibfield  {journal} {\bibinfo  {journal}
  {arXiv preprint arXiv:2301.06788}\ } (\bibinfo {year} {2023})}\BibitemShut
  {NoStop}%
\bibitem [{\citenamefont {Bortone}\ \emph {et~al.}(2024)\citenamefont
  {Bortone}, \citenamefont {Rath},\ and\ \citenamefont
  {Booth}}]{bortone_autoregressive_caveat}%
  \BibitemOpen
  \bibfield  {author} {\bibinfo {author} {\bibfnamefont {M.}~\bibnamefont
  {Bortone}}, \bibinfo {author} {\bibfnamefont {Y.}~\bibnamefont {Rath}},\ and\
  \bibinfo {author} {\bibfnamefont {G.~H.}\ \bibnamefont {Booth}},\ }\bibfield
  {title} {\bibinfo {title} {Impact of conditional modelling for a universal
  autoregressive quantum state},\ }\href
  {https://doi.org/10.22331/q-2024-02-08-1245} {\bibfield  {journal} {\bibinfo
  {journal} {{Quantum}}\ }\textbf {\bibinfo {volume} {8}},\ \bibinfo {pages}
  {1245} (\bibinfo {year} {2024})}\BibitemShut {NoStop}%
\bibitem [{\citenamefont {Abadi}\ \emph {et~al.}(2015)\citenamefont {Abadi},
  \citenamefont {Agarwal}, \citenamefont {Barham}, \citenamefont {Brevdo},
  \citenamefont {Chen}, \citenamefont {Citro}, \citenamefont {Corrado},
  \citenamefont {Davis}, \citenamefont {Dean}, \citenamefont {Devin},
  \citenamefont {Ghemawat}, \citenamefont {Goodfellow}, \citenamefont {Harp},
  \citenamefont {Irving}, \citenamefont {Isard}, \citenamefont {Jia},
  \citenamefont {Jozefowicz}, \citenamefont {Kaiser}, \citenamefont {Kudlur},
  \citenamefont {Levenberg}, \citenamefont {Man\'{e}}, \citenamefont {Monga},
  \citenamefont {Moore}, \citenamefont {Murray}, \citenamefont {Olah},
  \citenamefont {Schuster}, \citenamefont {Shlens}, \citenamefont {Steiner},
  \citenamefont {Sutskever}, \citenamefont {Talwar}, \citenamefont {Tucker},
  \citenamefont {Vanhoucke}, \citenamefont {Vasudevan}, \citenamefont
  {Vi\'{e}gas}, \citenamefont {Vinyals}, \citenamefont {Warden}, \citenamefont
  {Wattenberg}, \citenamefont {Wicke}, \citenamefont {Yu},\ and\ \citenamefont
  {Zheng}}]{tensorflow}%
  \BibitemOpen
  \bibfield  {author} {\bibinfo {author} {\bibfnamefont {M.}~\bibnamefont
  {Abadi}}, \bibinfo {author} {\bibfnamefont {A.}~\bibnamefont {Agarwal}},
  \bibinfo {author} {\bibfnamefont {P.}~\bibnamefont {Barham}}, \bibinfo
  {author} {\bibfnamefont {E.}~\bibnamefont {Brevdo}}, \bibinfo {author}
  {\bibfnamefont {Z.}~\bibnamefont {Chen}}, \bibinfo {author} {\bibfnamefont
  {C.}~\bibnamefont {Citro}}, \bibinfo {author} {\bibfnamefont {G.~S.}\
  \bibnamefont {Corrado}}, \bibinfo {author} {\bibfnamefont {A.}~\bibnamefont
  {Davis}}, \bibinfo {author} {\bibfnamefont {J.}~\bibnamefont {Dean}},
  \bibinfo {author} {\bibfnamefont {M.}~\bibnamefont {Devin}}, \bibinfo
  {author} {\bibfnamefont {S.}~\bibnamefont {Ghemawat}}, \bibinfo {author}
  {\bibfnamefont {I.}~\bibnamefont {Goodfellow}}, \bibinfo {author}
  {\bibfnamefont {A.}~\bibnamefont {Harp}}, \bibinfo {author} {\bibfnamefont
  {G.}~\bibnamefont {Irving}}, \bibinfo {author} {\bibfnamefont
  {M.}~\bibnamefont {Isard}}, \bibinfo {author} {\bibfnamefont
  {Y.}~\bibnamefont {Jia}}, \bibinfo {author} {\bibfnamefont {R.}~\bibnamefont
  {Jozefowicz}}, \bibinfo {author} {\bibfnamefont {L.}~\bibnamefont {Kaiser}},
  \bibinfo {author} {\bibfnamefont {M.}~\bibnamefont {Kudlur}}, \bibinfo
  {author} {\bibfnamefont {J.}~\bibnamefont {Levenberg}}, \bibinfo {author}
  {\bibfnamefont {D.}~\bibnamefont {Man\'{e}}}, \bibinfo {author}
  {\bibfnamefont {R.}~\bibnamefont {Monga}}, \bibinfo {author} {\bibfnamefont
  {S.}~\bibnamefont {Moore}}, \bibinfo {author} {\bibfnamefont
  {D.}~\bibnamefont {Murray}}, \bibinfo {author} {\bibfnamefont
  {C.}~\bibnamefont {Olah}}, \bibinfo {author} {\bibfnamefont {M.}~\bibnamefont
  {Schuster}}, \bibinfo {author} {\bibfnamefont {J.}~\bibnamefont {Shlens}},
  \bibinfo {author} {\bibfnamefont {B.}~\bibnamefont {Steiner}}, \bibinfo
  {author} {\bibfnamefont {I.}~\bibnamefont {Sutskever}}, \bibinfo {author}
  {\bibfnamefont {K.}~\bibnamefont {Talwar}}, \bibinfo {author} {\bibfnamefont
  {P.}~\bibnamefont {Tucker}}, \bibinfo {author} {\bibfnamefont
  {V.}~\bibnamefont {Vanhoucke}}, \bibinfo {author} {\bibfnamefont
  {V.}~\bibnamefont {Vasudevan}}, \bibinfo {author} {\bibfnamefont
  {F.}~\bibnamefont {Vi\'{e}gas}}, \bibinfo {author} {\bibfnamefont
  {O.}~\bibnamefont {Vinyals}}, \bibinfo {author} {\bibfnamefont
  {P.}~\bibnamefont {Warden}}, \bibinfo {author} {\bibfnamefont
  {M.}~\bibnamefont {Wattenberg}}, \bibinfo {author} {\bibfnamefont
  {M.}~\bibnamefont {Wicke}}, \bibinfo {author} {\bibfnamefont
  {Y.}~\bibnamefont {Yu}},\ and\ \bibinfo {author} {\bibfnamefont
  {X.}~\bibnamefont {Zheng}},\ }\href {https://www.tensorflow.org/} {\bibinfo
  {title} {{TensorFlow}: Large-scale machine learning on heterogeneous
  systems}} (\bibinfo {year} {2015}),\ \bibinfo {note} {software available from
  tensorflow.org}\BibitemShut {NoStop}%
\bibitem [{\citenamefont {Bradbury}\ \emph {et~al.}(2018)\citenamefont
  {Bradbury}, \citenamefont {Frostig}, \citenamefont {Hawkins}, \citenamefont
  {Johnson}, \citenamefont {Leary}, \citenamefont {Maclaurin}, \citenamefont
  {Necula}, \citenamefont {Paszke}, \citenamefont {Vander{P}las}, \citenamefont
  {Wanderman-{M}ilne},\ and\ \citenamefont {Zhang}}]{jax}%
  \BibitemOpen
  \bibfield  {author} {\bibinfo {author} {\bibfnamefont {J.}~\bibnamefont
  {Bradbury}}, \bibinfo {author} {\bibfnamefont {R.}~\bibnamefont {Frostig}},
  \bibinfo {author} {\bibfnamefont {P.}~\bibnamefont {Hawkins}}, \bibinfo
  {author} {\bibfnamefont {M.~J.}\ \bibnamefont {Johnson}}, \bibinfo {author}
  {\bibfnamefont {C.}~\bibnamefont {Leary}}, \bibinfo {author} {\bibfnamefont
  {D.}~\bibnamefont {Maclaurin}}, \bibinfo {author} {\bibfnamefont
  {G.}~\bibnamefont {Necula}}, \bibinfo {author} {\bibfnamefont
  {A.}~\bibnamefont {Paszke}}, \bibinfo {author} {\bibfnamefont
  {J.}~\bibnamefont {Vander{P}las}}, \bibinfo {author} {\bibfnamefont
  {S.}~\bibnamefont {Wanderman-{M}ilne}},\ and\ \bibinfo {author}
  {\bibfnamefont {Q.}~\bibnamefont {Zhang}},\ }\href
  {http://github.com/google/jax} {\bibinfo {title} {{JAX}: composable
  transformations of {P}ython+{N}um{P}y programs}} (\bibinfo {year}
  {2018})\BibitemShut {NoStop}%
\end{thebibliography}%

\onecolumngrid
\newpage %

\begin{center}
    \textbf{\large Supplementary material: Neural quantum states and peaked molecular wave functions: curse or blessing?}
\end{center}
\vspace{1em} %

\setcounter{section}{0}
\renewcommand{\thesection}{S\arabic{section}}

\setcounter{figure}{0}
\renewcommand{\thefigure}{S\arabic{figure}}

\setcounter{table}{0}
\renewcommand{\thetable}{S\arabic{table}}

This supplementary consists of two main parts.
In Section~\ref{sec:ablation_studies} we cover the ablation studies mentioned in the main text. Namely, we ablate autoregressive sampling without replacement, grouping qubit into qudits and stochastic reconfiguration.
The remaining sections are dedicated to GPU implementation of \Elocvar{} calculation using primitives of PyTorch software library~\cite{pytorch}.
In Section~\ref{sec:model_hamiltonian} we introduce a toy problem which we use as illustration in the remaining sections.
In Section~\ref{sec:general_principles} we cover the general principles behind our implementation: contiguous storing of information and pointer arithmetic.
In Section~\ref{sec:main_functions} we explain implementations of \CoupleAllToAll{} and \MatrixElement{}.
In Section~\ref{sec:prefix_tree_primitives} we discuss how to perform prefix tree operations on a GPU and provide implementation for \CoupleViaPrefixTree{}.
Finally, the considered higher-level routines require auxiliary lower-level functions not readily available in PyTorch, and we overview their implementation in Section~\ref{sec:aux_functions}.

\section{Ablation studies}\label{sec:ablation_studies}

\subsection{Autoregressive Gumbel sampling}
\begin{figure*}[b]
	\centering
	\includegraphics[width=\linewidth]{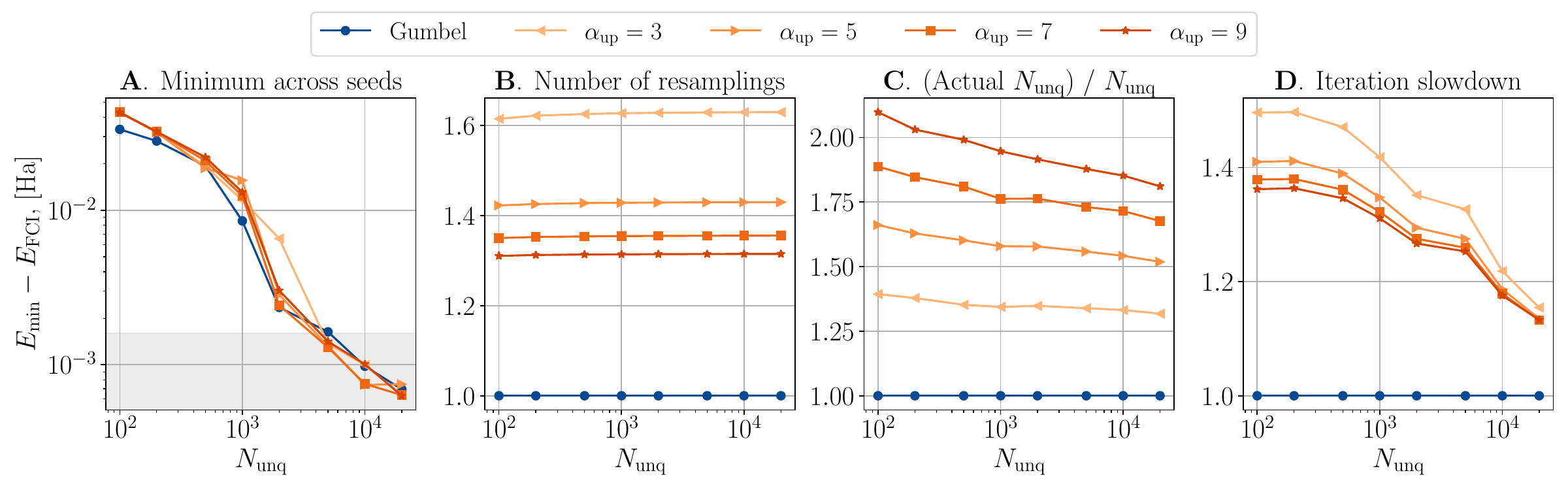}
	\caption{An ablation study comparing Gumbel sampling to adaptive statistics sampling strategies for several values of $\alpha_{\rm up}$. \textbf{A.} Achieved energy differences with respect to $E_{\rm FCI}$; the grey shaded area corresponds to the energy differences within chemical accuracy of 1.6 mHa. \textbf{B.} Number of resamplings required to obtain the desired number of unique samples. \textbf{C.} Ratio of the actual number of unique samples produced during the last resampling to the desired \Nunq{}. \textbf{D.} Slowdown in the total iteration time brought by adaptive sampling as compared to the Gumbel sampling iteration time.}
	\label{fig:gumbel_study}
\end{figure*}
In this ablation study we compare autoregressive Gumbel sampling to adaptive autoregressive statistics sampling.
As discussed in the main text, autoregressive Gumbel sampling is advantageous due to its ability to produce the number of unique samples \Nunq{} \emph{exactly} as specified by the user.
However, it is possible to obtain the desired number of samples at every iteration with the conventional autoregressive statistics sampling \cite{barrett_autoregressive_qchem,malyshev_anqs_with_quantum_numbers} too.
To that end one has to resort to adaptive sampling: if the number of produced unique samples is less than \Nunq{}, then one increases \Ns{} according to some batch size schedule and samples again.

Specifically, we consider the following adaptive scheme: if the number of produced unique samples is less than \Nunq{}, we increase \Ns{} by a factor of $\alpha_{\rm up}$ and sample again.
As soon as the number of obtained unique samples exceeds \Nunq{}, we stop sampling and select the first \Nunq{} unique samples with the highest probabilities $p(\xvec)$ produced by the ansatz.
In addition, to avoid soaring values of \Ns{}, we \emph{decrease} the batch size by a factor of $\alpha_{\rm down}=2$ before \emph{the first} sampling attempt of each iteration.
One might consider various other adaptive sampling approaches, however this remains outside the scope of this paper.

For our benchmark  we optimise an ANQS corresponding to the \ce{Li2O} molecule for $2\cdot 10^4$ iterations using plain Adam optimiser across a range of \Nunq{}. We choose five possible values of $\alpha_{\rm up} = \left\lbrace 3, 5, 7, 9\right\rbrace$.
In addition, we run calculations using only autoregressive Gumbel sampling.

The results of optimisation are presented in Fig.~\ref{fig:gumbel_study}.
One can see in Fig.~\ref{fig:gumbel_study}A that the energy errors achieved by all sampling strategies are similar to a good degree of accuracy, with no sampling scheme having a competitive edge over others.
At the same time all sampling approaches apart from Gumbel require on average more than one sampling per iteration as displayed in Fig.~\ref{fig:gumbel_study}B.
Evidently, the number of resamplings decreases as $\alpha_{\rm up}$ grows, since it takes fewer attempts to reach higher values of \Ns{}.
However, this comes at a cost of increased computational burden as shown in Fig.~\ref{fig:gumbel_study}C: the actual number of unique samples produced during each successful sampling attempt is larger than \Nunq{} for all adaptive sampling strategies.
As a result, every adaptive sampling scheme increases iteration time with respect to Gumbel sampling as presented in Fig.~\ref{fig:gumbel_study}D.
Notably, lower values of $\alpha_{\rm up}$ slow down the computation more, since restarting sampling multiple times is more expensive than producing more unique samples as long as the batch of unique samples fits in the GPU RAM. 

Finally, let us note that while the slowdowns displayed in Fig.~\ref{fig:gumbel_study}D are rather moderate, this is mainly due to sampling taking less than 15\% of the total iteration time as shown in Fig.~\ref{fig:local_energy_fractions} of the main text.
However, we expect the advantage of Gumbel sampling to become more pronounced for larger molecules, with sampling taking a larger part of iteration time.

\subsection{Grouping qubits into qudits}

\begin{figure*}[h!]
    \centering
    \begin{minipage}[c]{0.5\textwidth}
        \centering
        \includegraphics[width=\linewidth]{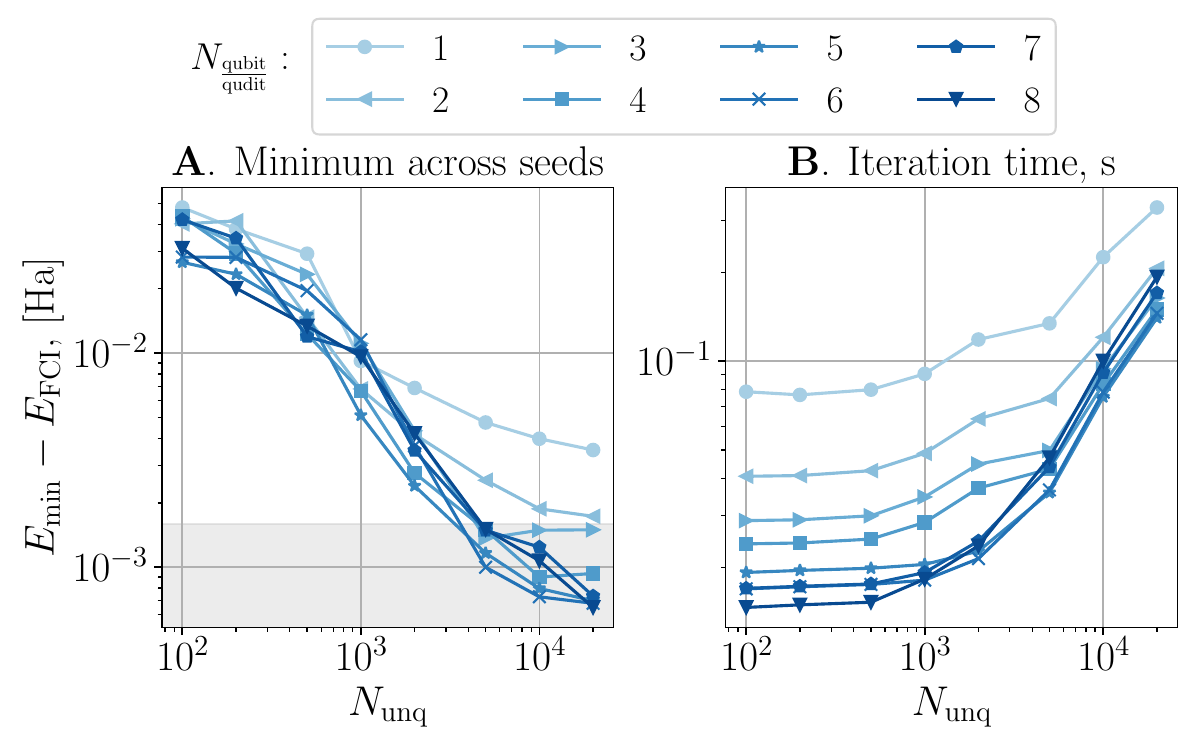}
    \end{minipage}
    \begin{minipage}[c]{0.4\textwidth}
        \centering
        \vspace*{\fill}
        \begin{tabular}{|c|c|}
            \hline
            \QuditDepth & \Np \\
            \hline
            1 & 317,048 \\
            \hline
            2 & 161,528 \\
            \hline
            3 & 112,288 \\
            \hline
            4 & 97,128 \\
            \hline
            5 & 85,376 \\
            \hline
            6 & 91,648 \\
            \hline
            7 & 118,408 \\
            \hline
            8 & 148,224 \\
            \hline
        \end{tabular}
        \vspace*{\fill}
    \end{minipage}%
    \caption{An ablation studying the impact of grouping qubits into qudits. \textbf{A.} Achieved energy differences with respect to $E_{\rm FCI}$; the grey shaded area corresponds to the energy differences within chemical accuracy of 1.6 mHa. \textbf{B.} Iteration time averaged across seeds.}
    \label{fig:qubit_per_qudit_study}
\end{figure*}
In this ablation study we investigate the impact of grouping qubits into qudits.
Similarly to above, we optimise the ANQS representing the \ce{Li2O} molecule for $2\cdot 10^4$ iterations without resorting to SR.
We consider every possible value of \QuditDepth{} from 1 to 8 and show the results in Fig.~\ref{fig:qubit_per_qudit_study}.
One can see in Fig.~\ref{fig:qubit_per_qudit_study}A that  low values of \QuditDepth{} such as 1 and 2 consistently perform the worst in terms of the achieved energy error, even though they have the largest number of parameters as shown in the table in Fig.~\ref{fig:qubit_per_qudit_study}.
This is in agreement with previous reports indicating that masking conditional wave functions required for correct symmetry-aware sampling reduces the expressivity of an ANQS~\cite{markus_review,bortone_autoregressive_caveat, malyshev_anqs_with_quantum_numbers}.
At the same time, increasing \QuditDepth{} results in steady improvement of the achieved accuracy, with $\QuditDepth{} = 5, 6$ providing the best energies at $\Nunq = 20000$.
Importantly, these values of $\QuditDepth{}$ also result in the least time spent on sampling and amplitude evaluation as depicted in Fig.~\ref{fig:qubit_per_qudit_study}.
This is due to the fact that larger \QuditDepth{} amount to fewer conditional wave functions $\psi(x_i|\xvec_{<i})$ constituting the ansatz (the number of conditional wave functions is given by $\left\lceil \frac{N}{\QuditDepth{}}\right\rceil$). 

Similar regularities were obtained in our preliminary experiments on other molecules. As a result, we chose $\QuditDepth{} = 6$ for all numerical experiments described in the main text. 

\subsection{Stochastic reconfiguration}
In the final ablation study we investigate how SR affects the accuracy and computational efficiency of ANQS quantum chemistry calculations. Prior to comparing the optimisation with and without SR, we select the optimal hyperparameter \NSR{}.
Similarly to the previous ablation studies, we consider the \ce{Li2O} molecule and run ANQS optimisation  with different values of $\NSR{} \in \left\lbrace 1, 10, 25, 50, 100, 200\right\rbrace$; as well we run optimisation instances which do not employ stochastic reconfiguration.
We run each optimisation configuration with 5 different seeds of the underlying pseudorandom number generator and in each run we measure the minimum achieved energy.

\begin{figure}[h]
  \centering
  \begin{minipage}[t]{0.46\textwidth}
    \centering
    \includegraphics[width=\textwidth]{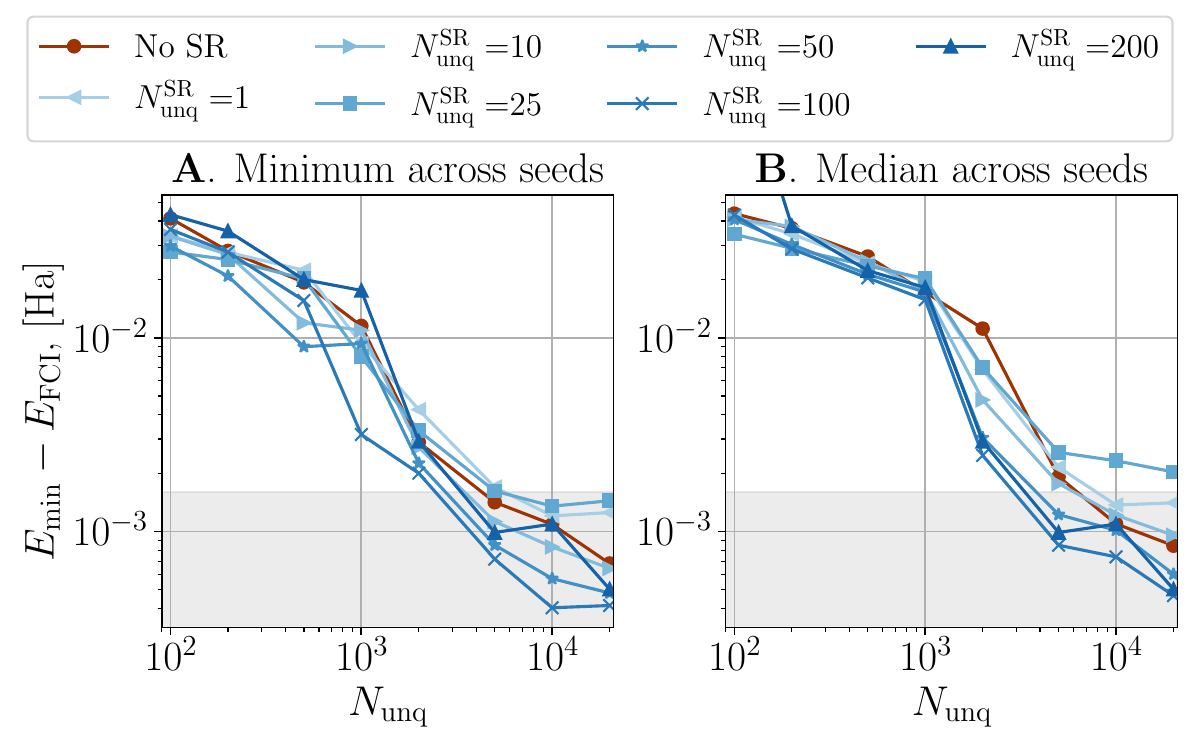}
    \caption{Impact of hyperparameter $\NSR$ in stochastic reconfiguration. We plot the achieved energy differences with respect to $E_{\rm FCI}$; the grey shaded area corresponds to  energy differences within the chemical accuracy of 1.6 mHa. Figures \textbf{A} and \textbf{B} show minimum and median statistics calculated across five seeds respectively.}
\label{fig:sr_hyperparameter_study}
  \end{minipage}
  \hfill
  \begin{minipage}[t]{0.46\textwidth}
    \centering
    \includegraphics[width=\textwidth]{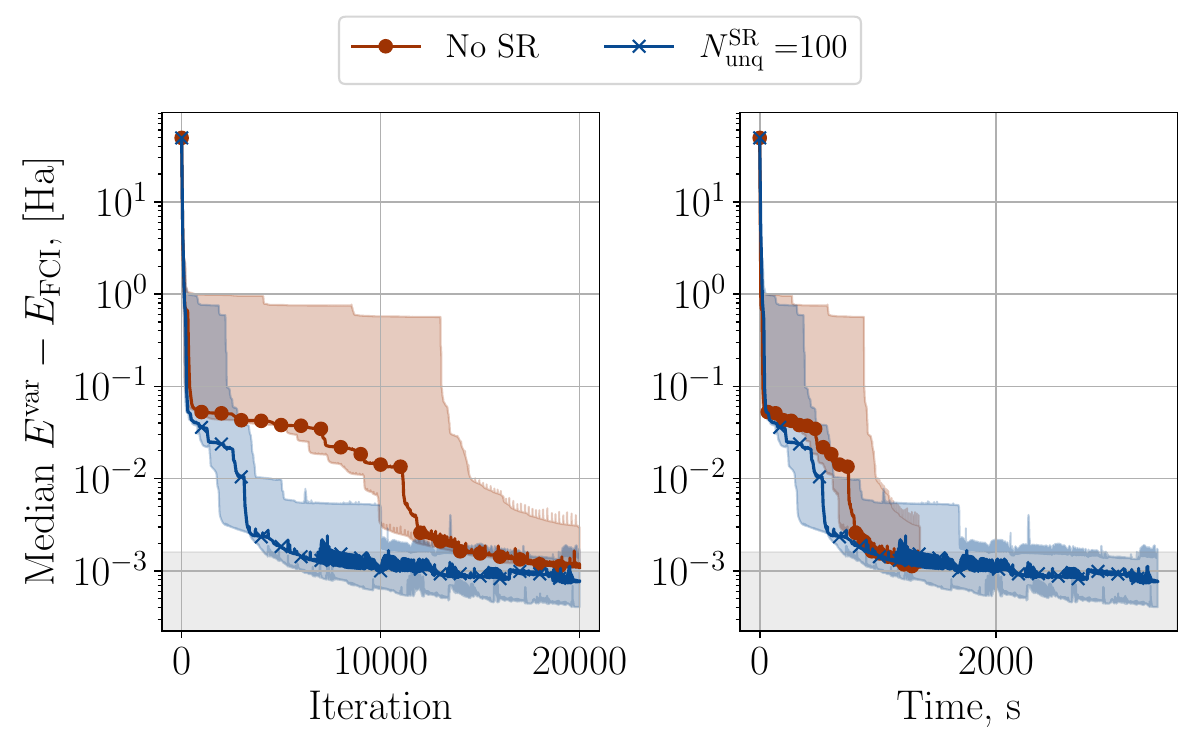}
    \caption{Example training curves for \ce{Li2O} molecule at $\Nunq = 10000$.  We show median energy differences to the exact diagonalisation (FCI) result as a function of (\textbf{A}) iteration number; (\textbf{B}) wall clock time. The coloured shaded areas depict the spread of values from minimum to maximum across seeds. The grey shaded areas correspond to the energy differences within chemical accuracy of 1.6 mHa.     }
	\label{fig:sr_learning_curves}
  \end{minipage}
\end{figure}

In Fig.~\ref{fig:sr_hyperparameter_study}A and Fig.~\ref{fig:sr_hyperparameter_study}B we show, respectively, the minimum and median achieved energies across the five seeds.
Generally, larger values of \NSR{} result in lower achieved energies, with $\NSR = 100$ providing the best energy starting from $\Nunq = 10^3$.
Rather surprisingly, stochastic reconfiguration does not always translate into the improved convergence: for example, $\NSR = 1$ and $\NSR = 25$ provide worse energies than purely Adam-based optimisation.
The larger value of \NSR{}, namely 200, does not result in better energies either, even though one might na\"ively expect it to provide a more accurate estimate of the metric tensor $S$. We attribute such behaviour to sharp singular spectrum of $S$ typical for NQS. A larger \NSR{} gives rise to multiple, very small singular values, introducing numerical noise which affects the accuracy of  the inversion of $S$.

In Fig.~\ref{fig:sr_learning_curves} we show exemplary training curves for optimisation of the \ce{Li2O} molecule with $\Nunq = 10^4$.
Specifically, we show how \Evar{} changes with the iteration number (Fig.~\ref{fig:sr_learning_curves}A) and elapsed time (Fig.~\ref{fig:sr_learning_curves}B) for two optimisations run (i) with $\NSR = 100$ and (ii) without SR.
It can be seen that the SR-driven optimisation achieves lower energies substantially earlier. 
However, once the required iteration time is taken into account, the optimisation performed without SR catches up and might eventually achieve similar energies in a similar \emph{time}.
The optimal strategy could therefore be to start the training with SR and switch it off as soon as the iterations ``break through'' to energies below a certain threshold. 

\section{GPU implementation: model problem}\label{sec:model_hamiltonian}
This section sets the stage for the subsequent discussion of GPU-based \Elocvar{} calculation.
First, we fill the gap left in the main text and discuss the subtleties of evaluating Hamiltonian matrix elements \Helem{}. 
Second, we introduce a toy problem and provide a detailed calculations of $\Elocvar(\xvec)$ values.
The results of this calculation will be used as an illustrative example in the sections covering GPU operations.

\subsection{Calculating the Hamiltonian matrix element}\label{sec:matrix_element}
Let \MatrixElement{} be a function which calculates \Helem{} for a given pair of \xvec{} and \xvecp{}.
To calculate the Hamiltonian matrix element between two basis vectors, one needs to sum the values $h_l \Phatlelem$ over the set of all \Phatl{} corresponding to the same \XYvec.
We define this set as follows:
\begin{equation}\label{equ:xy_to_yz_set}
    \mathcal{P}_{\XYvec} \coloneqq \Set{\Phatl \equiv  \left(h_l, \Xvecl, \Yvecl, \Zvecl \right )|  \XYvecl = \XYvec}.
\end{equation}
In this case, the matrix element \Helem{} can be calculated using the following algorithm:
\nextalgo
\begin{algorithm}[H]
    \caption{Calculating the Hamiltonian matrix element}\label{algo:matrix_element}
    \begin{algorithmic}[1]
        \Function{MatrixElement}{\xvec, \xvecp}
            \State $\Helem \coloneqq 0$
            \State $\XYvec \coloneqq \xvec \XORop \xvecp$
            \For{$\left(h_l, \Xvecl, \Yvecl, \Zvecl \right )$ \PIn{} \XYPauliSet}
                \State $\phi \coloneqq \frac{\pi}{2}\left|\Yvecl \right| + \pi \left|\xvecp \ANDop \YZvecl \right|$
                \State $\Helem \mathrel{+\!\!=} h_l \cdot e^{\imagi \phi}$
            \EndFor
            \State \textbf{return} \Helem
        \EndFunction

    \end{algorithmic}
\end{algorithm}
The set \XYPauliSet{} for every unique  \XYvec{} can be precalculated at the start of optimisation and accessed from the memory on demand.

\subsection{Toy Hamiltonian}
We consider the following ``toy'' Hamiltonian acting on the Hilbert space of four qubits:
\begin{equation}
	\Hhat = \WeightOne \cdot \underbrace{\PhatOne}_{\Phat^{(0)}} + \WeightTwo \cdot \underbrace{\PhatTwo}_{\Phat^{(1)}}  \WeightThree \cdot \underbrace{\PhatThree}_{\Phat^{(2)}}  \WeightFour \cdot \underbrace{\PhatFour}_{\Phat^{(4)}} + \WeightFive \cdot \underbrace{\PhatFive}_{\Phat^{(4)}}.
\end{equation}
This Hamiltonian does not correspond to any particular molecule; we select it exclusively for the purpose of illustration.  
Table~\ref{tab:toy_ham_breakdown} contains information about each of five Hamiltonian terms, including the bit vectors \Yvec{}, \XYvec{} and \YZvec{} introduced above.
For each bit vector we provide both its binary and decimal representations.
\begin{table}
	\centering
	\newcolumntype{Y}{>{\centering\arraybackslash}X}
	\begin{tabularx}{\linewidth}{|c|c|c|*{3}{Y|Y|}}
		\hline
		\multirow{2}{*}{$l$} & \multirow{2}{*}{$h_l$} & \multirow{2}{*}{$\Phatl$} & \multicolumn{2}{c|}{$\Yvec^{(l)}$} &  \multicolumn{2}{c|}{$\XYvec^{(l)}$} & \multicolumn{2}{c|}{$\YZvec^{(l)}$} \\ \cline{4-9}
		& & & \textsc{Bin} & \textsc{Dec} & \textsc{Bin} & \textsc{Dec} & \textsc{Bin} & \textsc{Dec} \\ \hline
		0 & \WeightOne  & \PhatOne &  0000 & \bf 0 &  0000 & \bf 0 & 0000 & \bf 0 \\ \hline
		1 & \WeightTwo & \PhatTwo &  0000 & \bf 0 &  0000 & \bf 0 & 0110 & \bf 6 \\ \hline
		2 & \WeightThree  & \PhatThree & 0000 & \bf 0 & 1010 & \bf 10 & 0000 & \bf 0 \\ \hline
		3 & \WeightFour  & \PhatFour &  0000 & \bf 0 &  0101 & \bf 5 & 0000 & \bf 0 \\ \hline
		4 & \WeightFive  & \PhatFive &  0110 & \bf 6 &  0110 & \bf 6 & 0110 & \bf 6 \\ \hline
	\end{tabularx}
	\caption{Decomposition of toy Hamiltonian terms into bit strings \Yvec{}, \XYvec{} and \YZvec{}. The columns \textsc{Bin} and \textsc{Dec} show binary and decimal representations of bit vectors correspondingly.}
	\label{tab:toy_ham_breakdown}
\end{table}

It can be seen that there are only \emph{four} unique \XYvec{} in Table~\ref{tab:toy_ham_breakdown} which we denote as follows\footnote{One should be careful to distinguish between three similarly looking notations: (i) $\XYvec{}^{(l)}$ denotes the \XYvec{} vector corresponding to $l$-th Hamiltonian term; (ii) $\underline{XY}_i$ denotes the $i$-th bit of a vector \XYvec{}; (iii) finally, $\XYvec_m$ is the $m$-th bit vector \XYvec{} in the set \UniqueXYSet{}.}:
\begin{equation}
    \begin{gathered}
	\XYvec_0 \coloneqq 0000;\quad\XYvec_1 \coloneqq 1010;\quad \XYvec_2 \coloneqq 0101;\quad\XYvec_3 \coloneqq 0110; \\
    \textsc{Dec}(\XYvec_0) = \textbf{0}; \quad \textsc{Dec}(\XYvec_1) = \textbf{10}; \quad \textsc{Dec}(\XYvec_2) = \textbf{5}; \quad \textsc{Dec}(\XYvec_3) = \textbf{6}; 
    \end{gathered}
\end{equation}
Thus, the set of unique \XYvec{} is as follows: $ \UniqueXYSet = \left[\XYvec_0, \XYvec_1, \XYvec_2, \XYvecN{3} \right]$. 
Finally, the sets \XYPauliSet{} of Hamiltonian terms corresponding to the same \XYvec{} are as follows:
\begin{equation}
	\begin{gathered}
		\PauliSet_{\XYvec_0} = \left[\PhatN{0}, \PhatN{1} \right]; \quad \PauliSet_{\XYvec_1} = \left[\PhatN{2} \right]; \quad \PauliSet_{\XYvec_2} = \left[\PhatN{3} \right]; \quad \PauliSet_{\XYvec_3} = \left[\PhatN{4} \right]. 
	\end{gathered}
\end{equation}
In other words, there are \emph{two} terms corresponding to \XYvecN{0}, while the rest of \XYvec{} in \UniqueXYSet{} are represented by single terms.

\subsection{Unique batch}
Suppose our batch of unique samples contains only three basis vectors $\Unique = \left[\xvec_0, \xvec_1, \xvec_2 \right]$ which are as follows:
\begin{equation}
	\begin{gathered}
		\xvec_0 \coloneqq 1100;\quad\xvec_1 \coloneqq 1001;\quad \xvec_2 \coloneqq 0110;\\
		\textsc{Dec}(\xvec_0) = \textbf{12}; \quad \textsc{Dec}(\xvec_1) = \textbf{9}; \quad \textsc{Dec}(\xvec_2) = \textbf{6}.
	\end{gathered}
\end{equation}
In addition, we assume that these basis vectors have the following (unnormalised) amplitudes:
\begin{equation}
	\psi(\xvec_0) = 2; \quad \psi(\xvec_1) = 1; \quad \psi(\xvec_2) = -1.
\end{equation}

The first step to evaluate local energies of the basis vectors is to find the coupled pairs.
To that end, we compose the following tables:
\begin{equation*}
	\begin{tabular}{|c|c|c|c|}
		\hline
		 \quad & $\xvec_0$ & $\xvec_1$ & $\xvec_2$ \\
		 \hline
		 $\xvec_0$ & 0000 & 0101 &1010 \\ 
		 \hline  
		 $\xvec_1$ & 0101 & 0000 &1111 \\
		 \hline  
		 $\xvec_2$ & 1010 & 1111 &0000 \\
		 \hline  
	\end{tabular}
	\equiv
	\begin{tabular}{|c|c|c|c|}
	\hline
	\quad & $\xvec_0$ & $\xvec_1$ & $\xvec_2$ \\
	\hline
	$\xvec_0$ & \bf 0 & \bf 5 & \bf 10 \\ 
	\hline  
	$\xvec_1$ & \bf 5 & \bf  0  & \bf 15 \\
	\hline  
	$\xvec_2$ & \bf 10  & \bf 15 & \bf 0 \\
	\hline  
	\end{tabular} \equiv
	\begin{tabular}{|c|c|c|c|}
	\hline
	\quad & $\xvec_0$ & $\xvec_1$ & $\xvec_2$ \\
	\hline
	$\xvec_0$ & $\XYvec_0$ & $\XYvec_2$ & $\XYvec_1$ \\ 
	\hline  
	$\xvec_1$ & $\XYvec_2$ & $\XYvec_0$  & \_ \\
	\hline  
	$\xvec_2$ & $\XYvec_1$  & \_ & $\XYvec_0$ \\
	\hline  
\end{tabular}
\end{equation*}
Here at the intersection of a row $\xvec_{i}$ with a column $\xvec_{j}$ we put the value of $\xvec_i \XORop \xvec_j$.
In the first table we display its binary representation, in the second table we put its decimal representation and in the third table we specify the corresponding \XYvec{} in \UniqueXYSet{}.
In our case the sets of coupled basis vectors are as follows:
\begin{equation}
		\SampledAndCoupled_{\xvec_0} = \left[\xvec_0, \xvec_1, \xvec_2 \right];\quad
		\SampledAndCoupled_{\xvec_1} = \left[\xvec_0, \xvec_1\right];\quad
		\SampledAndCoupled_{\xvec_2} = \left[\xvec_0, \xvec_2 \right].\\
\end{equation}
In other words, each basis vector is coupled to itself and in addition the basis vector $\xvec_0$ is coupled to both $\xvec_1$ and $\xvec_2$.

\subsection{Matrix elements}
The corresponding matrix elements are obtained with straightforward algebraic calculations:
\begin{equation*}
	\begin{gathered}
		\Hhat_{\xvec_0 \xvec_0} = 0.9 \braket{\xvec_0|\PhatN{0}|\xvec_0} + 0.1 \braket{\xvec_0|\PhatN{1}|\xvec_0} =  0.9 \underbrace{\braket{1100|1100}}_{+1} + 0.1 \underbrace{\braket{1100|\PhatTwo|1100}}_{-1} = \underline{0.8};\\
		\Hhat_{\xvec_0 \xvec_1} = -0.2 \braket{\xvec_0|\PhatN{3}|\xvec_1} =  -0.2 \underbrace{\braket{1100|\PhatFour|1001}}_{1} =  \underline{-0.2};\\
		\Hhat_{\xvec_0 \xvec_2} = -0.2 \braket{\xvec_0|\PhatN{2}|\xvec_2} = -0.2 \underbrace{\braket{1100|\PhatThree|0110}}_{1} =  \underline{-0.2};\\
	\end{gathered}
\end{equation*}
\begin{equation*}
	\begin{gathered}
		\Hhat_{\xvec_1 \xvec_1} = 0.9 \braket{\xvec_1|\PhatN{0}|\xvec_1} + 0.1 \braket{\xvec_1|\PhatN{1}|\xvec_1} =  0.9 \underbrace{\braket{1001|1001}}_{+1} + 0.1 \underbrace{\braket{1001|\PhatTwo|1001}}_{+1} =  \underline{1.0};\\
		\Hhat_{\xvec_1 \xvec_0} = -0.2 \braket{\xvec_1|\PhatN{3}|\xvec_0} =  -0.2 \underbrace{\braket{1001|\PhatFour|1100}}_{1} =  \underline{-0.2};\\
	\end{gathered}
\end{equation*}
\begin{equation*}
	\begin{gathered}
		\Hhat_{\xvec_2 \xvec_2} = 0.9 \braket{\xvec_2|\PhatN{0}|\xvec_2} + 0.1 \braket{\xvec_2|\PhatN{1}|\xvec_2} =  0.9 \underbrace{\braket{0110|0110}}_{+1} + 0.1 \underbrace{\braket{0110|\PhatTwo|0110}}_{1} =  \underline{1.0};\\
		\Hhat_{\xvec_2 \xvec_0} = -0.2 \braket{\xvec_2|\PhatN{2}|\xvec_0} =  -0.2 \underbrace{\braket{0110|\PhatThree|1100}}_{1} = \underline{-0.2};\\
	\end{gathered}
\end{equation*}

\subsection{Local energies}
Finally, we bring together the computed matrix elements and the amplitudes produced by the ansatz to calculate the local energy values:
\begin{equation*}
	\begin{gathered}
		\Elocvar(\xvec_0) = \frac{1}{\psi(\xvec_0)}\left(\psi(\xvec_0) \Hhat_{\xvec_0 \xvec_0} + \psi(\xvec_1) \Hhat_{\xvec_0 \xvec_1} + \psi(\xvec_2) \Hhat_{\xvec_0 \xvec_2} \right) = 0.8 - \frac{1}{2} \cdot 0.2 + \frac{1}{2}\cdot 0.2  = \underline{0.8};\\
		\Elocvar(\xvec_1) = \frac{1}{\psi(\xvec_1)} \left(\psi(\xvec_0) \Hhat_{\xvec_1 \xvec_0} + \psi(\xvec_1)\Hhat_{\xvec_1 \xvec_1} \right)= -2 \cdot 0.2 + 1.0 = \underline{0.6};\\
		\Elocvar(\xvec_2) = \frac{1}{\psi(\xvec_2)} \left(\psi(\xvec_0) \Hhat_{\xvec_2 \xvec_0} + \psi(\xvec_2)\Hhat_{\xvec_2 \xvec_2}\right) = 2\cdot 0.2 + 1.0 = \underline{1.4}.\\
	\end{gathered}
\end{equation*}

\section{GPU implementation: general principles}\label{sec:general_principles}
\begin{figure*}
	\centering
	\includegraphics{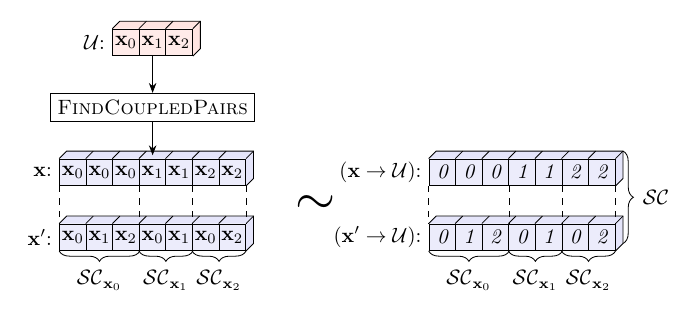}
	\caption{An example of storing the relevant data structures as contiguous pointers.}
	\label{fig:guiding_principles}
\end{figure*}
At the core of every deep learning framework such as PyTorch~\cite{pytorch}, TensorFlow~\cite{tensorflow} and Jax~\cite{jax} lies a library of GPU-accelerated tensor algebra primitives.
These primitives operate on contiguous arrays of known size that store relevant numerical quantities (e.g. network parameters and real-world data).
Such arrays are often manipulated in a vectorised (batched) manner so that the number of explicit loops is reduced to a minimum.

The toy example of previous section suggests that implementing our optimisation procedure using tensor algebra primitives is not straightforward.
For example, the size of each \SampledAndCoupledx{} set is \emph{dynamic} and depends on the batch \Unique{} sampled at the current iteration.
In addition, finding out coupled pairs requires a search operation to determine which of $\XYvec = \xvec \XORop \xvecp$ corresponds to a Hamiltonian term.

In principle, one might still leverage GPU parallelism by writing a specialised CUDA code implementing the variational optimisation.
However, this is beyond our ability.
Instead, in the remainder of this Supplementary we discuss how to implement key components of ANQS variational optimisation using only a stringent set of tensor algebra primitives, specifically that of PyTorch software library.

In this Section we outline the main principles behind our implementation, while specific routines such as \FindSampledAndCoupled{} and \MatrixElement{} are covered in the subsequent sections.
We illustrate every non-trivial operation with the arrays corresponding to the model problem considered in Section~\ref{sec:model_hamiltonian}.

\subsection{Contiguous arrays}
The first guiding principle of our implementation is storing information \emph{contiguously}, as illustrated in Fig.~\ref{fig:guiding_principles}A, which enables the use of vectorised PyTorch primitives.
As discussed in the main text, we represent $N$-bit basis vectors as tuples of $\IntPerVec \coloneqq \left\lceil\frac{N}{\BitDepth}\right\rceil$ integers, where \BitDepth{} is the number of bits contained in a single integer number.
In our case $N=4$ and thus we assume only one integer is needed per basis vectors.
As a result, the batch of three unique samples is stored as an integer array \Unique{} of length 3, or, equivalently, of shape $[\Nunq, 1]$. 
For the sake of clarity, in our diagrams, the first and second dimensions of each tensor are depicted horizontally and vertically, respectively. 
This representation is opposite to the conventional notion where a tensor with shape $[N, 1]$ is viewed as a column vector.
If more than one integer is required to represent a basis vector, \Unique{} becomes an array of shape $\left[\Nunq, \IntPerVec\right]$.
In this case all procedures described in this supplementary can be generalised to account for the extra array dimension (see our code for more details~\cite{anqs_qchem_github_repo}).

In Fig.~\ref{fig:guiding_principles} we emphasise that all procedures return contiguous arrays too.
For example, one can see that the \FindSampledAndCoupled{} procedure takes an array \Unique{} as input and outputs two arrays \xvec{} and \xvecp{}.
These arrays store concatenated first and second elements in each pair of every \SampledAndCoupledx{} respectively.

\subsection{Pointer arithmetic}
\begin{figure*}
	\begin{minipage}{0.45\linewidth}
		\vspace*{\fill}
		\vbox{
			\vfil
			\includegraphics{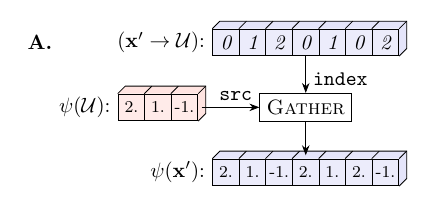}
			\vfil
		}
		\vspace*{\fill}
	\end{minipage}
	\begin{minipage}{0.45\linewidth}
		\vspace*{\fill}
		\vbox{
			\vfil
			\includegraphics{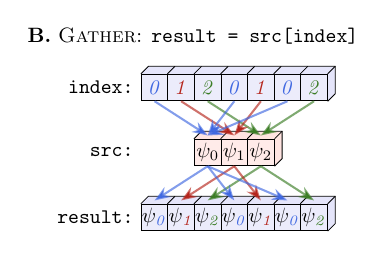}
			\vfil
		}
		\vspace*{\fill}
	\end{minipage}
	\caption{\textbf{A.} Efficient evaluation of ansatz amplitudes using \textsc{Gather} routine. The amplitudes are calculated only for unique basis vectors and then distributed as necessary. \textbf{B.} An example \Gather{} operation.}
	\label{fig:gather_amps}
\end{figure*}
The second key aspect of our code is the use of pointer arithmetic.
Suppose one has an array \Attt{} of length $L_{\Attt}$.
We assume that an array $\pointer{\Bttt}{\Attt}$ is an array of pointers to \Attt{} if $\pointer{\Bttt}{\Attt}$ is an array of integer numbers ranging from $0$ to $L_{\Attt} - 1$ inclusively.
In this case we interpret the $i$-th element of $\pointer{\Bttt}{\Attt}$ as a \emph{pointer} to the $\pointer{\Bttt}{\Attt}[i]$-th element of \Attt{}.
Arrays \xvecptr{} and \xvecpptr{} depicted in Fig.~\ref{fig:guiding_principles} are specific examples of pointer arrays.
They represent arrays \xvec{} and \xvecp{} by storing not $\xvec_0, \xvec_1, \ldots$ themselves, but their indices instead.
For example, $\xvec[3] = \xvec_1$ and thus $\pointer{\xvec}{\Unique}[3]= 1$.

The pointer arithmetic is crucial for our implementation for two reasons. 
First, it allows one to avoid redundant computation, as illustrated in Fig.~\ref{fig:gather_amps}A.
One obtains the amplitudes for each \xvecp{} in \SampledAndCoupled{} by reusing the known values of amplitudes for basis vectors in \Unique{}, rather than evaluating them directly.
The substitution is performed with a standard tensor algebra routine \Gather{} depicted in Fig.~\ref{fig:gather_amps}B.
As input, this routine takes a \src{} array and an array \indexttt{} of pointers to \src{}.
As output, it produces the \result{} array defined as $\result[i] = \src[\indexttt[i]]$.
We call this process \emph{dereferencing} of \indexttt{} with respect to \src{}.
Note that the same pointer array might be dereferenced with respect to different source arrays.
For example, $\Gather(\src:\psi(\Unique), \indexttt:\xvecpptr) = \psi(\xvecp)$, while $\Gather(\src:\Unique, \indexttt:\xvecpptr) = \xvecp$.

\begin{figure*}
	\begin{minipage}{0.45\linewidth}
		\vspace*{\fill}
		\vbox{
			\vfil
			\includegraphics{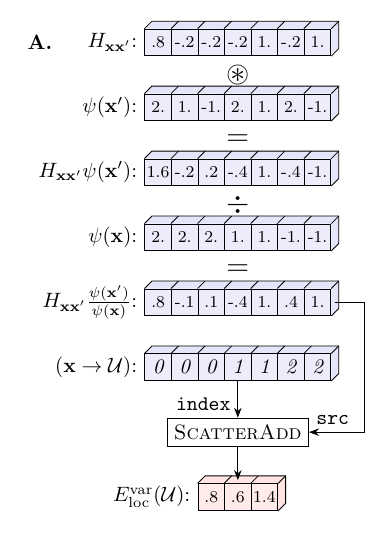}
			\vfil
		}
		\vspace*{\fill}
	\end{minipage}
	\begin{minipage}{0.45\linewidth}
		\vspace*{\fill}
		\vbox{
			\vfil
			\includegraphics{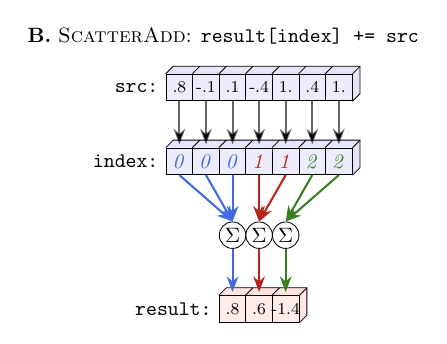}
			\vfil
		}
		\vspace*{\fill}
	\end{minipage}
	\caption{{\textbf{A.} Efficient evaluation of \Elocvar{} where the dynamic range summations are performed with \ScatterAdd{} routine. \textbf{B.} An example \ScatterAdd{} operation.}}
	\label{fig:scatter_local_energies}
\end{figure*}
The second advantage of pointer arithmetic is that it enables summation over dynamically defined ranges of indices as illustrated in Fig.~\ref{fig:scatter_local_energies}A.
Suppose we employed \MatrixElement{} function to obtain the array \Helem{} of Hamiltonian matrix elements.
To obtain $\Elocvar(\xvec)$ one has to perform three following steps:
\begin{enumerate}
	\item Multiply it elementwise by an array $\psi(\xvecp)$;
	\item Divide it elementwise by an array $\psi(\xvec)$;
	\item Sum the values $\Helem \frac{\psi(\xvecp)}{\psi(\xvec)}$ corresponding to the same unique \xvec{}.	
\end{enumerate}
The summation at the last stage can be performed with another tensor algebra routine known as \ScatterAdd{}.
This routine adds the $i$-th element of an array \src{} to the $\indexttt[i]$-th element of array \result{} and is depicted in Fig.~\ref{fig:scatter_local_energies}B.

\subsection{Notation remarks}\label{sec:notation_remarks}
Let us make two remarks regarding the notation and the diagrams representing GPU operations.
First, the names of arrays which store pointers of any kind will necessarily include the symbol $\rightarrow$.
Second, we use different font styling to display the content of arrays of different nature: (i)
we depict \textbf{in bold} the content of arrays storing values corresponding to $N$-bit vectors, e.g. \xvec{} and \xvecp{}; (ii) we display \textit{in italics} the content of arrays storing pointers of any kind, e.g. \xvecptr; (iii) we use normal styling for the content of arrays storing genuine floating point or integer values, e.g. $\psi(\xvec)$ or $\Elocvar(\xvec)$.

\subsection{Preparation}\label{sec:gpu_preparation}
\begin{figure*}
	\centering
	\includegraphics{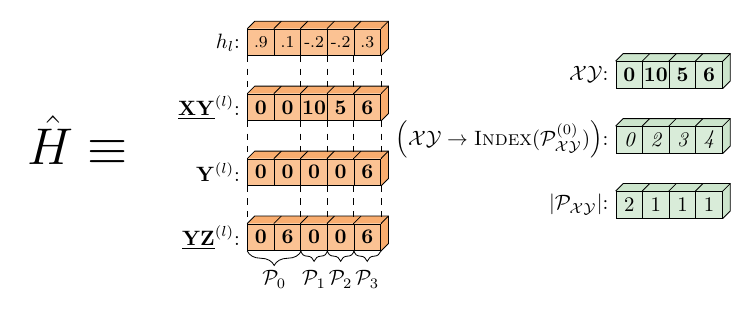}
	\caption{Storing the Hamiltonian as a set of tensors.}
	\label{fig:ham_prep}
\end{figure*}
Before the start of each simulation we preprocess the Hamiltonian to represent it as a set of tensors stored on a GPU as depicted in Fig.~\ref{fig:ham_prep}.
For example, $h_l$ is a contiguous array storing the weights of all Hamiltonian terms, i.e. $h_l[0] = h_0$.
The arrays \XYvecl{}, \Yvecl{} and \YZvecl{} are constructed in a similar way.

We also introduce three auxiliary arrays to represent the fact that each unique \XYvecl{} might correspond to several terms constituting the set \XYPauliSet{}.
In Fig.~\ref{fig:ham_prep} we show how elements of the arrays $h_l$, \Yvecl{} and \YZvecl{} are grouped into the corresponding sets.
For example, the zeroth and first elements of these arrays correspond to $\mathcal{P}_{\XYvec_0}$, which we denote for brevity as $\mathcal{P}_0$.
The first auxiliary array \UniqueXYSet{} is an array of unique \XYvec{} vectors themselves.
Second, \HamYZStarts{} stores pointers to the first elements in \XYPauliSet{} for each \XYvec{} in \UniqueXYSet{}. 
For example, $\HamYZStarts{}[1] = 2$ since the set $\mathcal{P}_1$ starts from the position 2 in arrays $h_l$, \Yvecl{} and \YZvecl{}.
Finally, the array \YZNums{} stores the number of terms in \XYPauliSet{} for each unique \XYvec{}.
These three arrays will prove instrumental for our implementation of \MatrixElement{} discussed further.

\section{GPU implementation: main functions}\label{sec:main_functions}
In this section we cover two key procedures of ANQS optimisation: \CoupleAllToAll{} and \MatrixElement{}.
The implementation of \CoupleViaUniqueXY{} is similar to that of \CoupleAllToAll{} and can be found in our code~\cite{anqs_qchem_github_repo}.
The discussion on implementation of \CoupleViaPrefixTree{} is left for Section~\ref{sec:prefix_tree_primitives}.

\subsection{\CoupleAllToAll{} implementation}\label{sec:all_to_all_implementation} 
\begin{figure*}	
	\begin{minipage}{0.45\linewidth}
		\vspace*{\fill}
		\vbox{
			\vfil
			\includegraphics{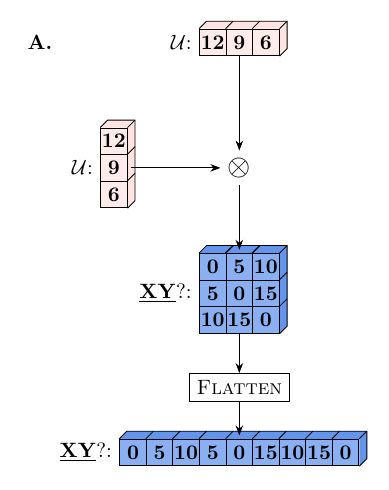}
			\vfil
		}
		\vspace*{\fill}
	\end{minipage}
	\begin{minipage}{0.45\linewidth}
		\vspace*{\fill}
		\vbox{
			\vfil
			\includegraphics{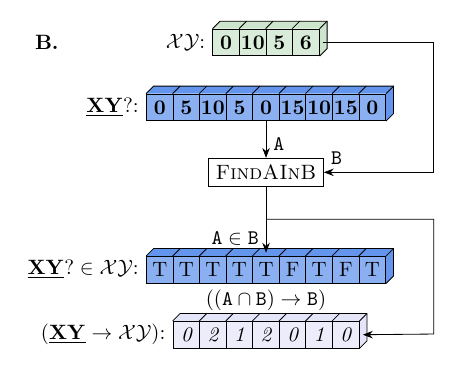}
			\vfil
		}
		\vspace*{\fill}
	\end{minipage}
	\caption{Two first stages of \CoupleAllToAll{} \textbf{A.} One evaluates candidate \XYvec{} bit vectors in a vectorised manner. \textbf{B}. One invokes custom a \FindAInB{} function to obtain a boolean mask indicating whether a candidate \XYvec{} belongs to \UniqueXYSet{}. As well,  one obtains an array of pointers from ``successful'' candidate \XYvec{} to \UniqueXYSet{}.}
	\label{fig:all_to_all_candidates}
\end{figure*}
We split GPU implementation of \CoupleAllToAll{} algorithm into three main stages.

\subsubsection*{Stage 1: Find candidate \XYvec{}}
The first stage is illustrated in Fig.~\ref{fig:all_to_all_candidates}A.
It in a vectorised way all possible candidate vectors $\XYvec = \xvec \XORop \xvecp,\ \forall \xvec, \xvecp \in \Unique$; we denote an array of such candidates as $\XYvec ?$. 
To that end, one applies the bitwise XOR operation to two copies of \Unique, with one reshaped into a column vector. 
The broadcasting rules of PyTorch ensure that the resulting 2D array $\XYvec?$ is a matrix of size $\Nunq \times \Nunq$.
At the end of this stage one flattens $\XYvec?$ into a 1D array by concatenating its rows.

\subsubsection*{Stage 2: Filter candidate \XYvec{}}
The second stage is shown in Fig.~\ref{fig:all_to_all_candidates}B.
It invokes a custom function \FindAInB{} to find which of the candidate $\XYvec{}?$ correspond to bit vectors in \UniqueXYSet{}.
 
To explain \FindAInB{} functioning, suppose we feed it two integer arrays, \Attt{} and \Bttt{}. 
We presume that there are no repeated values in the array \Bttt{}, while there might be some in \Attt{}.
\FindAInB{} produces two output arrays.
The first output array $\Attt \in \Bttt$ is a boolean mask indicating whether the elements of \Attt{} can be found in \Bttt{}.
In our example $(\Attt \in \Bttt)[1] = \textsc{True}$ since $\XYvec?[1] = \mathbf{5}$ belongs to \UniqueXYSet{}. At the same time $(\Attt \in \Bttt)[5] = \textsc{False}$ since $\XYvec?[5] = \mathbf{15}$ \emph{does not} belong to \UniqueXYSet

The second output array $\pointer{(\Attt \cap \Bttt)}{\Bttt}$ contains the pointers from those elements of \Attt{} which belong to \Bttt{} (i.e. $\Attt \cap \Bttt$) to their corresponding positions in $\Bttt$.
Its length is equal to the number of elements in \Attt{} which belong to \Bttt{} and the order between elements is the same as in the initial array \Attt{}. 
For example,  $(\pointer{(\Attt \cap \Bttt)}{\Bttt})[1] = \it{2}$ because the second element in $\XYvec?$ corresponds to the third element in \UniqueXYSet{}.

In what follows we refer to the array $\Attt \in \Bttt$ as $\XYvec?\in\UniqueXYSet$, and to the array $\pointer{(\Attt \cap \Bttt)}{\Bttt}$ as $\pointer{\XYvec}{\UniqueXYSet}$.
Since $\pointer{\XYvec}{\UniqueXYSet}$ corresponds to all correctly coupled \xvec{} and \xvecp{}, its length is equal to the length of  \xvecptr{}  and \xvecpptr{} representing \SampledAndCoupled{}.

\subsubsection*{Stage 3: Obtain \SampledAndCoupled{}}
\begin{figure*}	
	\centering
	\includegraphics{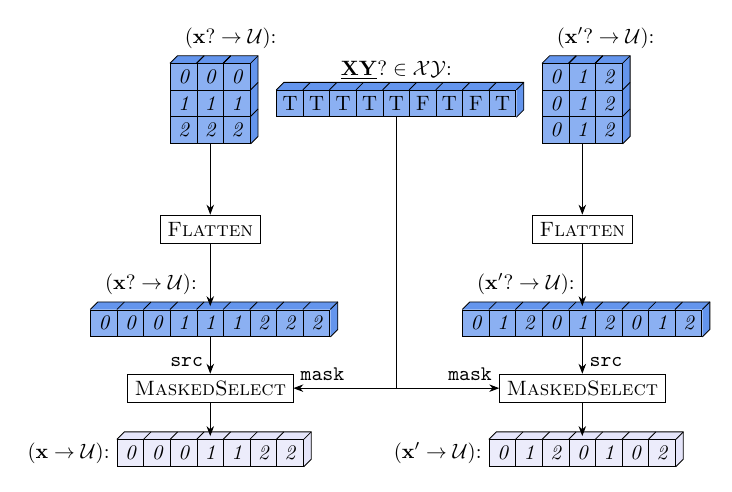}
	\caption{The last stage of \CoupleAllToAll{}. One obtains \SampledAndCoupled{} by masking off the pointers to uncoupled pairs of \xvec{} and \xvecp{}.}
	\label{fig:mask_pointers}
\end{figure*}
The third stage is depicted in Fig.~\ref{fig:mask_pointers}.
During this stage one uses the boolean mask $\XYvec?\in\UniqueXYSet$ to obtain \xvecptr{} and \xvecpptr{}.
To that end, one forms a 2D array of candidate pointers from \xvec{} to \Unique{} which we denote as \candxvecptr{}. 
The first row  of this array contains only zeros since it corresponds to all $\XYvec{}?$ formed with $\xvec_0$ being the first element in the pair.
Correspondingly, the second row contains only ones, the third contains only twos and so on.
The 2D array \candxvecpptr{} is constructed in a similar way, except for its \emph{columns} consisting of the same values.
One flattens \candxvecptr{} and \candxvecpptr{} and applies the boolean mask \candxyvecinxy{} to filter out only those pointers which correspond to the actually coupled pairs $(\xvec, \xvecp)$.

\subsection{\MatrixElement{} implementation}
The GPU implementation of \MatrixElement{} consists of three consecutive stages.
\subsubsection*{Stage 1: Expand the pointers}
\begin{figure*}	
	\centering
	\includegraphics{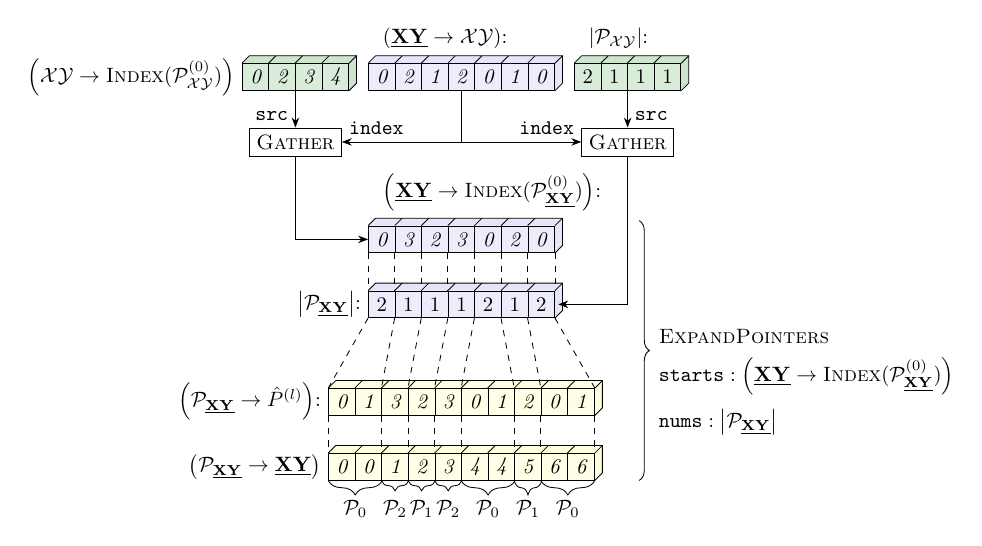}
	\caption{The first stage of \MatrixElement{}. For each coupling \XYvec{} one substitutes its pointer to \UniqueXYSet{} with a contiguous sequence of pointers \Phatl{} in \XYPauliSet. To that end, one employs a custom \textsc{ExpandPointers} function described fin Section~\ref{sec:aux_functions}.}
	\label{fig:expand_xy}
\end{figure*}
The first stage is illustrated in Fig.~\ref{fig:expand_xy}. 
Since each \XYvec{} might correspond to several terms in Hamiltonian, this stage substitutes every pointer in \xyvecptr{} with a sequence of pointers to \YZvec{} corresponding to the given \XYvec{}.
Let us provide an example.
The first element of \xyvecptr{} is $\it 0$ and thus it points to the first element of \UniqueXYSet{}.
This element $\UniqueXYSet{}[0]$ corresponds to $\XYvec = \bf 0$ which is represented with Hamiltonian terms $\Phat_0$ and $\Phat_1$.
Thus, $(\xyvecptr)[0]$ should be expanded to pointers $\it 0$ and $\it 1$.
We store these pointers contiguously in a larger array \xypaulisettermptr{}.
In addition, we keep the reverse pointers from each element in $\mathcal{P}_{\XYvec}$ to its parent \XYvec{} in an array \xypaulisetxyptr{} of the same size as \xypaulisettermptr{}.
\begin{figure*}	
	\centering
\includegraphics{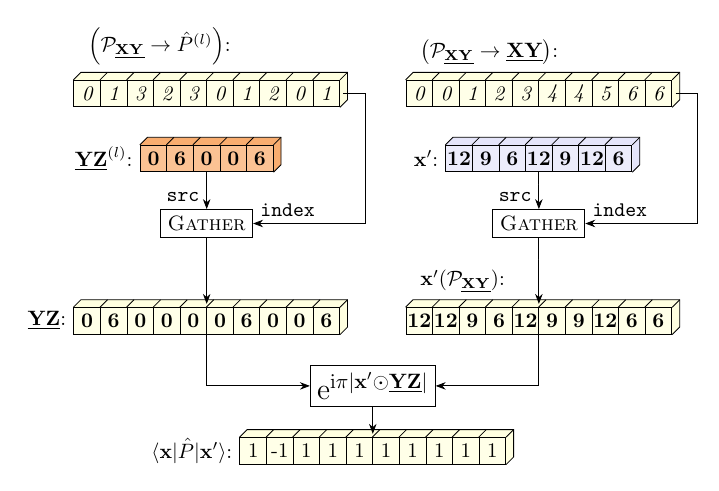}
\caption{The second stage of \MatrixElement{}. One employs the pointers to \Phatl{} to evaluate the matrix elements $\braket{\xvec|\Phat|\xvecp}$ for all \Phat{} coupling each pair $(\xvec, \xvecp)$.}
\label{fig:pauli_expectation}
\end{figure*}
\begin{figure*}	
	\centering
	\includegraphics{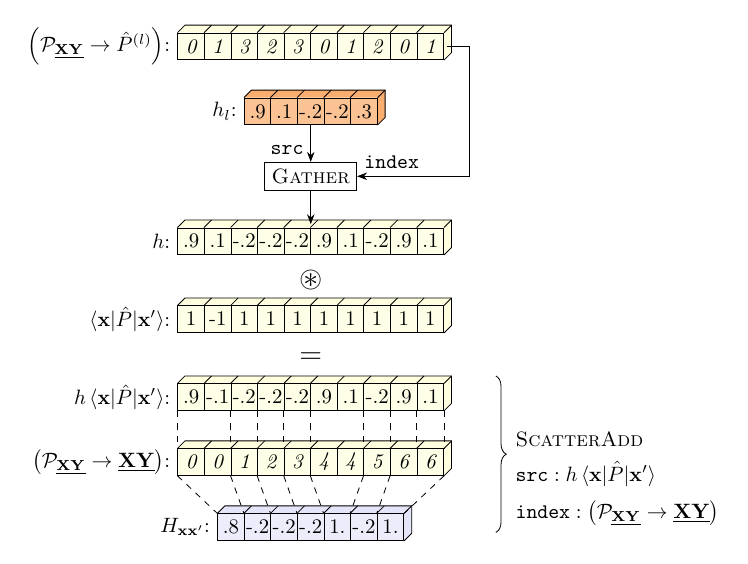}
	\caption{The third stage of \MatrixElement{}. One employs \ScatterAdd{} function to sum all $h \braket{\xvec|\Phat|\xvecp}$ corresponding to each pair $(\xvec, \xvecp)$ into \Helem{}.}
	\label{fig:scatter_matrix_elements}
\end{figure*}
In more details this stage proceeds as follows. 
\begin{enumerate}
	\item First, one uses the \Gather{} operation to obtain arrays \YZStarts{} and \YZNums{} by dereferencing the pointers \xyvecptr{} with respect to the arrays \HamYZStarts{} and \HamYZNums{}.
    Thus, each element of \YZStarts{} stores a pointer to the first Hamiltonian term in the set \XYPauliSet{}.
    Similarly, the array \YZNums{} stores the \emph{sizes} of the corresponding \XYPauliSet{} sets.

	\item As a second step, one expands pointers to the first \XYPauliSet{} elements (which we further refer to as \emph{starts}) and the sizes of \XYPauliSet{} into a contiguous array containing pointers to \emph{all} elements in \XYPauliSet{}.
	This is achieved with a custom function \ExpandPointers{}, which is explained in more details in Section~\ref{sec:aux_functions}.
	The resulting pointers are stored in the array \xypaulisettermptr{}

	\item Finally, one obtains the \emph{reverse} pointers \xypaulisetxyptr{} by repeating the \emph{index} of each \XYvec{} element \XYPauliSetSize{} times as illustrated in the last line of Fig.~\ref{fig:expand_xy}. This can be easily achieved with the standard \textsc{RepeatInterleave} routine of the PyTorch software library.
\end{enumerate}

\subsubsection*{Stage 2: Calculate term expectation values}

The second stage of \MatrixElement{} calculation is shown in Fig.~\ref{fig:pauli_expectation}. 
First, one dereferences the pointers \xypaulisettermptr{} with respect to the array \YZvecl{} to obtain an array of \YZvec{} corresponding to each element in \XYPauliSet{}.
Second, one dereferences the pointers \xypaulisetxyptr{} with respect to the array \xvecp{}.
Thus, one obtains an array $\xvecp(\XYPauliSet)$ which contains all \xvecp{} required to evaluate $\braket{\xvec|\Phat|\xvecp}$.
Finally, one feeds both \YZvec{} and $\xvecp(\XYPauliSet)$ into a simple arithmetic function calculating $\expe^{\imagi \pi \left|\xvecp \ANDop \YZvec \right|}$ elementwise.
As a result, one obtains the array containing the values of $\braket{\xvec|\Phat|\xvecp}$.

\subsubsection*{Stage 3: Scatter term expectation values}

The third stage is depicted in Fig.~\ref{fig:scatter_matrix_elements}.
First, one dereferences the pointers \xypaulisettermptr{} with respect to the array $h_l$.
Thus, one fetches the weights of Hamiltonian terms for corresponding $\braket{\xvec|\Phat|\xvecp}$.
Second, one multiplies $h$ and $\braket{\xvec|\Phat|\xvecp}$ elementwise to obtain an array $h_l \braket{\xvec|\Phat|\xvecp}$.
Finally, one employs the \ScatterAdd{} routine to sum the values of $h_l \braket{\xvec|\Phat|\xvecp}$ corresponding to the same \XYvec{}.
During \ScatterAdd{} the array \xypaulisetxyptr{} serves as the \indexttt{} input argument.

\section{GPU implementation: prefix tree operations}\label{sec:prefix_tree_primitives}
\begin{figure*}	
	\begin{minipage}{0.325\linewidth}
		\vspace*{\fill}
		\vbox{
			\vfil
			\includegraphics{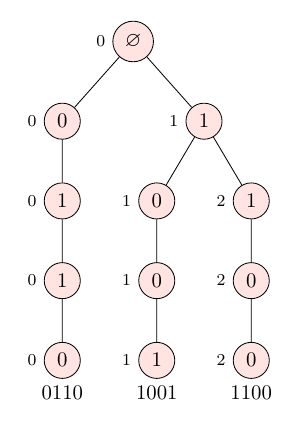}
			\vfil
		}
		\vspace*{\fill}
	\end{minipage}
	\begin{minipage}{0.6\linewidth}
		\vspace*{\fill}
		\vbox{
			\vfil
			\includegraphics{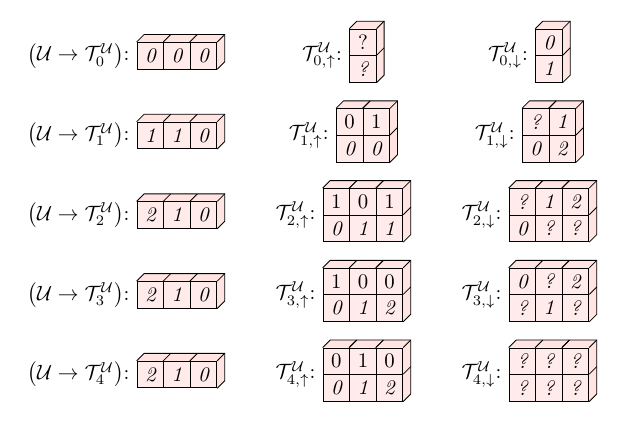}
			\vfil
		}
		\vspace*{\fill}
	\end{minipage}
	\caption{The prefix tree data structures for the batch of unique samples \Unique{}.}
	\label{fig:unique_data_structures}
\end{figure*}
\begin{figure*}	
	\begin{minipage}{0.325\linewidth}
		\vspace*{\fill}
		\vbox{
			\vfil
			\includegraphics{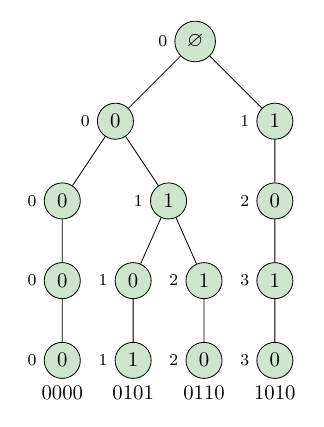}
			\vfil
		}
		\vspace*{\fill}
	\end{minipage}
	\begin{minipage}{0.6\linewidth}
		\vspace*{\fill}
		\vbox{
			\vfil
			\includegraphics{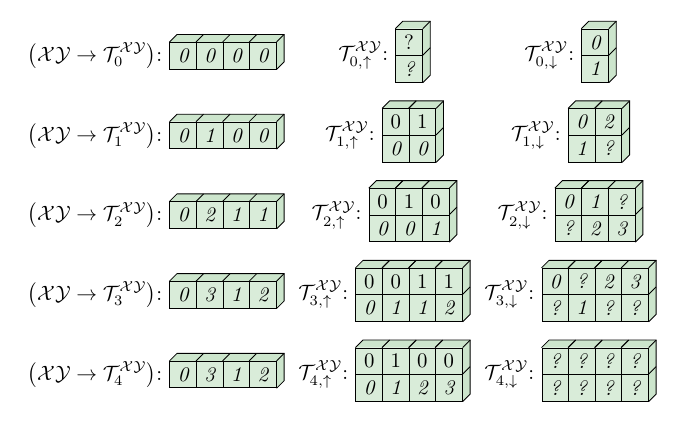}
			\vfil
		}
		\vspace*{\fill}
	\end{minipage}
	\caption{The prefix tree data structures for the set \UniqueXYSet{} of unique \XYvec{} strings representing Hamiltonian terms.}
	\label{fig:unique_xy_data_structures}
\end{figure*}
\subsection{Prefix tree data structures}
In our GPU implementation we store the prefix tree nodes as contiguous arrays, as opposed to storing them as individual data structures introduced in the main text.
As an example, let us consider the batch of unique samples \Unique{} and its prefix tree \UniqueTree{}.
For the $i$-th level of the prefix tree we store \emph{three} arrays denoted as \UniqueTreeArrOne{i}, \UniqueTreeArrUp{i} and \UniqueTreeArrDown{i} as depicted in Fig.~\ref{fig:unique_data_structures}.
\begin{enumerate}
	\item The array \UniqueTreeArrOne{i} has a shape $[\Nunq{}, 1]$. Its $k$-th element points to a node at the $i$-th level of the prefix tree that corresponds to the $k$-th basis vector in \Unique.
    For example, at the level 2 of the prefix tree $\UniqueTreeArrOne{2}[0] = \it{2}$ since the first two symbols of $\xvec_0$ are $11$, and this prefix path leads to the node 2 of level   2.
    At the same time, $\UniqueTreeArrOne{0}[k] = 0\ \forall k$ since \emph{all} vectors in \Unique{} start from an empty string $\varnothing$.
	
	\item The array \UniqueTreeArrUp{i} has a shape $\left[\ \left|\UniqueTreeLevel_i\right|, 2\right]$, where $\left|\UniqueTreeLevel_i\right|$ is the number of nodes at the $i$-th prefix tree level.
    It stores in a contiguous way the fields \Value{} and \Parent{} of the \textsc{Node} data structure introduced in the main text.
    Specifically, $\UniqueTreeArrUp{i}[l, 0]$ stores $x_i$, while $\UniqueTreeArrUp{i}[l, 1]$ keeps the pointer to the parent node at the previous level.
    For example, the bit value of node 2 at level 2 is equal to 1, and thus $\UniqueTreeArrUp{2}[2, 0] = 1$.
    At the same time, its parent is node 1 at level 1, and thus $\UniqueTreeArrUp{2}[2, 1] = \it 1$.
    Let us note that $\UniqueTreeArrUp{0}$ always has a shape $[1, 2]$ and we fill it with \textsc{NaN}s, since this node does not have a definite value and does not have a parent node.
	
    \item The array \UniqueTreeArrDown{i} has a shape $\left[\ \left|\UniqueTreeLevel_i\right|, 2\right]$ too. 
    It stores in a contiguous way the \Children{} field of the \textsc{Node} data structure.
    In particular, $\UniqueTreeArrDown{i}[l, 0]$ stores the pointer to a child with the value of 0 of the node $l$.
    Similarly, $\UniqueTreeArrDown{i}[l, 1]$ stores the pointer to a child with the value 1.
    In our diagrams these children nodes are to the left and to the right of the parent node correspondingly.
    If a node does not have a child, we store \textsc{NaN} instead of a pointer.
    For example, the node 0 at level 2 does not have a child with the value 0, and thus  $\UniqueTreeArrDown{i}[2, 0] = \it{?}$.
    At the same time, the same node \emph{does} have a child with the bit value 1, which is a node 0 at level 3.
    Thus, $\UniqueTreeArrDown{i}[2, 0] = \it{0}$.
    At the last prefix tree level nodes do not have children and thus \UniqueTreeArrDown{N} is a redundant array filled with \textsc{NaN}s.	
\end{enumerate}
For the sake of completeness, in Fig.~\ref{fig:unique_xy_data_structures} we provide the prefix tree data structures for the set \UniqueXYSet{}.

\subsection{\ConstructPrefixTree{} implementation}
\begin{figure*}	
	\centering
	\includegraphics{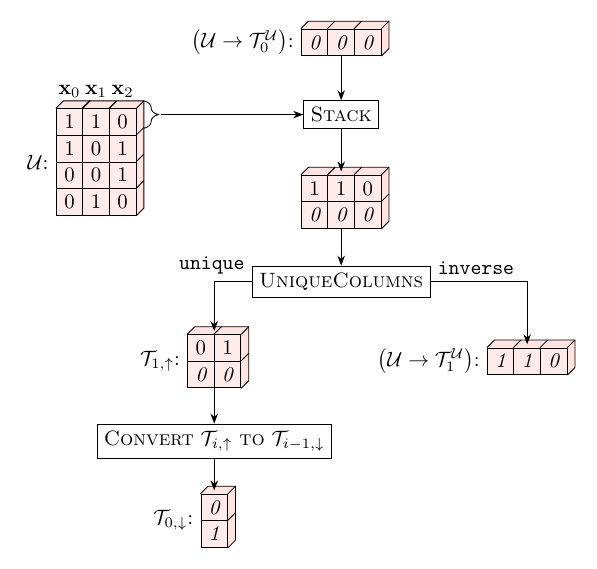}
	\caption{An iteration of the \ConstructPrefixTree{} algorithm.}
	\label{fig:construct_iteration}
\end{figure*}
The idea of \ConstructPrefixTree{} is to obtain the relevant prefix tree structures level by level starting from \UniqueTreeArrOne{0} and \UniqueTreeArrUp{0}.
We illustrate how the arrays of level 1 can be obtained from the arrays of level 0 in Fig.~\ref{fig:construct_iteration}.
More generally, the same procedure is applicable to obtain the arrays of level $i$ from the arrays of level $i-1$.
\begin{enumerate}
	\item First, one stacks the array containing $i$-th bits of all vectors in \Unique{} on top of the array \UniqueTreeArrOne{i-1} to form an array of shape $[\Nunq, 2]$.
	This array essentially contains all \emph{non-unique} prefixes at $i$-th level: the first row stores node values, while the second row contains pointers to the parents.
	\item Two non-unique prefixes that have the same bit value and a parent are considered identical and should be merged into a single prefix. 
	Thus, one feeds non-unique prefixes into a \UniqueFunc{} function which produces two output arrays. 
	The first output is a 2D array of shape $\left[\ \left|\UniqueTreeLevel_i\right|, 2\right]$ containing only unique prefixes; by inspection it can be seen that this array is equivalent \UniqueTreeArrUp{i}.
	The second output is an array containing pointers from non-unique prefixes to the unique ones.
	Since each non-unique prefix is associated with a basis vector in \Unique{}, this array represents \UniqueTreeArrOne{i}.
	The \UniqueFunc{} is our custom lower-level function and we cover it in Section~\ref{sec:aux_functions}.
	\item Finally, one uses the obtained arrays to fill the array \UniqueTreeArrDown{i - 1} at the \emph{previous} level. 
	First, one creates an array filled with \textsc{NaN}s of the same shape as \UniqueTreeArrUp{i - 1}. Then, one scatters into it the values from array \UniqueTreeArrDown{i} according to the following rule: 
	\begin{equation}
		\UniqueTreeArrDown{i-1}[\UniqueTreeArrUp{i}[l, 1], \UniqueTreeArrUp{i}[l, 0]] = l\ \forall l \in 1..\left|\UniqueTreeLevel_i\right|.
	\end{equation}
	In other words, for each node in $\UniqueTreeLevel_i$ we find its parent node $\UniqueTreeArrUp{i}[l, 1]$ and indicate that this node has a child $\it l$ having the bit value $\UniqueTreeArrUp{i}[l, 0]$
\end{enumerate}

\subsection{\CoupleViaPrefixTree{} implementation}
\begin{figure*}	
	\centering
	\includegraphics{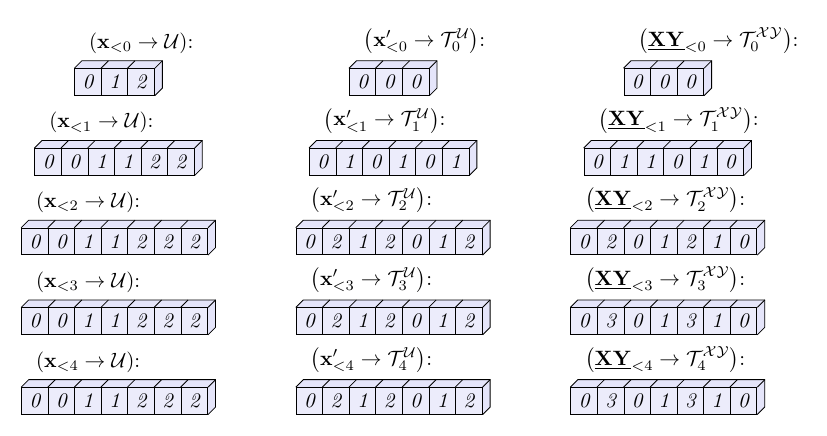}
	\caption{Arrays obtained during one full run of \CoupleViaPrefixTree{}.}
	\label{fig:loop_over_trie_data_structures}
\end{figure*}
The \CoupleViaPrefixTree{} algorithm described in Methods builds coupled pairs $(\xvec, \xvecp)$ bit by bit.
For a given \xvec{} and prefix tree level $i$ it calculates all acceptable prefixes $\xvecp_{<i}$ and the corresponding coupling $\XYvec_{<i}$.
The idea of our GPU implementation is to store the \emph{pointers} to valid $\xvecp_{<i}$ and $\XYvec_{<i}$ in an array for each \xvec{} and update them level by level.
In addition, we concatenate all such pointer arrays to vectorise the algorithm across all \xvec{} in \Unique{}.

Specifically, for each prefix tree level we obtain three arrays shown in Fig.~\ref{fig:loop_over_trie_data_structures}:
\begin{enumerate}
	\item The array \LoopTrieArrX{i} serves the vectorisation purpose and indicates chunks of other arrays which correspond to different \xvec{} being input	to \ConstructPrefixTree{}.
	For example, at the level 0 we start with empty prefixes $\varnothing$ corresponding to every vector in \Unique{} and thus $\LoopTrieArrX{0} = [\it 0, \it 1, \it 2]$.
	\item The array \LoopTrieArrXp{i} stores pointers to valid $\xvecp_{<i}$ prefixes for every \xvec{} in \Unique{}.
	For example, at level 0 any vector in \Unique{} can be coupled only to a single prefix $\varnothing$ and thus $\LoopTrieArrXp{0} = [\it 0, \it 0, \it 0]$.
	
	\item The array \LoopTrieArrXY{i} stores pointers to valid $\XYvec_{<i}$ prefixes in a way similar to \LoopTrieArrXp{i}.
\end{enumerate}

\begin{figure*}	
	\centering
	\includegraphics{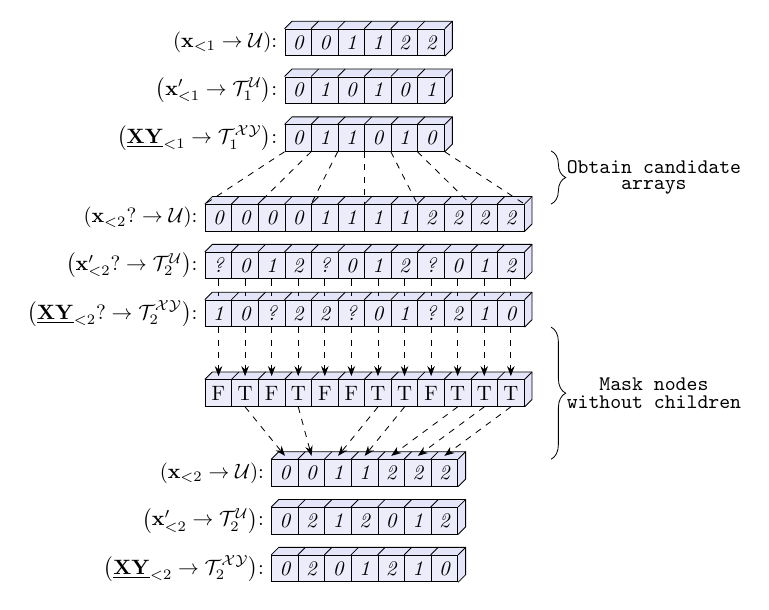}
	\caption{An iteration of \CoupleViaPrefixTree{} GPU implementation.}
	\label{fig:loop_over_trie_iteration}
\end{figure*}

A single iteration of the algorithm which produces level 2 arrays given the arrays at level 1 is shown in Fig.~\ref{fig:loop_over_trie_iteration}.
More generally, obtaining arrays for level $i$ from the arrays of level $i-1$ involves the following steps.
\begin{enumerate}
	\item First, we substitute every pointer to a valid prefix $\xvecp_{<i-1}$ ($\XYvec_{<i-1}$) with the pointers to its possible children (including \textsc{NaN}s for non-existing children).
	This produces the arrays of candidate pointers \LoopTrieArrXpCand{i} and \LoopTrieArrXYCand{i} twice the length of the initial arrays. 
	Specifically, entries of the array \LoopTrieArrXpCand{i} are as follows:	
	\begin{equation}
		\begin{gathered}
			\LoopTrieArrXp{i}[2k] = \UniqueTreeArrDown{i-1}[k, 0]; \\
			\LoopTrieArrXp{i}[2k + 1] = \UniqueTreeArrDown{i-1}[k, 1];
		\end{gathered}
	\end{equation}
	In other words, the $k$-th valid prefix pointer in the array \LoopTrieArrXp{i-1} is substituted with the pointers to its left and right children at the positions $2k$ and $2k+1$ respectively.
	The array \LoopTrieArrXYCand{i} is obtained in a similar way, however, with a subtlety: for a given $\XYvec_{<i-1}$ the order of its children in the array \LoopTrieArrXYCand{i} depends on the $i$-th bits of \emph{both} $\xvec{}$ and $\xvecp{}$. 
	For example, $\LoopTrieArrXp{i}[2k]$ corresponds to moving to the \emph{left} child of $\xvecp_{<i-1}$, i.e. appending 0 to it.
	However, if $x_i = 1$, then $XY_i = x_i \XORop 0 = 1$, and thus $\LoopTrieArrXYCand{i}[2k]$ should contain the pointer to the \emph{right} child of $\LoopTrieArrXY{i-1}[k]$.
	This is expressed with the following equations:
	\begin{equation}
		\begin{gathered}
			\LoopTrieArrXY{i}[2k] = \UniqueXYTreeArrDown{i-1}[k, 0 \XORop \xvec_k[i]]; \\
			\LoopTrieArrXY{i}[2k + 1] = \UniqueXYTreeArrDown{i-1}[k, 1 \XORop \xvec_k[i]];
		\end{gathered}
	\end{equation}
	
	Finally, to keep track of the vectorised \xvec, we double each element in $\LoopTrieArrX{i-1}$:
	\begin{equation}
		\LoopTrieArrX{i}[2k] = \LoopTrieArrX{i}[2k + 1] = \LoopTrieArrX{i-1}[k]
	\end{equation}
	The new arrays can be obtained using a combination of \textsc{Tile}, \textsc{Gather} and \textsc{Reshape} PyTorch primitives and we refer the reader to our code for more details~\cite{anqs_qchem_github_repo}.
	
	\item Once the arrays are produced, we calculate a boolean mask which is equal to \textsc{True} only if child pointers are not \textsc{\textsc{NaN}} for \emph{both} \LoopTrieArrXp{i} and \LoopTrieArrXY{i}.
	
	\item Finally, we use this boolean mask to obtain arrays of valid prefixes for the current level $i$. 
\end{enumerate}
The array $\LoopTrieArrX{N}$ obtained at the last level of \CoupleViaPrefixTree{} is clearly equivalent to the array $\pointer{\xvecp}{\Unique}$ produced by \FindSampledAndCoupled{} procedure.
The arrays $\pointer{\xvecp}{\Unique}$ and $\pointer{\XYvec}{\UniqueXYSet}$ can be obtained from $\LoopTrieArrXp{N}$ and $\LoopTrieArrXY{N}$ by using the correspondence between the indices of the nodes at the last level of a prefix tree, and their indices in the initial bit vector set.
As described above, this correspondence is expressed with the arrays \UniqueTreeArrOne{N} and \UniqueXYTreeArrOne{N} obtained during \ConstructPrefixTree{}.

\section{GPU implementation: auxiliary functions}\label{sec:aux_functions}
\subsection{FindAInB implementation}
\begin{figure*}	
	\centering
	\includegraphics{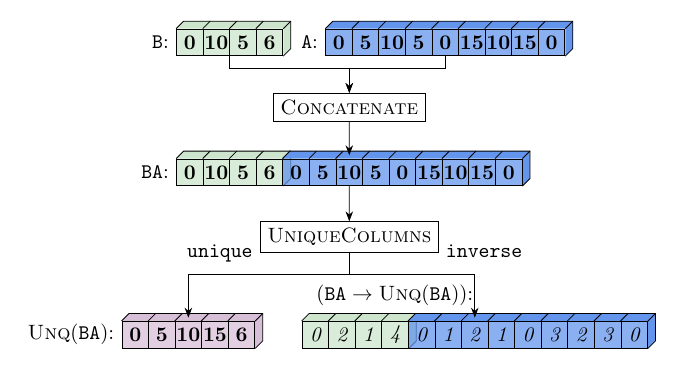}
	\caption{The first stage of \FindAInB. One concatenates arrays \Attt{} and \Bttt{} to find unique elements in their union. As well, one obtains the positions of those unique elements in the initial arrays.}
	\label{fig:cat_and_unique}
\end{figure*}
As mentioned in Section~\ref{sec:all_to_all_implementation} \FindAInB{} is a function which takes two arrays \Attt{} and \Bttt{} as input and produces two arrays $\Attt \in \Bttt$ and $\pointer{(\Attt \cap \Bttt)}{\Bttt}$ as output.
Crucially, the array \Bttt{} should not contain any repeated values.
We split its implementation into three stages.

\subsubsection*{Stage 1: Find unique elements in \Attt{} and \Bttt{} union}
We concatenate the arrays \Attt{} and \Bttt{} into a single array and apply our custom function \UniqueFunc{} to the concatenated array. 
We use \UniqueFunc{} instead of default PyTorch \textsc{Unique} function because, in general, when several integers are used per \xvec{}, the arrays \Attt{} and \Bttt{} will be two-dimensional, with integer values for each \xvec{} stored in columns.
Since the default PyTorch \textsc{Unique} function works only with 1D arrays, we have to employ our custom \UniqueFunc{} function explained further in this section.

As depicted in Fig.~\ref{fig:cat_and_unique}, \UniqueFunc{} returns two arrays. The first array $\textsc{Unq}(\Bttt\Attt)$ contains unique elements found in the union of \Attt{} and \Bttt{} arrays.
The second array $\pointer{\Bttt\Attt}{\textsc{Unq}(\Bttt\Attt)}$ can be considered as a concatenation of two arrays $\pointer{\Bttt}{\textsc{Unq}(\Bttt\Attt)}$ and $\pointer{\Attt}{\textsc{Unq}(\Bttt\Attt)}$, each containing pointers from elements in \Bttt{} and \Attt{} to the found unique values $\textsc{Unq}(\Bttt\Attt)$.

\subsubsection*{Stage 2: Finding the reverse pointers to \Bttt{}}
\begin{figure*}	
\begin{minipage}{0.45\linewidth}
	\vspace*{\fill}
	\vbox{
		\vfil
		\includegraphics{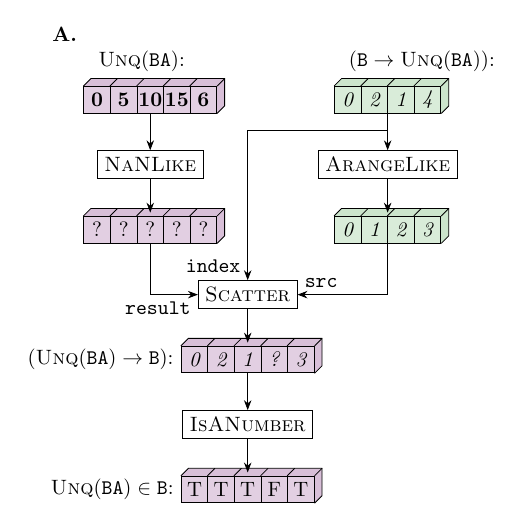}
		\vfil
	}
	\vspace*{\fill}
\end{minipage}
\begin{minipage}{0.45\linewidth}
	\vspace*{\fill}
	\vbox{
		\vfil
		\includegraphics{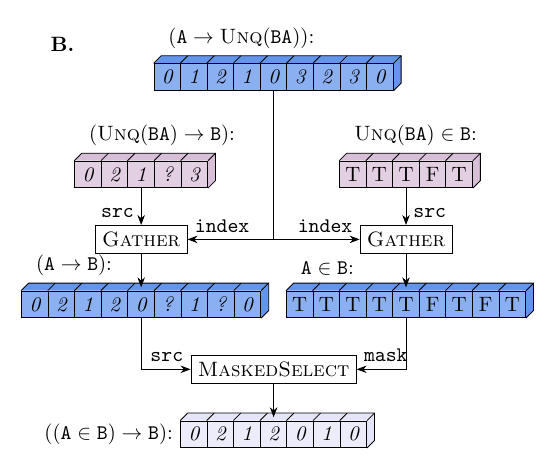}
		\vfil
	}
	\vspace*{\fill}
\end{minipage}
\caption{The two last stages of \FindAInB. \textbf{A.} Having the pointers from \Bttt{} to $\textsc{Unq}(\Bttt \Attt)$, one employs \textsc{Scatter} to find the \emph{reverse} pointers from  $\textsc{Unq}(\Bttt \Attt)$ to \Bttt{}. As well, one obtains a boolean mask indicating whether an element in $\textsc{Unq}(\Bttt \Attt)$ can be found in \Bttt. \textsc{B.} One uses \Gather{} to fetch the pointers from $\textsc{Unq}(\Bttt \Attt)$ to \Bttt{} for every element in \Attt{} and thus obtains the pointers $\pointer{\Attt}{\Bttt}$. In addition, one obtains a boolean mask indicating whether an element in \Attt{} can be found in \Bttt{}.}
\label{fig:get_b_pointers}
\end{figure*}

The purpose of this stage is to find the reverse pointers from $\textsc{Unq}(\Bttt\Attt)$ to \Bttt{}.
To that end, we create an array $\pointer{\textsc{Unq}(\Bttt\Attt)}{\Bttt}$ of the same shape as $\textsc{Unq}(\Bttt\Attt)$ filled with \textsc{NaN}s.
Then, we \textsc{Scatter} into this array \emph{indices} of elements in \Bttt{}.
For example, the second element in the $\pointer{\Bttt}{\textsc{Unq}(\Bttt\Attt)}$ array is $\textit{2}$, which means that $\Bttt[1]$ is the \emph{third} element of $\textsc{Unq}(\Bttt\Attt)$.
Thus, we scatter $\textit{2}$ to the third position in the array $\pointer{\textsc{Unq}(\Bttt\Attt)}{\Bttt}$ as depicted in Fig.~\ref{fig:get_b_pointers}.
If, after this operation, some values are still equal to \textsc{NaN}, it means that there are values in  $\textsc{Unq}(\Bttt\Attt)$ that can be found in \Attt{} but not in \Bttt{}.
Hence, one can apply a standard \textsc{IsANumber} function to $\pointer{\textsc{Unq}(\Bttt\Attt)}{\Bttt}$ and obtain a boolean mask indicating which elements of $\textsc{Unq}(\Bttt\Attt)$ belong to \Bttt{}.
 
\subsubsection*{Stage 3: Finding pointers from \Attt{} to \Bttt{}}
At the final stage we \Gather{} pointers $\pointer{\textsc{Unq}(\Bttt \Attt)}{\Bttt}$ using the array $\pointer{\Attt}{\textsc{Unq}(\Bttt\Attt)}$ as \indexttt.
As shown in Fig.~\ref{fig:get_b_pointers}B, we thus obtain an array $\pointer{\Attt}{\Bttt}$ with pointers from \Attt{} to \Bttt{} (if some element in $\pointer{\Attt}{\textsc{Unq}(\Bttt\Attt)}$ is \textsc{NaN}, it means that the corresponding element in \Attt{} does not belong to \Bttt).
In a similar way, we can \Gather{} boolean mask $\AInB{\textsc{Unq}(\Bttt\Attt)}{\Bttt}$ and obtain $\AInB{\Attt}{\Bttt}$.
Finally, we apply the boolean mask $\AInB{\Attt}{\Bttt}$ to $\pointer{\Attt}{\Bttt}$ to keep only the valid pointers to \Bttt{}, as required by the use of \FindAInB{} in \FindSampledAndCoupled{}.

\subsubsection*{Computational complexity}
The dominating cost of \FindAInB{} is finding unique columns with \UniqueFunc{}.
As discussed further in Section~\ref{sec:unique_func_implementation} it of complexity $\mathcal{O}\left(\left(\left|\Attt\right| + \left|\Bttt\right|\right) \log\left(\left|\Attt\right| + \left|\Bttt\right|\right)\right)$, which matches the figures given in Methods.
 
\subsection{ExpandPointers implementation}
\begin{figure*}	
	\centering
	\includegraphics{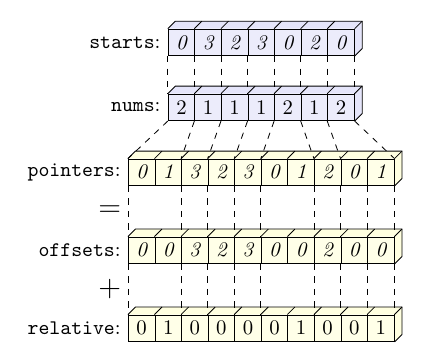}
	\caption{An example of the \textsc{ExpandPointers} routine. A $\texttt{starts}[i]$ pointer is substituted with a set of $\texttt{nums}[i]$ consecutive (i.e $\texttt{starts}[i], \texttt{starts}[i] + 1, \ldots, \texttt{starts}[i] + \texttt{nums}[i] - 1$). For the sake of simplicity, the final array of pointers is represented as a sum of two arrays \texttt{offsets} and \texttt{saw}.}
	\label{fig:expand_pointers}
\end{figure*}
\begin{figure*}		
	\begin{minipage}{0.45\linewidth}
		\vspace*{\fill}
		\vbox{
			\vfil
			\includegraphics{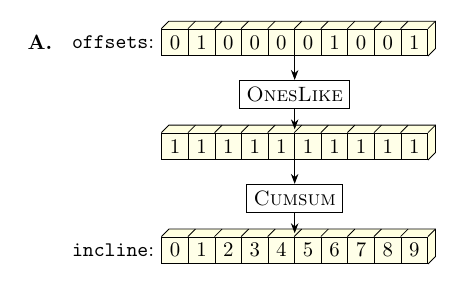}
			\vfil
		}
		\vspace*{\fill}
	\end{minipage}
	\begin{minipage}{0.45\linewidth}
		\vspace*{\fill}
		\vbox{
			\vfil
			\includegraphics{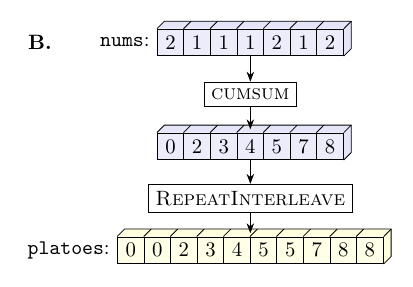}
			\vfil
		}
		\vspace*{\fill}
	\end{minipage}
	\caption{\textbf{A.} Calculation of the \texttt{incline} array. \textbf{B.} Calculation of the \texttt{platoes} array.}
	\label{fig:incline_and_platoes}
\end{figure*}

The \ExpandPointers{} procedure serves the following purpose.
Suppose one has an array \memoryttt{} whose elements are grouped into contiguous chunks of various length.
This is akin to such Hamiltonian data arrays as $h_l$, $\Yvecl$ and $\YZvecl$ which can be grouped according to their \XYvec{}.
\ExpandPointers{} is a function which allows one to retrieve those contiguous chunks of \memoryttt{} by specifying their starting positions \startsttt{} and their lengths \numsttt{}.
As output \ExpandPointers{} produces an array \pointersttt{} which replaces each starting pointer $\startsttt[i]$  with a \emph{sequence} of $\numsttt[i]$ consecutive pointers to each element in the corresponding contiguous chunk.
An example of the \ExpandPointers{} action is given in Fig.~\ref{fig:expand_pointers}.
We split its implementation into the following steps:
\begin{figure*}	
	\begin{minipage}{0.45\linewidth}
		\vspace*{\fill}
		\vbox{
			\vfil
			\includegraphics{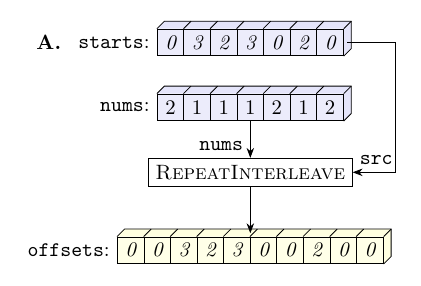}
			\vfil
		}
		\vspace*{\fill}
	\end{minipage}
	\begin{minipage}{0.45\linewidth}
		\vspace*{\fill}
		\vbox{
			\vfil
			\includegraphics{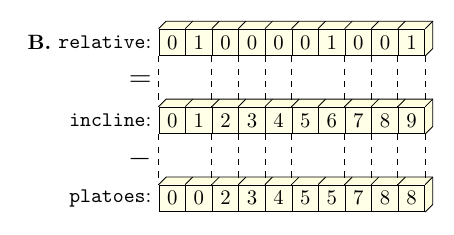}
			\vfil
		}
		\vspace*{\fill}
	\end{minipage}
	\caption{\textbf{A.} Calculation of \texttt{starts} array. \textbf{B.} Decomposition of \texttt{saw} array into a sum of \texttt{incline} and \texttt{platoes} arrays.}
	\label{fig:offsets_and_saw}
\end{figure*}
\begin{enumerate}
	\item We represent \pointersttt{} as a sum of two arrays: \offsetsttt{} and \sawttt{}.
	For each \emph{expanded} pointer in \pointersttt{}, the \offsetsttt{} array contains the starting position of the chunk to which the expanded pointer belongs to. As illustrated in Fig.~\ref{fig:offsets_and_saw}A it is trivially obtained with the   standard PyTorch routine \textsc{RepeatInterleave}: each element of $\startsttt[i]$ array is repeated $\numsttt[i]$ times.
	The \sawttt{} array contains the relative position of each extended pointer with respect to the chunk start. Obtaining \sawttt{} is slightly more involved and is performed in the next steps.
	\item  To calculate \sawttt{} we represent it as a difference of two arrays: \inclinettt{} and \platoesttt{}.
	The \inclinettt{} array is an arithmetic progression of integers with the step 1 starting from 0.
	This progression represents the fact that the neighbouring expanded pointers belonging to the same chunk should differ by one. 
	This array can be easily obtained using the standard \textsc{Cumsum} PyTorch function as shown in Fig.~\ref{fig:incline_and_platoes}A.
	\item However, \inclinettt{} does not account for the fact that this progression breaks down at the chunk borders, where increments should start from scratch. The array \platoesttt{} corrects for it by counting the number of increments that are accumulated by the start of each chunk and belong to the \emph{previous} chunks.
	The calculation of \platoesttt{} is shown in Fig.~\ref{fig:incline_and_platoes}B.
\end{enumerate}

\subsection{\UniqueFunc{} implementation}

As discussed, both \ConstructPrefixTree{} and \FindAInB{} require finding unique elements in an array, where each element is represented with more than one integer value.
Since the standard PyTorch \textsc{Unique} routine operates only with 1D arrays, we develop a custom function \UniqueFunc{} to find unique elements of a 2D array, where integer values representing each value are stored along the second dimension (i.e. in columns). 
Our implementation is based on the following idea: we devise an algorithm to assign to each column an integer label so that identical columns have identical labels.
Then, we apply conventional PyTorch \textsc{Unique} function to these labels and find unique among them.
Finally, we map the unique labels back to columns.
More specifically, let is consider an example of columns of size 2; its generalisation to larger column sizes is straightforward and can be found in our code.
\begin{enumerate}
	\item Suppose we are given an array \arrttt{} of shape $[L, 2]$. Let $\rowonettt{} \coloneqq \arrttt{}[:, 0]$ and $\rowtwottt{} \coloneqq \arrttt{}[:, 1]$ be its first and second row correspondingly.
    \item We apply the PyTorch \textsc{Unique} function to \rowonettt{} and obtain an array of its unique values $\unqrowonettt \coloneqq  \textsc{Unique}(\rowonettt{})$. We also obtain an array of pointers from \rowonettt{} to  \unqrowonettt{}, which we refer to as $\unqrowoneinvttt \coloneqq \pointer{\rowonettt}{\unqrowonettt}$.
	We apply the PyTorch \textsc{Unique} function to \rowtwottt{} to obtain the analogously defined \unqrowtwottt{} and \unqrowtwoinvttt{}.
	\item We note that there are $\left|\unqrowonettt\right| \cdot \left|\unqrowtwottt\right|$ possible pairs of unique elements in the first and second rows, where $\left|\cdot\right|$ denotes the array length.
	We assign to a pair $(\unqrowonettt[i], \unqrowtwottt[j])$ an integer label $\left|\unqrowtwottt\right| \cdot i  + j$.
	\item The arrays \unqrowoneinvttt{} and \unqrowtwoinvttt{} store indices $i$ and $j$ of the corresponding unique elements.
	Thus, for non-unique columns of \arrttt{} we calculate the array of labels as follows: $\labelsttt \coloneqq \left|\unqrowtwottt\right| \cdot \unqrowoneinvttt + \unqrowtwoinvttt{}$.
	\item We employ the PyTorch \textsc{Unique} function to obtain unique labels \unqlabelsttt{}, as well as the array \unqlabelsinvttt{} of pointers from \labelsttt{} to \unqlabelsttt{}. 
	Since there is a one-to-one correspondence between labels and \arrttt{} columns, \unqlabelsinvttt{} is actually an array of pointers from \arrttt{} to its unique columns: $\unqlabelsinvttt \equiv \pointer{\arrttt{}}{\textsc{UniqueColumns}(\arrttt)}$.
	\item Finally, we obtain the unique columns themselves.
	To that end, for each unique label we calculate the indices $i$ and $j$ introduced above.
	This is achieved as follows: $\texttt{i} \coloneqq \unqlabelsttt{}\ \texttt{//}\ \left|\unqrowtwottt\right|$; $\texttt{j} \coloneqq \unqlabelsttt{}\ \mod  \left|\unqrowtwottt\right|$,
	where $\texttt{//}$ denotes integer division.
	Having \texttt{i} and \texttt{j} for each unique column, it is straightforward to obtain the columns by applying the \Gather{} operation to the arrays \unqrowonettt{} and \unqrowtwottt{} and stacking the results.
\end{enumerate}
\subsubsection*{Computational complexity}
The dominating cost of \UniqueFunc{} corresponds to the repeated calls of the \textsc{Unique} PyTorch routine, which finds the unique elements of an array by sorting it with $\mathcal{O}\left(L \log{ L}\right)$ complexity.

\end{document}